\documentclass[prx,aps,twocolumn,groupedaddress]{revtex4-2}

\usepackage[pdftex]{graphicx}
\usepackage{amsbsy,amssymb,amsmath,bm,mathtools}

\usepackage{bm}
\usepackage{color}
 
\usepackage[mathscr]{euscript}

\usepackage{url}
\usepackage{placeins}
\usepackage[colorlinks=true,linkcolor=blue,citecolor=blue]{hyperref}

\begin{document}


\title{Adiabatic Variational Principle: \\
Born--Oppenheimer Theory for Emergent Electronic-Order Dynamics}

\author{Yang Yang}

\author{Ali Rayat}

\author{Ho Jang}

\author{Yunhao Fan}

\author{Sankha Subhra Bakshi}

\author{Gia-Wei Chern}
\affiliation{Department of Physics, University of Virginia, Charlottesville, VA 22904, USA}

\date{\today}

\begin{abstract}
Recent advances in microscopic many-body methods have enabled increasingly accurate descriptions of equilibrium quantum states, yet large-scale simulations of emergent-order dynamics typically rely on phenomenological equations with prescribed free-energy functionals. We introduce the Adiabatic Variational Principle (AVP), a framework for deriving order-parameter dynamics directly from an interacting many-body Hamiltonian. When microscopic relaxation is fast compared with order-parameter evolution, the many-body state follows the collective configuration adiabatically, in analogy with the Born--Oppenheimer separation of electronic and nuclear motion. This assumption makes explicit the adiabatic closure implicit in conventional phase-field descriptions. Each instantaneous order-parameter configuration then defines a constrained quasi-stationary many-body state, and the Lagrange multipliers enforcing these constraints directly provide the thermodynamic forces driving the collective evolution. Nonlinear Mori--Zwanzig projection theory further connects the reversible dynamics to the microscopic algebra of the collective variables and relates dissipation and fluctuations to the eliminated degrees of freedom. Applications to charge- and spin-density-wave dynamics recover Allen--Cahn domain growth, resolve evolving domain networks together with their microscopic electronic structure, and yield generalized Landau--Lifshitz dynamics for itinerant magnetism. The AVP thus provides a solver-independent microscopic foundation for phase-field-like simulations of the nonequilibrium dynamics of emergent electronic order.
\end{abstract}

\maketitle

\section{Introduction}

\label{sec:intro}

Collective electronic order driven by interactions among itinerant electrons is a central organizing principle of correlated quantum matter. Prominent examples include charge-density waves, magnetism, and superconductivity, which arise across a wide range of quantum materials, from transition-metal oxides and heavy-fermion compounds to moir\'e heterostructures~\cite{Imada1998,Dagotto2005,Keimer2015,Stewart1984,Cao2018,Balents2020,Mak2022}. Over the past several decades, substantial progress has been made in understanding the equilibrium properties of such phases, including how competing interactions, dimensionality, and electronic structure stabilize distinct ordered states and phase transitions~\cite{Kivelson2003,Fradkin2015}. A broad array of many-body techniques---including quantum Monte Carlo, dynamical mean-field and quantum-embedding methods, and tensor-network approaches---has provided increasingly powerful microscopic descriptions of ground-state and finite-temperature properties of interacting electron systems~\cite{Foulkes2001,Georges1996,Kotliar2006,Schollwock2011,Orus2014,Prokofev2008,Bartlett2007,Metzner2012}.

Extending microscopic many-body methods to excitation and nonequilibrium dynamics is considerably more demanding, despite substantial progress with a variety of time-dependent approaches~\cite{Aoki2014,White2004,Haegeman2011,Paeckel2019,Becca2017}. An additional difficulty arises for symmetry-breaking phases, where the relevant order parameters are not independent classical fields but emergent collective coordinates of the underlying many-body state. Their evolution is therefore intrinsically coupled to the microscopic electronic structure that stabilizes the ordered phase and is itself modified as the order develops. Capturing this mutual dependence between collective order and microscopic electronic degrees of freedom is a central problem in the theory of nonequilibrium correlated systems.

\begin{figure*}[t]
\centering
\includegraphics[width=1.7\columnwidth]{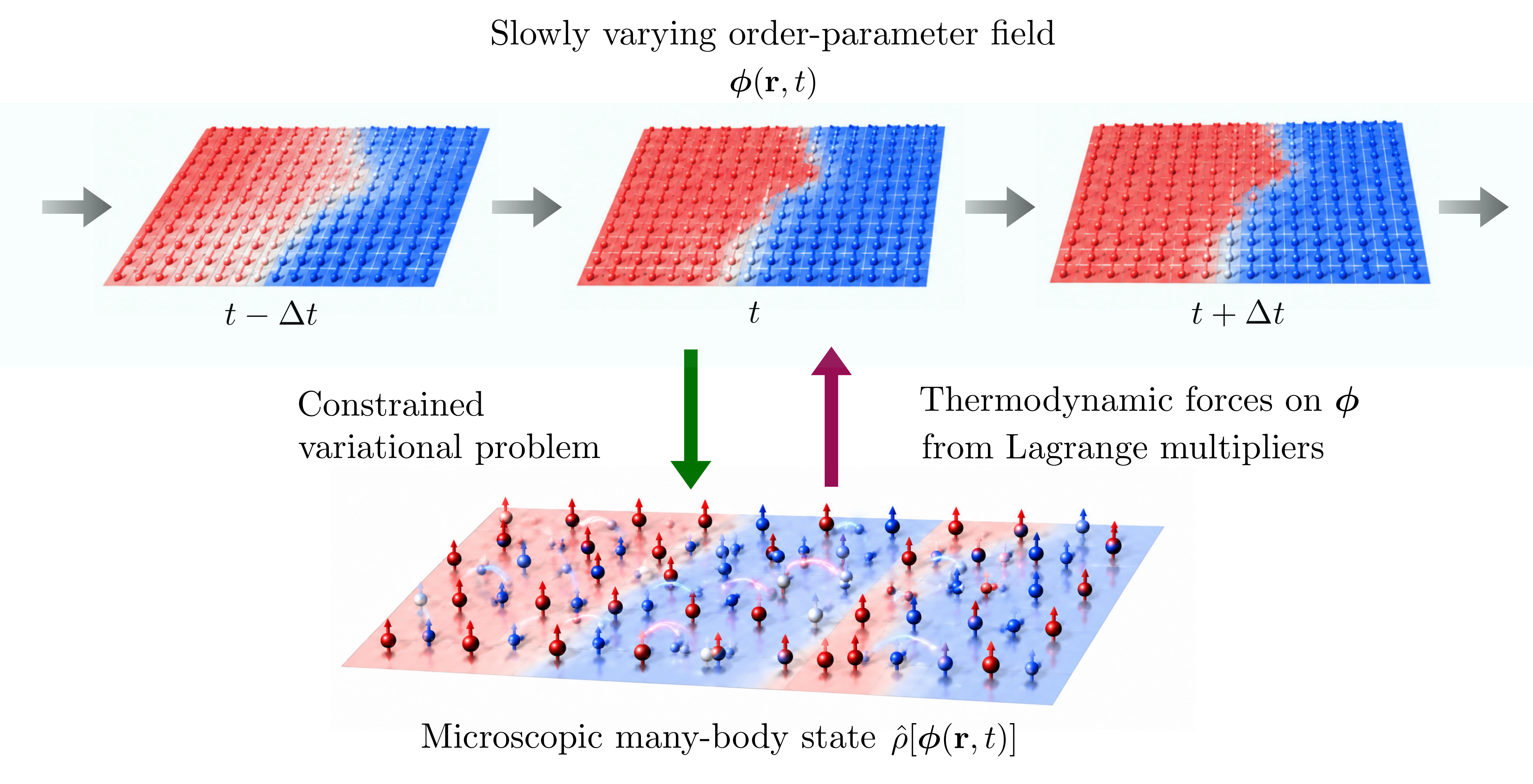}
\caption{
Schematic of the Adiabatic Variational Principle (AVP). In analogy with the Born--Oppenheimer separation of slow and fast degrees of freedom, a slowly evolving order-parameter field $\phi(\mathbf r,t)$ defines, at each instant, a constrained microscopic many-body state $\hat{\rho}[\phi]$. Here, however, the slow variable is an emergent collective observable rather than an independent microscopic coordinate. The associated Lagrange multipliers provide the thermodynamic forces that drive the collective evolution, while the microscopic state follows the changing order-parameter configuration adiabatically.
}
\label{fig:AVP_picture}
\end{figure*}

This difficulty is amplified in spatially inhomogeneous electronic states, where collective evolution unfolds over extended length and time scales and gives rise to complex real-space patterns and textures~\cite{Kivelson1998,Fradkin2010,Lang2002,Kohsaka2004,Gomes2007,Wise2009,Qazilbash2007,McLeod2017}. Spatially resolved probes such as scanning tunneling microscopy, spectroscopic imaging, and near-field optical spectroscopy have revealed such structures on nanometer scales, including ordered domains separated by domain walls and other defects~\cite{Yazdani2016,Liu2017}. Their formation and relaxation naturally connect the dynamics of electronic order to the broader problem of \emph{phase-ordering kinetics}, encompassing domain growth, the development of spatial correlations, and the relaxation of order-parameter textures following a nonequilibrium perturbation~\cite{Lifshitz61,Wagner61,Bray1994,Puri2009,Onuki2002,Cugliandolo15}. For electronic phases, however, these evolving patterns remain embedded in a microscopic many-body state, so that domain growth, defect motion, and changes in the local amplitude or orientation of the order parameter can be accompanied by substantial reconstruction of the underlying electronic structure.

At a coarse-grained level, the dynamics of textured electronic states is often described using continuum theories such as time-dependent Ginzburg--Landau and related phase-field approaches~\cite{Hohenberg1977,CahnHilliard1958,Chen2002,Boettinger2002}. These frameworks treat the order parameter as an effective classical field evolving under a prescribed free-energy functional and kinetic coefficients, and provide a powerful description of universal aspects of domain growth and defect motion. Although microscopic calculations can determine these parameters in simplified settings or controlled expansion regimes, such descriptions generally do not determine the underlying many-body state for each instantaneous order-parameter texture. Establishing such a connection would enable phase-field-like simulations in which the electronic structure and thermodynamic driving forces are determined directly from the microscopic interacting model.

In this work, we address this challenge by introducing the Adiabatic Variational Principle (AVP), a microscopic framework for the dynamics of emergent electronic order. The construction is motivated by an adiabatic separation of time scales analogous to the Born--Oppenheimer picture~\cite{BornOppenheimer1927,BornHuang1954,Marx2009}, but with a crucial distinction: the slow variables are emergent collective observables rather than independent microscopic coordinates. The central physical picture is illustrated schematically in Fig.~\ref{fig:AVP_picture}. For each instantaneous order-parameter configuration, the AVP associates a constrained many-body state and its conjugate thermodynamic forces, thereby extending equilibrium many-body methods to spatially inhomogeneous collective dynamics.  Nonlinear projection theory provides the corresponding dynamical foundation~\cite{Zwanzig1960,Mori1965,Grabert1978}, relating reversible evolution to the microscopic commutator structure of the collective observables and dissipation to the relaxation of the remaining microscopic degrees of freedom. The AVP thus establishes a microscopic bridge between interacting many-body theory and phase-field-like simulations, without requiring an empirical free-energy landscape or a phenomenological prescription for the reversible dynamics.

Variational methods are central to modern many-body physics, providing tractable representations of interacting quantum states within manifolds tailored to the problem at hand~\cite{RingSchuck1980,BlaizotRipka1986,Gutzwiller1963,Schollwock2011,Carleo2017}. Beyond their widespread use in equilibrium and ground-state calculations, this strategy also extends naturally to quantum dynamics. A prominent example for isolated systems is the time-dependent variational principle (TDVP)~\cite{McLachlan1964,KramerSaraceno1981,Haegeman2011,Haegeman2016,Paeckel2019}, in which the many-body wavefunction is restricted to a chosen manifold---such as Slater determinants, tensor-network states, or neural-network quantum states---and the Schr\"odinger equation is projected onto its tangent space. When the ansatz permits broken symmetry, electronic order parameters can naturally emerge as collective coordinates characterizing the evolving state. The resulting projected equations then provide a microscopic description of their dynamics within the chosen manifold.

Similar ideas extend to open quantum systems by combining Lindblad master equations~\cite{Gorini1976,Lindblad1976} with TDVP or related variational methods. The density matrix is represented within a restricted manifold---for example, using matrix-product operators or neural-network density operators---and evolved by projecting the Lindblad equation onto its tangent space, while steady states can be obtained through variational optimization~\cite{Weimer2015,Nagy2019,Hartmann2019,Vicentini2019}. These approaches have enabled studies of dissipative many-body dynamics, nonequilibrium steady states, and dissipation-driven phase transitions. Accurately representing the evolving density matrix nevertheless remains computationally demanding for large, spatially inhomogeneous systems, with the growth of operator-space entanglement posing a particular challenge for tensor-network methods.

The AVP exploits an emergent separation of time scales between collective order and the remaining microscopic degrees of freedom. Such a hierarchy may arise from relaxation induced by a thermal environment, intrinsic scattering and dephasing of microscopic excitations, or the slow evolution of collective modes. When the microscopic sector relaxes sufficiently rapidly toward a quasi-stationary state compatible with the instantaneous order-parameter configuration, this state can be determined variationally subject to that configuration as a constraint. The resulting constrained states define a manifold on which the slow collective variables evolve.

This construction differs fundamentally from TDVP--Lindblad approaches, which seek to approximate the real-time evolution of the many-body density matrix itself. The computational cost of following that evolution limits the spatial and temporal scales accessible for collective pattern formation and defect dynamics. By determining the electronic state conditionally on the evolving collective configuration, the AVP provides a route for adapting equilibrium many-body methods to these processes. This requires extending the underlying solvers to impose collective constraints and extract their conjugate forces, with an implementation that depends on the chosen many-body representation. Such constrained calculations retain the microscopic electronic response without explicitly resolving the fast electronic evolution.

The adiabatic premise of this reduction is not unique to the AVP. Widely used phenomenological theories of order-parameter dynamics, including time-dependent Ginzburg--Landau, Cahn--Hilliard, Landau--Lifshitz--Gilbert, and phase-field models~\cite{Hohenberg1977,CahnHilliard1958,Chen2002,Boettinger2002}, likewise close the evolution on a restricted set of slow variables without retaining independent dynamics for the microscopic degrees of freedom that have been eliminated. In their Markovian forms, these theories implicitly rely on a Born--Oppenheimer-like separation of time scales, in which the microscopic sector adjusts sufficiently rapidly to the instantaneous configuration of the slow variables. The AVP adopts the same physical premise, but replaces the phenomenological closure by an explicit microscopic construction of the constrained many-body state and its thermodynamic forces. In this sense, it provides a Born--Oppenheimer theory of emergent electronic order.

An important feature of the AVP is that its construction is not tied to a particular approximation for the underlying many-body problem. The constrained microscopic state and its conjugate thermodynamic forces may be obtained using a broad range of many-body solvers, from mean-field and quasiparticle approaches to tensor-network, Monte Carlo, and neural-network methods. This solver-independent structure allows the microscopic description to be systematically improved without altering the form of the reduced collective dynamics. The computational cost nevertheless depends on the chosen solver and can remain substantial.

As a proof of principle, we formulate the constrained many-body problem within a mean-field approximation and study the nonequilibrium dynamics of charge- and spin-density-wave order in interacting lattice models. Even at this level, the calculation is computationally demanding: the constrained mean-field state must be determined iteratively at every dynamical time step, requiring repeated solution of a large quadratic single-particle problem, for example through explicit diagonalization. Nevertheless, GPU acceleration enables us to follow charge-ordering dynamics on lattices containing roughly 4,000 sites, resolving extended domain patterns together with their thermodynamic forces and microscopic electronic structure. The applications illustrate domain growth and defect dynamics in charge ordering, as well as the rapid formation of local magnetic moments followed by slower development of antiferromagnetic orientational correlations.

The remainder of this paper is organized as follows. In Sec.~\ref{sec:AVP}, we introduce the general AVP construction and formulate the constrained many-body state and its conjugate thermodynamic forces. In Sec.~\ref{sec:nonlinear-MZ}, we derive the dynamical structure of the AVP using nonlinear projection theory, including its reversible and dissipative components. Section~\ref{sec:MF-AVP} develops the mean-field AVP (MF-AVP). In Sec.~\ref{sec:CDW}, we apply the resulting MF-AVP to charge-density-wave (CDW) ordering in the spinless $t$--$V$ model. Section~\ref{sec:Hubbard_SDW} considers spin-density-wave (SDW) dynamics in the Hubbard model. We conclude in Sec.~\ref{sec:discussion} with a discussion of the broader implications, limitations, and future directions. Technical details of the numerical implementations and classical reference models are provided in the Appendixes.

\section{Adiabatic Variational Principle for Emergent Order Dynamics}

\label{sec:AVP}

The central assumption underlying the AVP is a hierarchy of time scales between slow collective variables and rapidly relaxing microscopic degrees of freedom. At each instant, the microscopic electronic state is regarded as quasi-stationary and parametrized by the instantaneous configuration of the retained slow variables. For a prescribed configuration, the remaining microscopic degrees of freedom relax before the collective configuration changes appreciably. Denoting the microscopic relaxation and collective evolution time scales by $\tau_{\rm qp}$ and $\tau_\phi$, respectively, the adiabatic regime requires~$\tau_{\rm qp}\ll\tau_\phi$.

The choice of slow variables follows the general principles established by Hohenberg and Halperin~\cite{Hohenberg1977}. In addition to the order-parameter fields $\bm\phi(\mathbf r,t)$, conserved densities and other long-lived modes must be retained whenever their characteristic time scales are comparable to or longer than that of the order-parameter dynamics. Such modes cannot be assumed to relax instantaneously to values determined by the order parameters alone. A consistent reduced description therefore requires that the retained variables exhaust the relevant slow sector.

For the open electronic systems considered here, coupling to a thermal environment, such as phonons or other dissipative baths, provides a natural mechanism for microscopic relaxation; intrinsic scattering and dephasing may also contribute. The microscopic state thus adjusts to the slowly evolving collective configuration, following the Born--Oppenheimer-like picture introduced above. This conditional relaxation does not require the collective configuration itself to be in equilibrium: domains and textures can continue to evolve while the microscopic degrees of freedom remain approximately relaxed subject to the instantaneous constraints.

We formulate this construction as the Adiabatic Variational Principle (AVP). Consider a general interacting lattice Hamiltonian $\hat{\mathcal H}$ coupled to a thermal bath. We introduce a set of local order-parameter fields $\bm\phi(\mathbf r,t)=(\phi_1,\phi_2,\ldots,\phi_M)$ through the expectation values of local operators $\{\hat A_i^\alpha\}$,
\begin{equation}
    \label{eq:phi-constraint1}
    \phi_i^\alpha = \langle \hat A_i^\alpha \rangle.
\end{equation}
The AVP itself does not require these operators to form a closed algebra. For the class of order parameters considered below, however, it is convenient to assume that
\begin{equation}
    \label{eq:commutator_A}
    [\hat A_i^\alpha,\hat A_j^\beta] = i\,\delta_{ij}\sum_\gamma C_{\alpha\beta\gamma}\hat A_i^\gamma ,
\end{equation}
where $C_{\alpha\beta\gamma}$ are the corresponding structure constants. Closure of the operator algebra allows the reversible dynamics to be expressed directly in terms of the commutation relations of the retained variables.

The description of slow collective dynamics has a long history in nonequilibrium thermodynamics. Relaxational dynamics goes back to Landau and Khalatnikov~\cite{Landau1954}, while Onsager's theory established the linear-response relations governing dissipative couplings~\cite{Onsager1931a,Onsager1931b}. Subsequent developments, including the classification of dynamical universality classes~\cite{Hohenberg1977} and unified thermodynamic formulations such as the general equation for the nonequilibrium reversible--irreversible coupling (GENERIC)~\cite{Grmela1997,Ottinger1997}, organize the dynamics into reversible and dissipative sectors.

Eliminating microscopic degrees of freedom generally produces memory effects in both sectors. For the formulation and applications in this section, we adopt the Markovian limit, in which the evolution is local in time. To leading order in the thermodynamic forces, the dynamics then takes the familiar reversible--dissipative form~\cite{Hohenberg1977,Grmela1997,Ottinger1997},
\begin{eqnarray}
    \label{eq:dphi_dt1}
    \frac{d \phi_i^\alpha}{dt} = \sum_{j,\beta} \mathcal{R}_{ij}^{\alpha\beta} \frac{\partial F}{\partial \phi_j^\beta}
    - \sum_{j,\beta} \Gamma_{ij}^{\alpha\beta} \frac{\partial F}{\partial \phi_j^\beta}
    + \eta_i^\alpha(t).
\end{eqnarray}
Here $F[\bm\phi]$ is the effective free-energy functional of the collective variables. The reversible tensor is antisymmetric, $\mathcal{R}_{ij}^{\alpha\beta}=-\mathcal{R}_{ji}^{\beta\alpha}$, whereas the dissipative tensor is symmetric and positive semidefinite, $\Gamma_{ij}^{\alpha\beta}=\Gamma_{ji}^{\beta\alpha}$. Consequently, the reversible dynamics leaves $F$ unchanged, while the deterministic dissipative contribution drives relaxation toward lower free energy, with
\[
\frac{dF}{dt}
= -\sum_{i,\alpha,j,\beta}
\frac{\partial F}{\partial\phi_i^\alpha}
\Gamma_{ij}^{\alpha\beta}
\frac{\partial F}{\partial\phi_j^\beta}
\leq 0
\]
in the absence of fluctuations. The fluctuating force $\eta_i^\alpha(t)$ represents the stochastic influence of degrees of freedom outside the retained slow sector. In thermal equilibrium, its statistics and the dissipative tensor $\Gamma$ are related by the fluctuation--dissipation relation.

Within the AVP, the reversible and dissipative tensors have distinct microscopic origins. For operators satisfying Eq.~(\ref{eq:commutator_A}), $\mathcal R$ follows from their commutation relations and acquires a Poisson-bracket-like structure. The tensor $\Gamma$ describes relaxation mediated by the eliminated degrees of freedom, including microscopic electronic excitations and the external environment. In the Markovian limit, their influence is represented by kinetic coefficients whose microscopic form depends on the dynamics and couplings of this complementary sector.

Equations of this form also underlie time-dependent Ginzburg--Landau descriptions of phase-ordering dynamics and, more broadly, phase-field approaches to mesoscale ordering and microstructure evolution~\cite{Bray1994,Chen2002,Boettinger2002}. In these theories, the thermodynamics is encoded in a coarse-grained continuum free-energy functional of the order-parameter fields and their spatial variations. Schematically, one writes
\begin{equation}
	F_{\rm GL}[\bm\phi]=\int d\mathbf r\left[V(\bm\phi)+\frac{1}{2}\sum_{\alpha\beta,\mu\nu}K_{\alpha\beta}^{\mu\nu}(\bm\phi)\,\partial_\mu\phi^\alpha\,\partial_\nu\phi^\beta+\cdots\right],
\end{equation}
where the local potential $V(\bm\phi)$ and gradient terms are constructed from symmetry-allowed invariants. The forces $-\delta F_{\rm GL}/\delta\phi^\alpha$, together with appropriate kinetic coefficients, generate the collective dynamics. This framework has been highly successful in describing domains, interfaces, competing phases, and microstructure on mesoscopic scales.

The coefficients of $F_{\rm GL}$ may be specified phenomenologically, fitted to experiment, or derived from microscopic models. Microscopic derivations provide a valuable connection to the underlying Hamiltonian, but generally rely on approximations valid in particular regimes. In particular, low-order amplitude and gradient expansions may become inadequate for sharp interfaces, defect cores, or textures that substantially reconstruct the electronic state. Extending the description to such configurations therefore calls for a way to retain the microscopic electronic response without relying on a prescribed low-order continuum functional.

The AVP provides such a construction within the adiabatic regime. Starting from a specified interacting lattice Hamiltonian, it associates each instantaneous configuration $\bm\phi$ with a constrained many-body state and thereby defines a microscopic free-energy functional $F_{\rm AVP}[\bm\phi]$. The functional and its conjugate forces are obtained from this constrained problem without requiring an amplitude or gradient expansion. This offers a route to phase-field-like simulations that retain the electronic response to evolving textures, with an accuracy that can in principle be systematically improved through the underlying many-body solver.

The connection to microscopic dynamics extends beyond the free-energy functional. In Sec.~\ref{sec:nonlinear-MZ}, we embed the AVP within nonlinear projection-operator theory, which yields the organized reversible motion, memory-dependent dissipation, and fluctuations of the retained variables. This provides a microscopic statistical-mechanical foundation for Eq.~(\ref{eq:dphi_dt1}) and clarifies the conditions under which its Markovian form emerges.

\subsection{Variational Formulation in Terms of Many-Body Wave Functions}

Constrained variational constructions are well established in many-body physics, including constrained Hartree--Fock and density-functional approaches. Here we use this established construction as the microscopic foundation for a description of spatially inhomogeneous, nonequilibrium order-parameter dynamics. Although the general AVP is formulated in terms of density matrices, we begin with pure many-body states $|\Psi\rangle$, which provide a familiar setting for introducing its variational foundation. For a prescribed order-parameter configuration, constrained energy minimization defines the instantaneous adiabatic state and makes the relation between the constraint fields and thermodynamic forces explicit, without specifying a particular variational ansatz. This wave-function formulation serves as a pedagogical and formal starting point; it does not assume that the electronic subsystem remains pure during its actual evolution. Coupling to an environment generally produces a mixed reduced state, even for a zero-temperature bath, motivating the density-matrix formulation developed in the following subsection.

Within this simpler wave-function setting, the central construction is to regard the order-parameter configuration $\bm\phi$ as specifying a set of macroscopic constraints on the underlying many-body state. For each~$\bm\phi$, we seek a normalized state $|\Psi\rangle$ consistent with Eq.~(\ref{eq:phi-constraint1}), while all remaining microscopic degrees of freedom are variationally optimized. Within this zero-temperature description, the adiabatic approximation identifies the corresponding instantaneous state with the minimum-energy state compatible with these constraints:
\begin{equation}
|\Psi[\bm\phi]\rangle = \operatorname*{argmin}_{\substack{|\Psi\rangle:\,\langle\Psi|\Psi\rangle=1\\ \langle\Psi|\hat{A}_i^\alpha|\Psi\rangle=\phi_i^\alpha}} \langle\Psi|\hat{\mathcal{H}}|\Psi\rangle.
\end{equation}
This constrained minimization can be implemented by introducing Lagrange multipliers $f_i^\alpha$ for the order-parameter constraints and $\Lambda$ for normalization, leading to the energy Lagrangian
\begin{equation}
\label{eq:functional1}
\begin{aligned}
&\mathcal{L}\!\left(|\Psi\rangle,\Lambda,\bm f;\bm\phi\right) = \langle\Psi|\hat{\mathcal{H}}|\Psi\rangle \\
&\quad + \sum_{i,\alpha} f_i^\alpha \left(\langle\Psi|\hat{A}_i^\alpha|\Psi\rangle-\phi_i^\alpha\right) + \Lambda\left(1-\langle\Psi|\Psi\rangle\right).
\end{aligned}
\end{equation}
The adiabatic state $|\Psi[\bm\phi]\rangle$, together with the conjugate fields $f_i^\alpha$, is determined by the stationarity conditions of~$\mathcal{L}$. Variation with respect to the Lagrange multipliers enforces $\langle\hat{A}_i^\alpha\rangle=\phi_i^\alpha$ and $\langle\Psi|\Psi\rangle=1$, while variation with respect to $|\Psi\rangle$ determines the optimal many-body state subject to these constraints. For unrestricted variations of the many-body state, the latter leads to the effective eigenvalue problem
\begin{equation}
\label{eq:eigen-eq1}
\hat{\mathcal{H}}^*[\bm f]|\Psi\rangle = \Lambda|\Psi\rangle,
\end{equation}
with the effective Hamiltonian
\begin{equation}
\label{eq:H_eff1}
\hat{\mathcal{H}}^*[\bm f] = \hat{\mathcal{H}} + \sum_{i,\alpha} f_i^\alpha \hat{A}_i^\alpha.
\end{equation}
Thus, the constrained adiabatic state satisfies a stationary eigenvalue equation for a Hamiltonian deformed by the conjugate fields $f_i^\alpha$, with the relevant solution selected by the original constrained minimization. Without the order-parameter constraints, these fields vanish and the construction reduces to the ordinary ground-state problem for $\hat{\mathcal{H}}$, with $\Lambda$ equal to the ground-state energy. For prescribed $\bm\phi$, the fields are determined self-consistently to enforce the corresponding expectation values, rather than imposed externally. As shown below, they also have a direct physical interpretation as the generalized thermodynamic forces conjugate to the order parameters.

In practice, solving the stationarity conditions depends on the chosen variational representation of $|\Psi\rangle$. Within a restricted variational class, stationarity is imposed with respect to the ansatz parameters, while the multipliers are adjusted until the constraints $\langle\hat{A}_i^\alpha\rangle=\phi_i^\alpha$ are satisfied. The constrained state and its conjugate fields are therefore determined together in the same optimization problem. Although the numerical procedure depends on the underlying many-body solver, the variational construction itself is broadly applicable across different many-body representations.

The microscopic free-energy functional introduced above as $F_{\rm AVP}[\bm\phi]$ is defined, within the present wave-function formulation, by the energy of the constrained adiabatic state,
\begin{equation}
F_{\rm AVP}[\bm\phi] = \langle\Psi[\bm\phi]|\hat{\mathcal{H}}|\Psi[\bm\phi]\rangle.
\end{equation}
For notational simplicity, we henceforth denote this functional simply by $F[\bm\phi]$. Thus, rather than postulating an effective functional of the order parameter, as in phenomenological Ginzburg--Landau theories, the AVP determines $F[\bm\phi]$ directly from the underlying microscopic Hamiltonian through the constrained many-body problem. The resulting energy landscape reflects the interactions and correlations captured by the chosen variational state, and its accuracy can in principle be systematically improved by refining the ansatz.

We next establish the physical meaning of the Lagrange multipliers $f_i^\alpha$. Consider neighboring optimized states along a smooth branch of the constrained manifold, $|\Psi[\bm\phi]\rangle \rightarrow |\Psi[\bm\phi+d\bm\phi]\rangle = |\Psi[\bm\phi]\rangle+|\delta\Psi\rangle$. Their expectation values differ by $\delta \langle\hat{A}_i^\alpha\rangle=\delta \phi_i^\alpha$. To apply the stationarity condition, we evaluate the variation of the energy Lagrangian with respect to the wave function while holding its explicit arguments $\bm\phi$, $\bm f$, and $\Lambda$ fixed at the original constrained solution. Equation~(\ref{eq:functional1}) then gives $0 = \delta_\Psi\mathcal{L} = \delta F + \sum_{i,\alpha} f_i^\alpha\,\delta \phi_i^\alpha$, where the normalization term vanishes because the neighboring states are normalized. It follows that
\begin{equation}
\label{eq:force1}
f_i^\alpha = -\frac{\partial F}{\partial\phi_i^\alpha}.
\end{equation}
Thus, the Lagrange multipliers $f_i^\alpha$ are precisely the generalized thermodynamic forces conjugate to the order parameters $\phi_i^\alpha$, quantifying the energetic cost of maintaining a prescribed order-parameter configuration. The AVP uses this variational relation to obtain the forces entering Eq.~(\ref{eq:dphi_dt1}) directly from the constrained many-body problem.

This identification also has an important practical consequence. The derivatives $\partial F/\partial\phi_i^\alpha$ need not be evaluated by numerical differentiation of the effective free-energy landscape, which would require repeated constrained calculations and can be susceptible to numerical noise. Instead, the driving forces are obtained directly from the Lagrange multipliers $f_i^\alpha$ already determined in the constrained optimization.

\subsection{Variational Formulation in Terms of Many-Body Density Matrix}

The wave-function formulation provides a transparent introduction to the constrained variational structure of the AVP, but the density matrix is the natural object for its general formulation. The rapid microscopic relaxation required by the adiabatic approximation, $\tau_{\rm qp}\ll\tau_\phi$, is generally facilitated by coupling to an external environment, so the reduced electronic state need not remain pure even when the environment is at zero temperature. The density-matrix formulation associates each instantaneous configuration of the slow variables with a quasi-stationary statistical state of the remaining microscopic degrees of freedom. This construction is distinct from solving for the full nonequilibrium evolution of the density matrix: the AVP determines the state conditioned on the instantaneous slow variables and, through it, the effective free energy and conjugate thermodynamic forces. We now establish the constrained statistical ensemble that implements this construction.

For a given order-parameter configuration $\bm\phi$, the adiabatic density matrix is obtained by minimizing the thermodynamic free energy over all microscopic states consistent with Eq.~(\ref{eq:phi-constraint1}). The microscopic degrees of freedom thus equilibrate subject to the instantaneous values of the slow collective variables, defining a manifold of constrained quasi-equilibrium states $\hat{\rho}[\bm\phi]$. At finite temperature, we therefore introduce the free-energy Lagrangian
\begin{equation}
\begin{aligned}
\mathcal{L}(\hat{\rho},\Lambda,\bm f;\bm\phi) &= \mathrm{Tr}(\hat{\rho}\,\hat{\mathcal{H}}) + T\,\mathrm{Tr}(\hat{\rho}\ln\hat{\rho}) \\
&\quad + \sum_{i,\alpha} f_i^\alpha\left[\mathrm{Tr}(\hat{\rho}\,\hat{A}_i^\alpha)-\phi_i^\alpha\right] + \Lambda\left[1-\mathrm{Tr}(\hat{\rho})\right],
\end{aligned}
\end{equation}
where $f_i^\alpha$ are Lagrange multipliers enforcing the order-parameter constraints, while $\Lambda$ enforces normalization. We use units in which $k_{\rm B}=1$. The first two terms constitute the Helmholtz free energy of the microscopic state. Explicitly, the constrained state is defined by
\begin{equation}
\hat{\rho}[\bm\phi] = \operatorname*{argmin}_{\substack{\hat{\rho}\geq 0,\;\mathrm{Tr}(\hat{\rho})=1\\ \mathrm{Tr}(\hat{\rho}\hat{A}_i^\alpha)=\phi_i^\alpha}} \left[\mathrm{Tr}(\hat{\rho}\,\hat{\mathcal{H}})+T\,\mathrm{Tr}(\hat{\rho}\ln\hat{\rho})\right].
\end{equation}
The auxiliary Lagrangian implements this constrained minimization through stationarity with respect to $\hat{\rho}$, $\Lambda$, and $\bm f$, providing the statistical-mechanical counterpart of the constrained ground-state construction introduced above.

Variation with respect to $\Lambda$ imposes $\mathrm{Tr}(\hat{\rho})=1$, while variation with respect to $f_i^\alpha$ gives
\begin{equation}
\mathrm{Tr}(\hat{\rho}\,\hat{A}_i^\alpha)=\phi_i^\alpha.
\end{equation}
Variation with respect to $\hat{\rho}$ then yields the constrained Gibbs state
\begin{equation}
\label{eq:constrained_Gibbs_rho}
\hat{\rho}[\bm\phi] = \frac{e^{-\beta\hat{\mathcal{H}}^*[\bm f]}}{Z[\bm f]}, \qquad Z[\bm f] = \mathrm{Tr}\,e^{-\beta\hat{\mathcal{H}}^*[\bm f]},
\end{equation}
where $\beta=1/T$ and $\hat{\mathcal{H}}^*[\bm f]$ is the effective Hamiltonian defined in Eq.~(\ref{eq:H_eff1}). The Lagrange multiplier $\Lambda$ is fixed by normalization, with $\beta\Lambda-1=-\ln Z[\bm f]$.

The constrained Gibbs construction gives the AVP adiabatic states a direct physical interpretation: the auxiliary Hamiltonian $\hat{\mathcal{H}}^*[\bm f]$ has a thermal equilibrium state that reproduces the instantaneous collective configuration $\bm\phi$. The fields $f_i^\alpha$ are determined self-consistently to enforce these values, rather than applied externally. Even a spatially inhomogeneous, globally nonequilibrium configuration can thus be represented by a constrained Gibbs ensemble, with the remaining microscopic degrees of freedom relaxed subject to the instantaneous slow variables. This conditional equilibration expresses the adiabatic separation underlying the AVP.

The effective free-energy functional $F[\bm\phi]$ is obtained by evaluating the free energy with the variationally optimized density matrix,
\begin{equation}
F[\bm\phi] = \mathrm{Tr}\bigl\{\hat{\rho}[\bm\phi]\,\hat{\mathcal{H}}\bigr\} + T\,\mathrm{Tr}\bigl\{\hat{\rho}[\bm\phi]\ln\hat{\rho}[\bm\phi]\bigr\}.
\end{equation}
It includes both energetic and entropic contributions and defines the thermodynamic landscape governing the slow collective variables.

The physical meaning of the constraint fields follows from the same variational argument used in the wave-function formulation. Consider an infinitesimal change of the order-parameter configuration, accompanied by the corresponding variation of the optimized density matrix along the constrained manifold. At fixed temperature, stationarity of the free-energy Lagrangian with respect to $\hat{\rho}$, with its explicit arguments $\bm\phi$, $\bm f$, and $\Lambda$ held fixed at the original solution, gives $0=\delta_{\rho}\mathcal{L}=\delta F+\sum_{i,\alpha}f_i^\alpha\,\delta \phi_i^\alpha$. Thus, also in the density-matrix formulation, the Lagrange multipliers introduced to enforce the order-parameter constraints are precisely the generalized thermodynamic forces conjugate to the slow variables, $f_i^\alpha=-\partial F/\partial\phi_i^\alpha$.

The relation between $\bm\phi$ and the constraint fields $\bm f$ can equivalently be understood in terms of a thermodynamic Legendre transformation. The partition function $Z[\bm f]$ defines the thermodynamic potential of the deformed Hamiltonian $\hat{\mathcal{H}}^*[\bm f]$,
\begin{equation}
G[\bm f] = -T\ln Z[\bm f], \qquad \frac{\partial G}{\partial f_i^\alpha} = \langle\hat{A}_i^\alpha\rangle = \phi_i^\alpha,
\end{equation}
with the fields $f_i^\alpha$ adjusted so that the corresponding expectation values reproduce the prescribed $\phi_i^\alpha$. Transforming from $\bm f$ to $\bm\phi$ gives
\begin{equation}
F[\bm\phi] = G[\bm f] - \sum_{i,\alpha} f_i^\alpha\phi_i^\alpha.
\end{equation}
At fixed temperature, $dG=\sum_{i,\alpha}\phi_i^\alpha\,df_i^\alpha$, and hence
\begin{equation}
dF = -\sum_{i,\alpha} f_i^\alpha\,d\phi_i^\alpha.
\end{equation}
This Legendre structure provides a complementary thermodynamic interpretation of Eq.~(\ref{eq:force1}) and makes clear that the constraint fields are not merely auxiliary variables, but the physical driving forces associated with variations of the order parameters. For unrestricted minimization over density matrices, the resulting functional is convex in the prescribed expectation values, since statistical mixtures remain admissible. This exact ensemble functional should be distinguished from a restricted variational landscape: the Gaussian manifold used in the MF-AVP applications below is not closed under statistical mixtures and can support nonconvex branches and barriers. The force relation remains valid along each smooth constrained variational branch.

As in the wave-function formulation, these forces are obtained directly from the Lagrange multipliers, without numerical differentiation of the free-energy landscape. Thus, for any instantaneous configuration $\bm\phi$, the AVP simultaneously provides the corresponding quasi-stationary microscopic state and the thermodynamic forces entering the collective dynamics. The constrained ensemble and its conjugate fields supply the statistical foundation for the reduced dynamical theory developed below.

\begin{figure}[t]
\centering
\includegraphics[width=0.95\columnwidth]{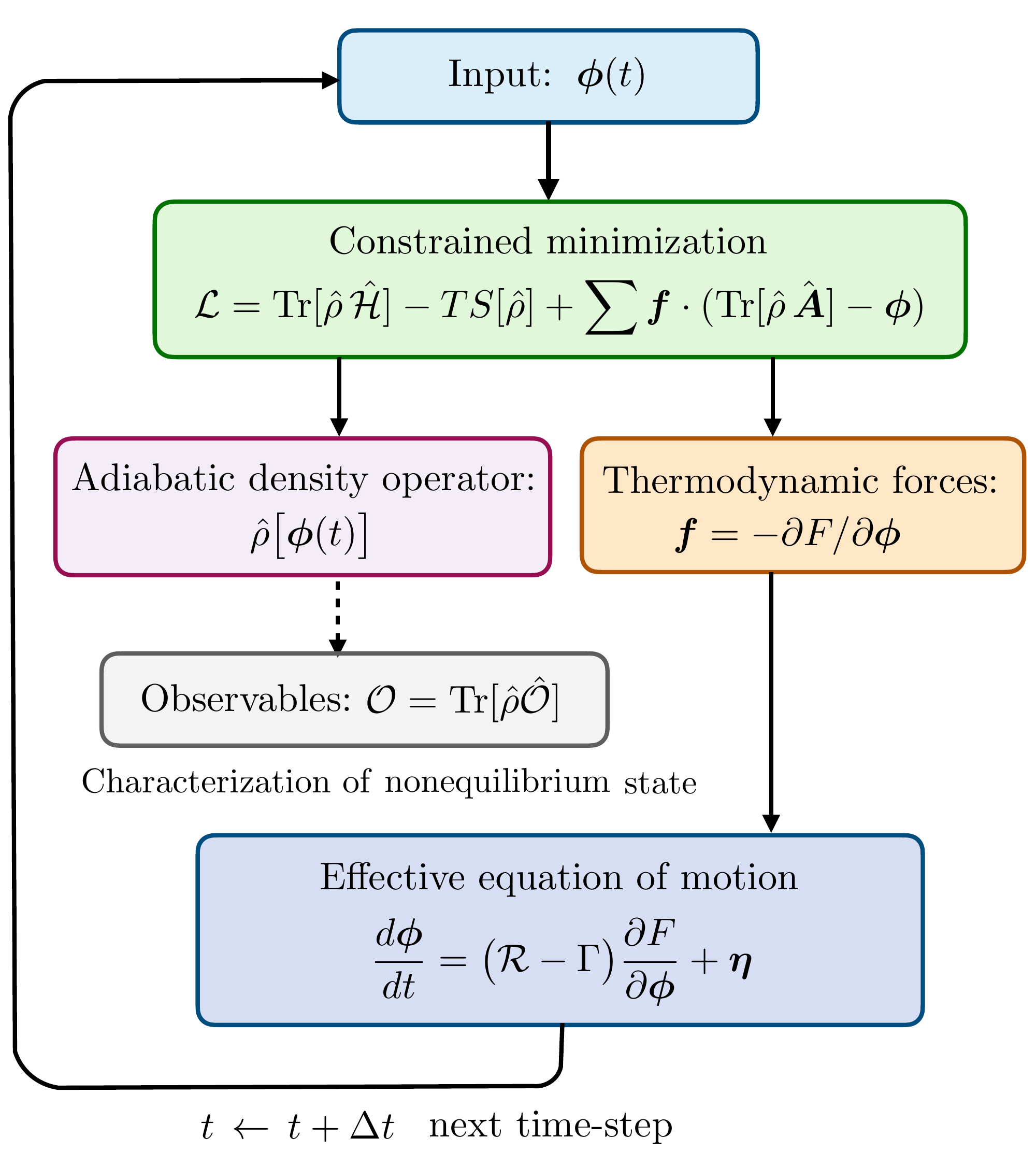}
\caption{Schematic illustration of adiabatic dynamics based on the AVP. For a given instantaneous configuration of the slow order parameters, the constrained variational problem determines the corresponding adiabatic microscopic state and the conjugate thermodynamic forces. The resulting state also provides the expectation values of microscopic observables characterizing the nonequilibrium system. The effective equation of motion, including both reversible and dissipative contributions, then advances the order parameters to the next time step. Repeating these two steps generates the coupled evolution of the slow collective variables and the microscopic many-body state.}
    \label{fig:AVP_scheme}
\end{figure}

\subsection{Adiabatic Dynamics Simulation Workflow}

Combining the above AVP construction with the effective equation of motion in Eq.~\eqref{eq:dphi_dt1} provides a practical framework for adiabatic dynamics simulations, as summarized schematically in Fig.~\ref{fig:AVP_scheme}. At a given time $t$, the instantaneous configuration $\boldsymbol{\phi}(t)$ is imposed as a set of constraints on the microscopic many-body state. The AVP then determines the corresponding quasi-stationary electronic state $\hat{\rho}[\boldsymbol{\phi}(t)]$ and its conjugate fields $\bm f(t)$. Other microscopic observables can also be evaluated in this state, allowing the electronic structure to be followed alongside the evolving order-parameter texture.

The conjugate forces are subsequently used in Eq.~\eqref{eq:dphi_dt1}, together with the reversible and dissipative tensors and the appropriate fluctuating forces, to propagate the slow variables over one time step. The resulting configuration $\boldsymbol{\phi}(t+\Delta t)$ defines a new constrained variational problem, and the electronic state and conjugate fields are recalculated. Repeating these steps generates the collective trajectory without explicitly propagating the fast microscopic electronic dynamics. This alternating procedure gives a practical realization of the Born--Oppenheimer-like picture: domains and textures evolve under forces determined by an electronic state that remains approximately relaxed subject to the instantaneous constraints. The microscopic dynamical foundation of this simulation scheme is developed in the next section.

\section{Order-parameter dynamics from nonlinear projection theory}
\label{sec:nonlinear-MZ}

We now turn to the microscopic foundation of the collective dynamics used in the AVP simulation scheme. Its Markovian form was introduced in Eq.~(\ref{eq:dphi_dt1}) based on thermodynamic, symmetry, and conservation-law considerations. Here we derive this dynamics using the nonlinear projection-operator formalism of Mori--Zwanzig and Grabert~\cite{Zwanzig1960,Mori1965,Robertson1966,KawasakiGunton1973,Grabert1978}. Within this framework, the constrained density matrix $\hat{\rho}[\bm\phi]$ is identified with the relevant statistical ensemble associated with the chosen slow variables, placing the AVP in a broader statistical-mechanical context.

The central idea of the Mori--Zwanzig projection formalism is to separate the microscopic description into a relevant sector associated with the selected collective variables and a complementary sector containing the remaining microscopic information. Introducing projection operators $\mathcal{P}$ and $\mathcal{Q}=1-\mathcal{P}$ onto these two sectors, respectively, the exact microscopic dynamics can be reorganized into a generalized Langevin equation for the retained variables. The eliminated degrees of freedom are not simply discarded: their influence reappears through memory-dependent and fluctuating contributions, while the evolution projected directly onto the relevant subspace describes the organized motion of the slow variables.

For adiabatic order-parameter dynamics, the relevant variables are the fields $\bm \phi(\mathbf r,t)$ introduced in Sec.~\ref{sec:AVP}, supplemented, when necessary, by conserved densities or other long-lived quantities required to exhaust the slow sector. Phase ordering generally explores spatially inhomogeneous and strongly nonequilibrium configurations, so the constrained ensemble associated with these variables changes along the collective trajectory. A fixed Mori projection constructed around an equilibrium reference state does not follow this evolving ensemble. This motivates a state-dependent projection $\mathcal{P}(t)=\mathcal{P}[\bm \phi(t)]$, which adapts to the instantaneous collective configuration and accommodates nonlinear order-parameter dynamics.

To formulate this construction, let $\{\hat{\mathcal A}_\mu\}$ denote a set of relevant observables, or macrovariables, whose expectation values $a_\mu(t)$ specify the instantaneous macroscopic state. The corresponding relevant ensemble is obtained by maximizing the entropy subject to these expectation values~\cite{Grabert1978,Meretukov2024,teVrugt2020},
\begin{equation}
	\hat{\rho}_{\rm rel}[a(t)] = \frac{1}{Z(t)} \exp\!\left[-\sum_\mu \lambda_\mu(t)\hat{\mathcal A}_\mu\right],
\end{equation}
where the conjugate fields $\lambda_\mu(t)$ are determined by the constraints $a_\mu(t)={\rm Tr}\{\hat{\rho}_{\rm rel}[a(t)]\hat{\mathcal A}_\mu\}$. At this stage, the density matrix and trace refer to the full closed microscopic system, including any environment incorporated into its Hamiltonian. The relevant ensemble reproduces the chosen macroscopic expectation values but need not coincide with the actual microscopic density matrix: it retains only the information specified by the macrovariables. The same maximum-entropy relevant ensemble underlies Zubarev's nonequilibrium statistical-operator approach, which constructs the nonequilibrium state from a weighted history of microscopically evolved relevant ensembles~\cite{Zubarev1981,Roepke2019}. This provides a closely related route to nonlinear transport equations; here we use the projection-operator formulation with the initial preparation specified below.

The projection constructed from this evolving ensemble yields equations for both the macroscopic means and their fluctuations. We summarize the essential results here; further details, including the fluctuation equation and the role of the initial-state assumption, are given in Appendix~\ref{app:nonlinear_projection}. For an initial state of constrained equilibrium, $\hat{\rho}(0)=\hat{\rho}_{\rm rel}[a(0)]$, the ensemble-averaged macrovariables obey an exact transport equation of the form~\cite{Grabert1978}
\begin{equation}
	\label{eq:da-dt}
	\frac{d a_\mu(t)}{dt} = v_\mu[a(t)] + \int_0^t ds\,\mathcal K_\mu[t,s;a],
\end{equation}
where $\mathcal K_\mu[t,s;a]$ is the memory contribution associated with dynamics in the complementary sector. It describes how the past collective evolution influences the rate of change of $a_\mu$ at time $t$; the notation $[t,s;a]$ indicates dependence on the two times and, in general, on the macroscopic trajectory $\{a_\nu(u):s\leq u\leq t\}$. The instantaneous organized drift is given by
\begin{equation}
	\label{eq:organized_v}
	v_\mu[a(t)] = {\rm Tr}\!\left\{\hat{\rho}_{\rm rel}[a(t)]\,\dot{\hat{\mathcal A}}_\mu\right\}.
\end{equation}
Here $\dot{\hat{\mathcal A}}_\mu$ denotes the Heisenberg time derivative of the corresponding microscopic observable under the full microscopic Hamiltonian. The organized drift is determined entirely by the instantaneous relevant state and describes reversible motion that preserves the entropy of the relevant ensemble.

The absence of an explicit random force in Eq.~(\ref{eq:da-dt}) does not imply that fluctuations are neglected. The initial-state assumption ensures that the fluctuating force has zero ensemble average, while its correlations remain nonzero. Fluctuations about the mean trajectory obey an accompanying generalized Langevin equation driven by this force~\cite{Grabert1978}. Its correlations are related to the dissipative transport kernels through a generalized fluctuation--dissipation relation, connecting relaxation of the mean variables with fluctuations about their evolution. Together, these equations provide the microscopic starting point for a stochastic description in which both the transport kernels and the fluctuating-force statistics depend on the evolving macroscopic state.

When correlations in the complementary sector decay rapidly compared with the evolution of the macrovariables, the memory integral can be approximated by a local-in-time contribution. The resulting Markovian transport equation retains coefficients that depend on the instantaneous macroscopic state. Together with the corresponding description of fluctuations, this reduction provides the microscopic basis for the local dynamical structure introduced in Eq.~(\ref{eq:dphi_dt1}). For the AVP, it is consistent with the assumed hierarchy $\tau_{\rm qp}\ll\tau_\phi$, provided the retained variables exhaust the relevant slow sector.

We now specialize this general framework to the AVP. We regard the interacting electronic system together with its environment as a closed microscopic system described by the total Hamiltonian
\begin{equation}
	\hat{\mathcal H}_{\rm tot}=\hat{\mathcal H}+\hat{\mathcal H}_{\rm B}+\hat{\mathcal H}_{\rm SB},
\end{equation}
where $\hat{\mathcal H}$ is the electronic Hamiltonian introduced in Sec.~\ref{sec:AVP}, $\hat{\mathcal H}_{\rm B}$ describes the bath, and $\hat{\mathcal H}_{\rm SB}$ denotes their coupling. The combined dynamics is unitary, while dissipation emerges upon eliminating electronic and environmental degrees of freedom outside the relevant sector.

The crucial step is the choice of macrovariables: we choose $\{\hat{\mathcal A}_\mu\}$ to consist of the system energy $\hat{\mathcal H}$, the bath energy $\hat{\mathcal H}_{\rm B}$, and the order-parameter operators $\{\hat A_i^\alpha\}$. Their expectation values are $E(t)=\langle\hat{\mathcal H}\rangle$, $E_{\rm B}(t)=\langle\hat{\mathcal H}_{\rm B}\rangle$, and $\phi_i^\alpha(t)=\langle\hat A_i^\alpha\rangle$, with conjugate fields $\beta(t)$, $\beta_{\rm B}(t)$, and $\lambda_i^\alpha(t)$, respectively. Retaining the system and bath energies separately allows energy exchange with the environment to be described alongside the order-parameter dynamics. With these constraints, the full relevant ensemble factorizes as $\hat{\rho}_{\rm rel}^{\rm tot}(t)=\hat{\rho}_{\rm rel}(t)\otimes\hat{\rho}_{{\rm B},{\rm rel}}(t)$, where $\hat{\rho}_{{\rm B},{\rm rel}}(t)=Z_{\rm B}^{-1}(t)e^{-\beta_{\rm B}(t)\hat{\mathcal H}_{\rm B}}$. In the reservoir limit, $\beta_{\rm B}$ is held fixed. The relevant density matrix of the electronic system is
\begin{equation}
	\label{eq:rho_rel}
	\hat{\rho}_{\rm rel}(t)=\frac{1}{Z(t)}
	\exp\!\left[-\beta(t)\hat{\mathcal H}
	-\sum_{i,\alpha}\lambda_i^\alpha(t)\hat A_i^\alpha\right],
\end{equation}
where $Z(t)$ normalizes the electronic trace. Introducing $f_i^\alpha=\lambda_i^\alpha/\beta$, this state assumes precisely the constrained Gibbs form in Eq.~\eqref{eq:constrained_Gibbs_rho}, obtained variationally in Sec.~\ref{sec:AVP}. The system's $\beta(t)$ may evolve as energy is exchanged with the bath; holding it fixed recovers the fixed-temperature density-matrix AVP.

\subsection{Energy Relaxation and the Thermal Adiabatic Limit}

We first consider the evolution of the electronic energy. Applying Eq.~(\ref{eq:da-dt}) to $\hat{\mathcal H}$ gives the mean transport equation $dE(t)/dt=v_E(t)+\int_0^t du\,\mathcal K_E(t,u)$, where $v_E$ is the organized drift and $\mathcal K_E$ is the energy memory contribution, with its trajectory dependence left implicit. Since $\hat{\mathcal H}$ commutes with both itself and $\hat{\mathcal H}_{\rm B}$, its microscopic time derivative arises entirely from the system--bath coupling. We assume that the system--bath coupling has zero average in the reference thermal bath state, ${\rm Tr}_{\rm B}\{\hat{\rho}_{{\rm B},{\rm rel}}\hat{\mathcal H}_{\rm SB}\}=0$, as occurs naturally for linear coupling to harmonic-bath displacements about equilibrium. This condition excludes an additional static bath-average field acting on the electrons, while retaining the fluctuations responsible for energy exchange and relaxation.  Together with the factorized relevant ensemble, the zero bath average implies $v_E=0$, leaving the memory term to govern mean energy relaxation. Appendix~\ref{app:bath_centering} derives this result and discusses its relation to the standard centering procedure.

The thermodynamic driving force follows from the entropy of the relevant ensemble. The electronic state in Eq.~(\ref{eq:rho_rel}) satisfies $dS=\beta\,dE+\sum_{i,\alpha}\beta f_i^\alpha\,d\phi_i^\alpha$. Neglecting changes in the system--bath interaction energy in the weak-coupling limit gives $dE_{\rm B}=-dE$. A reservoir at inverse temperature $\beta_{\rm B}=1/T_{\rm B}$ therefore contributes $dS_{\rm B}=-\beta_{\rm B}dE$, so that
\begin{equation}
	dS_{\rm tot}=(\beta-\beta_{\rm B})\,dE+\sum_{i,\alpha}\beta f_i^\alpha\,d\phi_i^\alpha.
	\label{eq:dStot}
\end{equation}
Here $S_{\rm tot}=S+S_{\rm B}$ is the entropy of the full relevant ensemble. The force conjugate to energy exchange at fixed order parameter is consequently $(\partial S_{\rm tot}/\partial E)_\phi=\beta(E,\phi)-\beta_{\rm B}$.

In the Markovian limit, the dissipative fluxes are linear in the entropy-conjugate forces to leading order, with kinetic coefficients evaluated at the instantaneous relevant state. Including fluctuations, and now interpreting $E(t)$ as a stochastic coarse-grained energy, we write
\begin{equation}
	\frac{dE}{dt}
	=L_{EE}\left[\beta(E,\phi)-\beta_{\rm B}\right]
	+\sum_{i,\alpha}L_{E,i\alpha}\,\beta f_i^\alpha+\eta_E(t).
	\label{eq:energy_relax}
\end{equation}
Here $L_{EE}>0$ characterizes direct thermal relaxation, while $L_{E,i\alpha}$ couples energy exchange to order-parameter relaxation. The zero-mean force $\eta_E(t)$ describes energy fluctuations; its correlations, including cross correlations with the order-parameter noise, are constrained by the fluctuation--dissipation relation. The diagonal term transfers energy from the hotter subsystem to the colder one, whereas the cross terms allow energy exchange driven by ordering even at $\beta=\beta_{\rm B}$. Nonnegative dissipative entropy production requires the full symmetric transport matrix to be positive semidefinite, rather than the energy contribution to be nonnegative separately.

For each fixed $\phi$, the condition $\beta(E_{\rm ad},\phi)=\beta_{\rm B}$ defines a thermal-balance energy $E_{\rm ad}(\phi,T_{\rm B})$. Writing $E=E_{\rm ad}+\delta E$ gives $\beta(E,\phi)-\beta_{\rm B}\simeq-\delta E/(T_{\rm B}^2C_\phi)$, where $C_\phi=(\partial E/\partial T)_\phi>0$ is evaluated at $T_{\rm B}$. The diagonal thermal relaxation scale is therefore
\begin{equation}
	\tau_E=\frac{T_{\rm B}^2 C_\phi}{L_{EE}}.
	\label{eq:tau_E}
\end{equation}
The thermal adiabatic reduction starts from the hierarchy
\begin{equation}
	\tau_E\ll\tau_\phi,
	\label{eq:adiabatic_condition}
\end{equation}
where $\tau_\phi$ characterizes the order-parameter evolution, while also accounting for the cross-coupled energy current.

For a prescribed slowly varying order-parameter trajectory, the deviation from the moving thermal-balance energy obeys
\begin{equation}
	\frac{d\delta E}{dt}
	\simeq-\frac{\delta E}{\tau_E}
	+J_{E\phi}-\frac{dE_{\rm ad}}{dt}+\eta_E(t),
\end{equation}
where $J_{E\phi}=\sum_{i,\alpha}L_{E,i\alpha}\beta f_i^\alpha$. After the initial thermal transient, averaging over fast fluctuations gives the leading systematic lag
\begin{equation}
	\overline{\delta E}(t)\simeq\tau_E\left[J_{E\phi}(t)-\frac{dE_{\rm ad}}{dt}\right],
\end{equation}
provided the remaining terms vary slowly on the scale $\tau_E$; the right-hand side is evaluated on the thermal-balance manifold. The mismatch between the two currents thus produces a lag proportional to $\tau_E$, without requiring their cancellation. When both currents remain on the collective scale, the associated temperature offset $\overline{\delta T}\simeq\overline{\delta E}/C_\phi$ is small.

To leading adiabatic order, the energy averaged over fast thermal fluctuations follows $E(t)\simeq E_{\rm ad}[\phi(t),T_{\rm B}]$ and no longer constitutes an independent slow variable. The constrained electronic state can therefore be evaluated at $\beta=\beta_{\rm B}$ for each instantaneous configuration. This fixes the temperature rather than the energy: the electronic energy changes as ordering proceeds, with a small systematic lag and thermal fluctuations about the evolving thermal-balance state.

\subsection{Effective Order-Parameter Dynamics}

We now turn to the order parameters. Applying Eq.~(\ref{eq:da-dt}) to the observables $\hat A_i^\alpha$ gives the mean transport equation
\begin{equation}
	\label{eq:dphi_dt2}
	\frac{d\phi_i^\alpha(t)}{dt} = v_i^\alpha(t)+\int_0^t du\,\mathcal K_i^\alpha(t,u),
\end{equation}
where $v_i^\alpha$ is the organized drift and $\mathcal K_i^\alpha$ is the corresponding memory contribution, with its trajectory dependence left implicit. Since $\hat A_i^\alpha$ commutes with $\hat{\mathcal H}_{\rm B}$, and the centered coupling has zero organized contribution under the factorized relevant ensemble, the drift reduces to
\begin{equation}
	\label{eq:organized_v1}
	v_i^\alpha(t)=i\,{\rm Tr}\{\hat{\rho}_{\rm rel}(t)[\hat{\mathcal H},\hat A_i^\alpha]\},
\end{equation}
where the trace is over the electronic Hilbert space; see Appendix~\ref{app:bath_centering}. The bath continues to influence the dynamics through memory and fluctuations.

With $f_i^\alpha=\lambda_i^\alpha/\beta$, the electronic relevant state can be written as $\hat{\rho}_{\rm rel}\propto\exp[-\beta\hat{\mathcal H}^*[f]]$, where $\hat{\mathcal H}^*[f]$ is the constrained Hamiltonian introduced in Eq.~(\ref{eq:H_eff1}). Since $\hat{\rho}_{\rm rel}$ commutes with $\hat{\mathcal H}^*$ and $\hat{\mathcal H}=\hat{\mathcal H}^*-\sum_{j,\beta}f_j^\beta\hat A_j^\beta$, we obtain
\begin{eqnarray}
	\label{eq:organized_v2}
	v_i^\alpha & = & i\,{\rm Tr}\!\left\{\hat{\rho}_{\rm rel}\left[\hat{\mathcal H}^*-\sum_{j,\beta}f_j^\beta\hat A_j^\beta,\hat A_i^\alpha\right]\right\} \nonumber\\
	& = & i\sum_{j,\beta}f_j^\beta\,{\rm Tr}\!\left\{\hat{\rho}_{\rm rel}[\hat A_i^\alpha,\hat A_j^\beta]\right\}.
\end{eqnarray}
Using the closed operator algebra in Eq.~\eqref{eq:commutator_A}, this becomes
\begin{equation}
	\label{eq:organized_v3}
	v_i^\alpha=-\sum_{\beta,\gamma} C_{\alpha\beta\gamma}\,\phi_i^\gamma f_i^\beta.
\end{equation}
Together with $f_i^\alpha=-\partial F/\partial\phi_i^\alpha$ from Eq.~\eqref{eq:force1}, this identifies the organized drift with the reversible contribution in Eq.~\eqref{eq:dphi_dt1}, with
\begin{equation}
	\label{eq:R_tensor}
	\mathcal R_{ij}^{\alpha\beta}=\delta_{ij}\sum_\gamma C_{\alpha\beta\gamma}\phi_i^\gamma.
\end{equation}
The antisymmetry of $\mathcal R$ ensures that this motion preserves the free energy at fixed temperature. Its Poisson-bracket--like structure follows directly from the microscopic commutation algebra. A complementary interpretation based on the constrained wave-function formulation is presented in Appendix~\ref{app:reversible_wavefunction}.

The memory contribution accounts for the influence of the eliminated electronic and environmental degrees of freedom. When their correlations decay rapidly compared with the order-parameter evolution, it can be approximated by a local transport response. The thermal adiabatic reduction additionally eliminates energy as an independent slow variable. As discussed above, this step must account for energy--order-parameter cross couplings: any surviving contributions from the small temperature lag enter the effective order-parameter dynamics, together with the corresponding corrections to its fluctuations.

On the thermal-balance manifold, $\beta\simeq\beta_{\rm B}$, the electronic relevant state becomes the fixed-temperature constrained density matrix $\hat{\rho}[\boldsymbol{\phi}]$ of Sec.~\ref{sec:AVP}. Equation~\eqref{eq:dStot} then gives the entropy-conjugate force $(\partial S_{\rm tot}/\partial\phi_i^\alpha)_E=\beta_{\rm B}f_i^\alpha=-\beta_{\rm B}\partial F/\partial\phi_i^\alpha$. We denote the effective dissipative tensor after eliminating the fast energy variable by $\Gamma_{ij}^{\alpha\beta}$, absorbing the factor $\beta_{\rm B}$ into its definition and including the effects of the cross couplings. The symmetric dissipative tensor must be positive semidefinite, ensuring nonnegative dissipative entropy production. Including fluctuations promotes $\boldsymbol{\phi}(t)$ to a stochastic collective field, with a force $\eta_i^\alpha(t)$ whose statistics are related to the effective dissipative response by the fluctuation--dissipation relation.

The reversible and dissipative sectors therefore require different microscopic information. For a specified electronic Hamiltonian $\hat{\mathcal H}$ and set of retained observables, the constrained state determines the thermodynamic forces, while their commutation algebra fixes the reversible tensor. The dissipative coefficients, by contrast, depend on the dynamics of the eliminated electronic modes and, for relaxation mediated by the environment, on the bath Hamiltonian and system--bath coupling. These environmental ingredients are not specified by $\hat{\mathcal H}$ alone. Treating $\Gamma$ phenomenologically is therefore a natural practical choice when the environment is not modeled explicitly, provided its form respects the relevant symmetries, conservation laws, and thermodynamic constraints. This retains microscopic thermodynamic forces while parametrizing the kinetics of relaxation.

Under these Markovian and thermal adiabatic approximations, the projected dynamics yields the fixed-temperature AVP structure introduced in Eq.~\eqref{eq:dphi_dt1}. We now turn to its practical implementation for interacting electronic systems.

\section{Mean-Field Formulation of the Adiabatic Variational Principle}
\label{sec:MF-AVP}

Practical application of the AVP to an interacting electronic Hamiltonian requires a many-body approximation or numerical solver for the constrained electronic problem, while the AVP construction itself is independent of this choice. The constrained variational problem can, in principle, be formulated within a broad class of many-body representations, including Slater determinants, variational Monte Carlo wave functions, tensor-network states, and neural-network quantum states. From this perspective, the AVP provides a route for extending established equilibrium many-body methods to the dynamics of emergent order. The microscopic solver is applied along a sequence of constrained states specified by the evolving order-parameter configuration and determines, for each $\bm\phi$, both the corresponding electronic state and the conjugate thermodynamic forces needed to advance the collective dynamics. The chosen solver therefore controls the accuracy and computational cost of the constrained-state calculation.

A particularly accessible realization is a self-consistent mean-field formulation closely related to conventional Hartree--Fock (HF) theory. Constrained variational calculations have a long history, including constrained HF and related self-consistent mean-field methods~\cite{Bassichis1971,Flocard1973,Berger1989,Bender2003}, as well as constrained density-functional theory (DFT)~\cite{Dederichs1984,WuVanVoorhis2005,Kaduk2012}. These methods impose prescribed collective observables through constraining fields and have been widely used to map energy landscapes and extract effective interactions. This established machinery provides a natural starting point for a mean-field implementation of the~AVP.

In this work, we adopt this mean-field realization of the AVP. Our aim is to simulate phase-ordering dynamics and the evolution of spatially extended order-parameter textures in large systems over many time steps, for which the comparatively modest computational cost of a mean-field treatment is essential. We formulate the constrained mean-field problem directly at finite temperature using number-conserving fermionic Gaussian density matrices, as employed in our simulations. The conventional zero-temperature Hartree--Fock wave-function formulation then follows as a special case.

Consider a general interacting fermion Hamiltonian
\begin{equation}
	\hat{\mathcal H}=\sum_{ij}\sum_{ab}t_{ij}^{ab} \hat c_{i,a}^{\dagger} \hat c_{j,b}^{\,}
	+\frac{1}{2}\sum_{ijkl}\sum_{abcd}V_{ijkl}^{abcd} \hat c_{i,a}^{\dagger} \hat c_{j,b}^{\dagger} \hat c_{k,c}^{\,} \hat c_{l,d}^{\,},
\end{equation}
where $\hat c_{i,a}^{\dagger}$ ($\hat c_{i,a}^{\,}$) creates (annihilates) a fermion at lattice site $i$ with internal label $a$, such as spin or orbital. The coefficients $t_{ij}^{ab}$ and $V_{ijkl}^{abcd}$ specify the hopping and two-body interactions, respectively. A number-conserving fermionic Gaussian density matrix $\hat{\rho}_{\rm G}$ is the normalized exponential of a number-conserving quadratic fermionic operator and is completely characterized by its one-particle density matrix
\begin{equation}
	G_{ij}^{ab}\equiv\mathrm{Tr}\!\left(\hat{\rho}_{\rm G}\,\hat c_{j,b}^{\dagger} \hat c_{i,a}^{\,}\right).
\end{equation}
The single-particle density matrix $G$ is Hermitian, with eigenvalues between $0$ and $1$ and $\mathrm{Tr}\,G=N_e$, where $N_e$ is the average number of electrons. Conversely, any $G$ satisfying these conditions specifies a number-conserving Gaussian state.

Restricting the expectation value of the interacting Hamiltonian to this Gaussian manifold defines the Hartree--Fock energy functional,
\begin{eqnarray}
	& & E_{\rm HF}[G]\equiv\mathrm{Tr}\!\left(\hat{\rho}_{\rm G}[G]\hat{\mathcal H}\right) \\
	& & \quad =\sum_{ij}\sum_{ab}t_{ij}^{ab}G_{ji}^{ba} +\frac{1}{2}\sum_{ijkl}\sum_{abcd}V_{ijkl}^{abcd}\left(G_{li}^{da}G_{kj}^{cb}-G_{ki}^{ca}G_{lj}^{db}\right). \nonumber
\end{eqnarray}
The second equality follows from Wick's theorem: because $\hat{\rho}_{\rm G}$ is Gaussian, all interaction expectation values reduce to products of elements of $G$, giving the usual direct and exchange contributions. The same Gaussian structure allows the many-body von Neumann entropy to be written entirely in the single-particle space as
\begin{equation}
	S_{\rm G}[G]=-\mathrm{Tr}\left[G\ln G+(1-G)\ln(1-G)\right],
\end{equation}
where the trace in this expression is over single-particle states.

For a prescribed instantaneous order-parameter configuration $\boldsymbol{\phi}=\{\phi_i^\alpha\}$, the finite-temperature MF-AVP free-energy Lagrangian is
\begin{equation}
	\mathcal{L}_{\rm MF}[G,\bm f;\boldsymbol{\phi}]=E_{\rm HF}[G]-T S_{\rm G}[G]+\sum_{i,\alpha}f_i^\alpha\left(A_i^\alpha[G]-\phi_i^\alpha\right).
\end{equation}
Here $A_i^\alpha[G]\equiv\mathrm{Tr}\{\hat{\rho}_{\rm G}[G]\hat A_i^\alpha\}$ is the expectation value of the corresponding collective observable, and $\bm f=\{f_i^\alpha\}$ are the conjugate fields enforcing its prescribed value. The first two terms constitute the finite-temperature Hartree--Fock free energy, while the last term imposes the order-parameter constraints. The energetic and constraint contributions define the effective single-particle Hamiltonian
\begin{equation}
	\label{eq:mf_hamiltonian}
	H_{ij}^{ab}[G,\bm f]\equiv\frac{\partial}{\partial G_{ji}^{ba}}\left[E_{\rm HF}[G]+\sum_{k,\alpha}f_k^\alpha A_k^\alpha[G]\right].
\end{equation}
It contains the usual Hartree--Fock contribution together with the single-particle fields associated with the prescribed order parameters.

At any temperature, stationarity with respect to unitary rotations of $G$ requires
\begin{equation}
	[H,G]=0.
\end{equation}
Such rotations preserve the eigenvalues of $G$ and hence its Gaussian entropy, while stationarity of the energetic and constraint contributions yields the commutation condition above. Consequently, $G$ and $H$ can be diagonalized in the same single-particle basis. In an orbital representation, the corresponding eigenvalue problem is
\begin{equation}
    \label{eq:mf_eigen}
    \sum_{jb}H_{ij}^{ab}[G,\bm f]\psi_n(j,b)=\epsilon_n\psi_n(i,a),
\end{equation}
where $\psi_n$ and $\epsilon_n$ are the single-particle eigenvectors and eigenvalues. The distinction between zero and finite temperature then enters through the allowed eigenvalues, or occupations, of $G$.

At $T=0$, a pure constrained Gaussian solution, when available, has the form $\hat{\rho}_{\rm G}=|\Psi_{\rm HF}\rangle\langle\Psi_{\rm HF}|$, where $|\Psi_{\rm HF}\rangle$ is a Slater determinant. Its one-particle density matrix is therefore the projector onto the occupied single-particle subspace. Beyond the general properties of $G$ stated above, the defining additional condition is its idempotency, $G^2=G$. For a gapped self-consistent solution, the $N_e$ lowest-energy states of Eq.~\eqref{eq:mf_eigen} are occupied, with $G$ given by the corresponding projector. Degeneracies at the chemical potential or constraints requiring fractional occupations may instead necessitate retaining a mixed Gaussian state at zero temperature.

At finite temperature, the eigenvalues of $G$ may vary continuously between $0$ and $1$. Minimization of the constrained free energy with respect to these occupations, at fixed average particle number, gives
\begin{equation}
	\label{eq:finiteT_G}
	G=\frac{1}{e^{\beta(H-\mu)}+1},
\end{equation}
where $\beta=1/T$ and the chemical potential $\mu$ enforces $\mathrm{Tr}\,G=N_e$. Thus, in the common eigenbasis defined by Eq.~\eqref{eq:mf_eigen}, the eigenvalues of $G$ are the Fermi--Dirac occupations $n_{\rm F}(\epsilon_n-\mu)$. Equation~\eqref{eq:finiteT_G} also makes the relation $[H,G]=0$ explicit, while its zero-temperature limit recovers the occupied-state projector when the self-consistent spectrum is gapped at the chemical potential.

\begin{figure}[t]
\centering
\includegraphics[width=0.85\columnwidth]{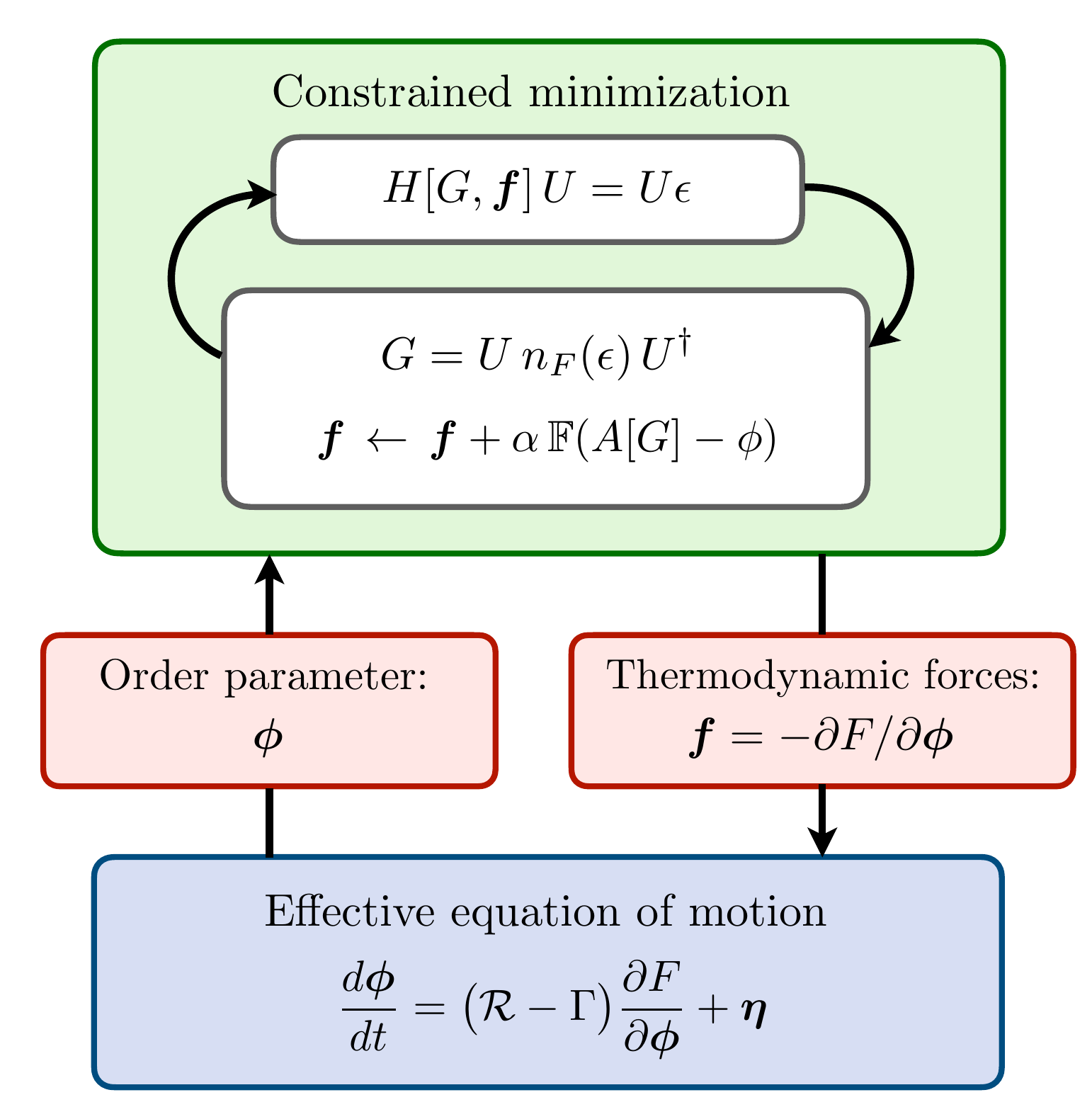}
\caption{
Schematic of the MF-AVP workflow. For a prescribed order-parameter configuration $\boldsymbol{\phi}$, the inner loop determines the single-particle density matrix $G$ and the Lagrange multipliers $\bm f$ enforcing the constraints. The map $\mathbb{F}$ converts constraint residuals $A[G]-\boldsymbol{\phi}$ into corrections to $\bm f$ and depends on the numerical implementation. At convergence, the multipliers provide the thermodynamic forces $\bm f=-\partial F/\partial\boldsymbol{\phi}$ used to advance $\boldsymbol{\phi}$ in the outer dynamical loop.
}
\label{fig:MF_AVP_scheme}
\end{figure}

Since $H[G,\bm f]$ generally depends on $G$, the effective Hamiltonian and density matrix must be determined self-consistently at either zero or finite temperature. At the same time, the fields $\bm f$ are adjusted until the prescribed constraints $A_i^\alpha[G]=\phi_i^\alpha$ are satisfied. Components of $G$ not fixed directly by these constraints remain variational and are determined as part of the same constrained solution. Once convergence is reached, the fields $\bm f$ provide the thermodynamic forces identified in Eq.~\eqref{eq:force1}.

Although Eq.~\eqref{eq:mf_eigen} gives a spectral representation of the mean-field problem, MF-AVP requires only the density matrix $G$ and the observables derived from it, without requiring explicit access to every single-particle eigenvector. Here we obtain these quantities by GPU-accelerated diagonalization of the effective single-particle Hamiltonian. Alternative implementations include the kernel polynomial method and related stochastic density-matrix techniques, rational or continued-fraction approximations to the Fermi--Dirac matrix function, and Green-function recursion or contour-integration methods~\cite{Weisse2006,Barros2013,Wang2018,Ozaki2007,Sidje2011,Rehr1977,Ozaki2010}. Depending on the Hamiltonian and constraints, only selected matrix elements or traces involving $G$ may be needed, avoiding construction and storage of the full density matrix. Pairing orders can also be treated using Hartree--Fock--Bogoliubov states, with a generalized density matrix that includes anomalous correlations.

The constrained mean-field calculation forms the inner loop of the MF-AVP simulation summarized in Fig.~\ref{fig:MF_AVP_scheme}. For a prescribed instantaneous configuration $\boldsymbol{\phi}(t)$, trial values of $G$ and $\bm f$ define the constrained effective Hamiltonian $H[G,\bm f]$. Solving the single-particle problem updates $G$ and the observables $A_i^\alpha[G]$, while any unconstrained mean fields are determined self-consistently. The fields $f_i^\alpha$ are adjusted using the constraint residuals $A_i^\alpha[G]-\phi_i^\alpha(t)$, and the iteration continues until mean-field self-consistency and the prescribed constraints are satisfied. Details of the constraint updates for CDW ordering in the $t$--$V$ model and SDW ordering in the Hubbard model are given in Appendices~\ref{app:tV_HF} and \ref{app:SDW_MFAVP}, respectively.

The converged fields $f_i^\alpha$ then enter the effective equation of motion as thermodynamic forces. The outer loop advances the order parameters to $\boldsymbol{\phi}(t+\Delta t)$, defining the next constrained electronic problem. Repeating this procedure generates a dynamical trajectory along which the microscopic state adjusts to the evolving order parameters. This nested scheme implements the AVP in the applications below. More sophisticated many-body solvers can replace the constrained mean-field calculation without changing the overall dynamical structure, as discussed further in the concluding section.

\section{Charge-Density-Wave Ordering Dynamics}
\label{sec:CDW}

CDW order is widespread in low-dimensional electronic materials~\cite{Gruner1988}, with layered transition-metal dichalcogenides providing prominent examples of commensurate and incommensurate ordering~\cite{Wilson1975,Rossnagel2011}. The formation and relaxation of these spatial modulations motivate microscopic studies of charge-ordering dynamics.  Here we apply MF-AVP to charge ordering on the square and triangular lattices at half filling and one-third filling, respectively. The square lattice provides a benchmark through conventional Allen--Cahn coarsening between two symmetry-related checkerboard states. The triangular lattice supports three-sublattice $\sqrt{3}\times\sqrt{3}$ order, whose three equivalent realizations give rise to a richer network of domain walls and junctions. These examples explore how ordering patterns and defect structures shape the collective dynamics and the electronic structure of the evolving textures.

Both systems are described by the spinless $t$--$V$ Hamiltonian,
\begin{equation}
	\hat{\mathcal H}=-t_{\rm nn}\sum_{\langle ij\rangle}\left(\hat c_i^\dagger \hat c_j^{\,}+{\rm h.c.}\right)+V\sum_{\langle ij\rangle}\hat n_i\hat n_j,
	\label{eq:tV_model}
\end{equation}
where $\hat c_i^\dagger$ ($\hat c_i^{\,}$) creates (annihilates) a spinless fermion at site $i$, $\hat n_i=\hat c_i^\dagger \hat c_i^{\,}$ is the local density operator, and $\langle ij\rangle$ denotes nearest-neighbor bonds, each counted once. The hopping amplitude $t_{\rm nn}$ sets the electronic energy scale, while $V>0$ describes nearest-neighbor repulsion.

\subsection{Charge-density-wave coarsening on the square lattice}
\label{sec:tV-square-CDW}

We first consider the square lattice at half filling, where the tendency toward checkerboard CDW order can be understood from complementary weak- and strong-coupling perspectives. In the weak-coupling limit, the noninteracting nearest-neighbor band has a perfectly nested Fermi surface with nesting vector $\bm Q=(\pi,\pi)$. The resulting divergence of the particle-hole susceptibility at zero temperature drives an instability toward checkerboard charge order. In the opposite limit $V/t_{\rm nn}\gg1$, the origin of this order is transparent in real space: at $t_{\rm nn}=0$, the repulsive interaction is minimized by occupying one sublattice and leaving the other empty, avoiding all occupied nearest-neighbor pairs. Finite hopping introduces charge fluctuations around these configurations, but the preference for two-sublattice order persists in the strong-coupling regime. The two limits thus lead to the same ordering pattern, with two symmetry-related states distinguished by which sublattice has the higher density.

Denoting the coarse-grained sublattice densities by $n_A$ and $n_B$, the scalar order parameter $\phi=n_A-n_B$ changes sign under sublattice interchange. The CDW therefore breaks a discrete $Z_2$ symmetry, with the same symmetry-breaking structure as the Ising model. Its long-wavelength energetics is conventionally described by the scalar Ginzburg--Landau functional
\begin{equation}
	F_{\rm GL}[\phi]=\int d^2r\left[\frac{\kappa}{2}|\nabla\phi|^2+\frac{a}{2}\phi^2+\frac{b}{4}\phi^4\right],
	\label{eq:square_GL}
\end{equation}
with $a<0$ and $b>0$ in the ordered phase. Purely relaxational dynamics, $\partial_t\phi=-\gamma\,\delta F_{\rm GL}/\delta\phi$, then yields curvature-driven coarsening of a nonconserved scalar order parameter. Here we retain a dissipative kinetic description but obtain the thermodynamic forces from the microscopic $t$--$V$ model through the MF-AVP, allowing us to follow both the evolving domains and their associated electronic structure without prescribing $F_{\rm GL}$.

\begin{figure*}[t]
\centering
\includegraphics[width=1.99\columnwidth]{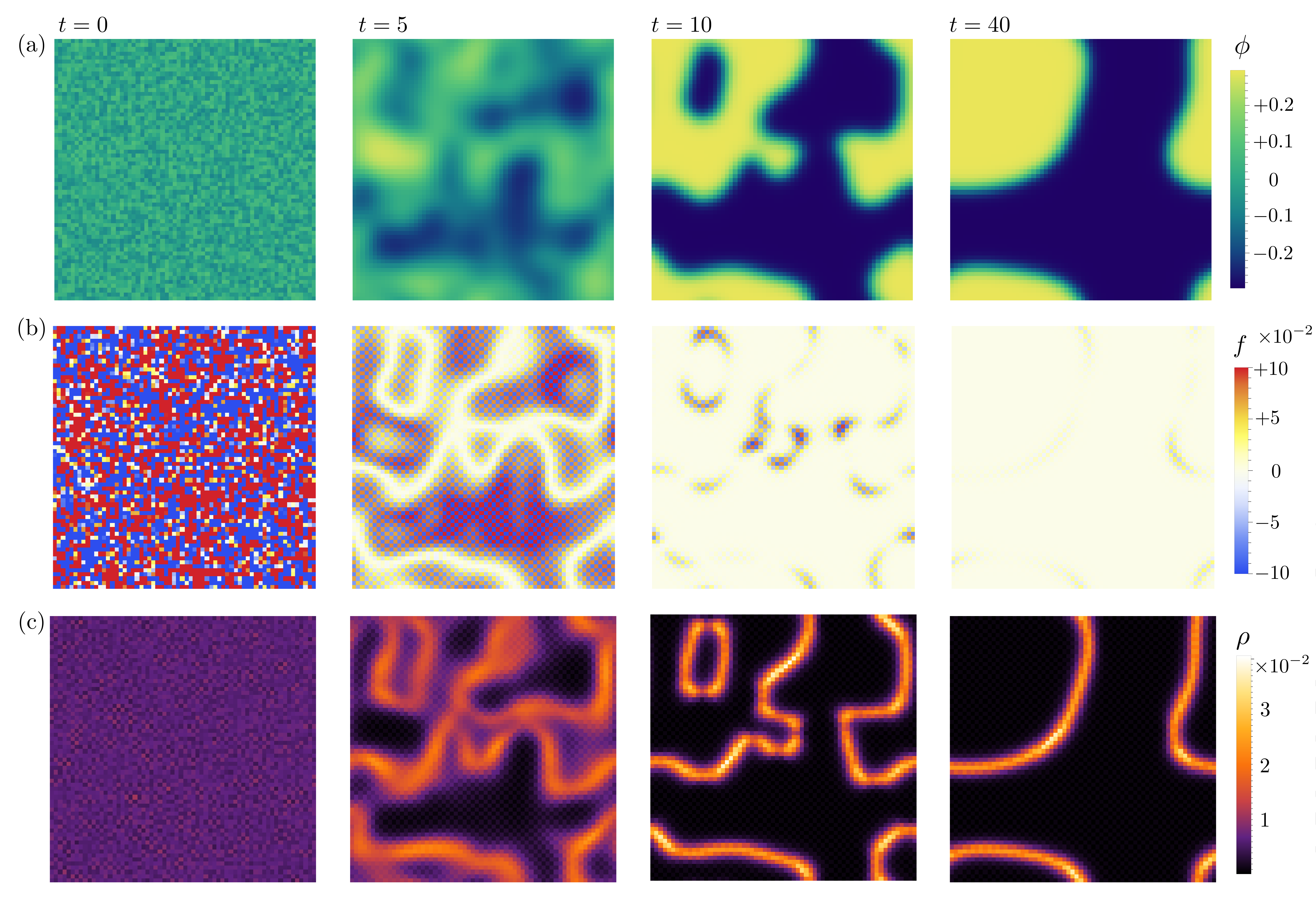}
\caption{Coarsening dynamics of the charge-density-wave state in the square-lattice $t$--$V$ model. The columns show snapshots at $t=0$, $5$, $10$, and $40$, with time measured in units of $(\Gamma t_{\rm nn})^{-1}$, where $\Gamma$ is the dissipative coefficient in Eq.~\eqref{eq:tV_modelA}. (a) Local checkerboard order parameter $\phi_i$ defined in Eq.~\eqref{eq:local_phi}, showing the formation and coarsening of domains whose opposite signs distinguish the two symmetry-related checkerboard states. (b) Conjugate thermodynamic force $f_i$ associated with the dynamical density variable $n_i$. (c) Local density of states $\rho_i$ near the Fermi energy, evaluated from the instantaneous constrained mean-field Hamiltonian.}
\label{fig:tV_coarsening}
\end{figure*}

For a spatially inhomogeneous configuration, we define the local checkerboard order parameter as
\begin{equation}
	\label{eq:local_phi}
	\phi_i=(-1)^{x_i+y_i}\left(n_i-\frac{1}{4}\sum_{j\in\mathcal N(i)}n_j\right),
\end{equation}
where $(x_i,y_i)$ are the integer lattice coordinates of site $i$ and $\mathcal N(i)$ denotes its four nearest neighbors. The density contrast measures the local charge modulation between opposite sublattices, while the factor $(-1)^{x_i+y_i}$ removes its checkerboard alternation. The resulting field has approximately uniform sign within each domain, with opposite signs distinguishing the two ordered states. For the dynamics, we retain the local densities $n_i$ themselves as the collective variables and evaluate $\phi_i$ from Eq.~\eqref{eq:local_phi}. This avoids reconstructing the density profile from the order-parameter field, which additionally requires specifying the uniform density component.

The simulations are performed at fixed chemical potential, chosen beforehand at the center of the single-particle gap of the reference, perfectly ordered CDW state and held fixed during the evolution. The total particle number is allowed to fluctuate, corresponding to an effective reservoir that permits particle exchange. In this setting, $F$ denotes the constrained grand potential. For a prescribed instantaneous density profile $\bm n=\{n_i\}$, the auxiliary MF-AVP Lagrangian is
\begin{equation}
	\mathcal L_{\rm MF}[G,\bm f;\bm n]=E_{\rm HF}[G]-T S_{\rm G}[G]-\mu\,\mathrm{Tr}\,G+\sum_i f_i(G_{ii}-n_i).
	\label{eq:tV_constrained_F}
\end{equation}
Here $G_{ii}$ is the on-site density of the trial mean-field state, whereas $n_i$ is the prescribed dynamical variable. The conjugate fields $\bm f=\{f_i\}$ enforce $G_{ii}=n_i$ at convergence and provide the thermodynamic forces $f_i=-\partial F/\partial n_i$.

Applying the general MF-AVP construction to Eq.~\eqref{eq:tV_constrained_F} gives the effective quadratic Hamiltonian
\begin{equation}
	\label{eq:tV_HF}
	\begin{aligned}
		\hat{\mathcal H}_{\rm HF}
		&=-\sum_{\langle ij\rangle}\left[(t_{\rm nn}+VG_{ij})\hat c_i^\dagger\hat c_j^{\,}+{\rm h.c.}\right]\\
		&\quad+\sum_i\left(f_i+V\sum_{j\in\mathcal N(i)}G_{jj}\right)\hat n_i,
	\end{aligned}
\end{equation}
where $G_{ij}=\langle\hat c_j^\dagger\hat c_i^{\,}\rangle$. The interaction generates a site-dependent Hartree potential and a Fock correction to the hopping, while $f_i$ enforces the prescribed local density. The chemical potential enters the occupations of this Hamiltonian. For each instantaneous $\bm n$, the constraint fields and unconstrained components of $G$ are determined self-consistently until $G_{ii}=n_i$. Details of this constrained HF calculation are given in Appendix~\ref{app:tV_HF}.

Because the density operators commute, $[\hat n_i,\hat n_j]=0$, the organized reversible contribution vanishes for this choice of retained variables. We adopt local dissipative dynamics,
\begin{equation}
	\label{eq:tV_modelA}
	\frac{dn_i}{dt}=-\Gamma\frac{\partial F}{\partial n_i}+\eta_i(t)=\Gamma f_i+\eta_i(t),
\end{equation}
where $\Gamma$ is a phenomenological kinetic coefficient and $\eta_i(t)$ is Gaussian thermal noise satisfying the fluctuation--dissipation relation. This local relaxation law allows the total density to fluctuate, consistently with the fixed-chemical-potential setting above. The forces are obtained directly from the constrained electronic state, without constructing or numerically differentiating a free-energy landscape. We retain $\Gamma$ phenomenologically to represent the dissipative response of the eliminated degrees of freedom and the reservoir; its microscopic determination would require specifying their dynamics and couplings. Together, the constrained HF calculation and Eq.~\eqref{eq:tV_modelA} implement the inner and outer loops of Fig.~\ref{fig:MF_AVP_scheme}, providing the dynamics examined below.

We now turn to the CDW coarsening dynamics summarized in Fig.~\ref{fig:tV_coarsening}. The simulations are performed on a $64\times64$ square lattice with $V/t_{\rm nn}=0.5$, electronic temperature $T/t_{\rm nn}=0.1$, and fixed chemical potential $\mu/t_{\rm nn}=1$. As discussed before, the nearest-neighbor hopping $t_{\rm nn}$ is used as the electronic energy unit, while time is measured in units of $(\Gamma t_{\rm nn})^{-1}$. The initial density profile is obtained from an electronic calculation with random on-site potentials. During the subsequent MF-AVP evolution, the constrained mean-field problem is solved using GPU-accelerated exact diagonalization, providing the thermodynamic forces and the instantaneous electronic structure at each step.

Figure~\ref{fig:tV_coarsening}(a) shows the local checkerboard order parameter $\phi_i$ defined in Eq.~\eqref{eq:local_phi}, whose sign distinguishes the two symmetry-related ordered states. The initially disordered configuration develops domains of both states, which subsequently grow and coalesce. The corresponding fields $f_i$, shown in Fig.~\ref{fig:tV_coarsening}(b), are conjugate to the dynamical densities $n_i$; their staggered pattern reflects the alternating density adjustments associated with checkerboard ordering. As coarsening proceeds, appreciable forces become increasingly concentrated near domain walls, while the interiors of well-formed domains are nearly relaxed. The microscopic force distribution thus identifies the interfaces as the regions responsible for the continuing collective evolution.

The constrained electronic calculation also reveals how the evolving domain pattern shapes the electronic spectrum. Figure~\ref{fig:tV_coarsening}(c) shows the local density of states near the Fermi level, obtained from the instantaneous mean-field eigenvalues $\epsilon_n$ and eigenvectors $\psi_{n,i}$ as $\rho_i(\omega)=\sum_n|\psi_{n,i}|^2\delta(\omega-\epsilon_n)$. A finite broadening is used numerically, and the Fermi level is located at $\omega=\mu$. The eigenvalues and eigenvectors are those of the constrained mean-field Hamiltonian in Eq.~\eqref{eq:tV_HF}, including the fields $f_i$ that enforce the instantaneous density configuration. As CDW order develops, low-energy spectral weight is strongly suppressed within the ordered domains but remains concentrated along their boundaries. These interfacial states lie within the bulk CDW gap, providing an electronic signature of the defects. MF-AVP thus follows both the coarsening of the domain pattern and the accompanying evolution of the electronic states supported by its interfaces.

\begin{figure}[t]
\centering
\includegraphics[width=0.98\columnwidth]{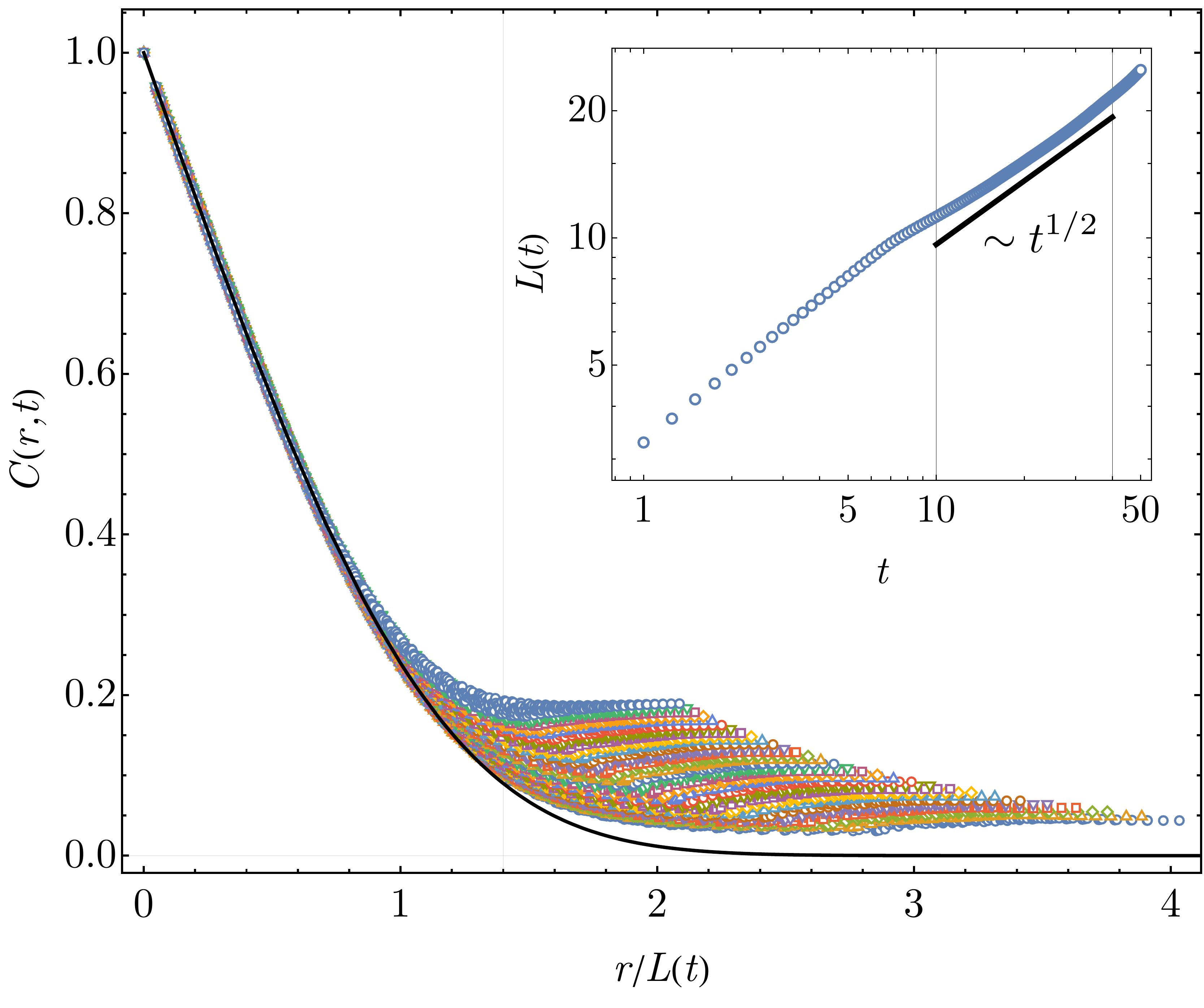}
\caption{Dynamical scaling of square-lattice CDW coarsening. The normalized equal-time correlation function $C(r,t)$ of the staggered CDW field is plotted against $r/L(t)$, where $C(L(t),t)=1/2$. Correlations at different times collapse at short rescaled distances, with deviations at larger distances as finite-size effects become important. The inset shows the characteristic domain size $L(t)$, whose growth is consistent with the Allen--Cahn law $L(t)\sim t^{1/2}$ over the coarsening regime.}
\label{fig:tV_scaling}
\end{figure}

To quantify the domain evolution, we use the equal-time correlation function $C(r,t)$ of the local checkerboard order parameter $\phi_i$, normalized so that $C(0,t)=1$. The sublattice factor incorporated in $\phi_i$ makes this correlation sensitive to the domain structure rather than the underlying checkerboard alternation. We extract a characteristic domain size from the half-height condition $C(L(t),t)=1/2$, without imposing a fitting form on the full correlation function. Figure~\ref{fig:tV_scaling} shows the correlations plotted against the rescaled distance $r/L(t)$. Their collapse at short rescaled distances supports a dynamical scaling description in which the evolving domain structure is characterized by a single growing length. Deviations at larger distances are consistent with the increasing influence of the finite simulation cell.

The growth of this characteristic length is approximately algebraic,
\begin{equation}
	L(t)\sim t^\alpha,
	\label{eq:coarsening_powerlaw}
\end{equation}
with an exponent consistent with the Allen--Cahn value $\alpha=1/2$ for a nonconserved scalar order parameter. This growth accompanies the smoothing of domain walls and the disappearance of smaller domains visible in Fig.~\ref{fig:tV_coarsening}(a). Together, the correlation collapse and the growth law support conventional curvature-driven coarsening over the accessible regime, before the domain size becomes comparable to the system dimensions.

This behavior can be understood in terms of interfaces whose local profiles relax more rapidly than the domain pattern evolves. The nearest-neighbor repulsion favors alternating charge densities, so disruptions of this pattern across a domain wall carry an interaction-energy cost. Electronic hopping, exchange, and thermal occupations also contribute to the wall profile and its free energy. Once the interfaces acquire approximately size-independent tension and mobility, their subsequent motion is governed primarily by curvature: the interfacial tension produces a driving force proportional to the local curvature, and a finite mobility converts this force into a wall velocity. For a domain pattern characterized by the scale $L$, the typical curvature is of order $1/L$, giving $dL/dt\propto 1/L$ and hence $L(t)\sim t^{1/2}$~\cite{Bray1994}. The observed scaling is consistent with this picture, in which the microscopic electronic structure sets the interfacial properties and the coarsening rate while leaving the large-scale growth exponent unchanged.

A complementary comparison is provided by the classical lattice-gas simulations in Appendix~\ref{app:classical_CDW}.  The model $\mathcal H_{\rm cl}=V\sum_{\langle ij\rangle}n_i n_j$ isolates the repulsion-driven preference for checkerboard order and also exhibits $L(t)\sim t^{1/2}$ growth under the nearest-neighbor particle-hopping kMC dynamics used there. Those updates conserve particle number, whereas the present MF-AVP simulations allow local density relaxation at fixed chemical potential; the two calculations therefore employ distinct microscopic dynamics. Their common growth law illustrates that different microscopic evolution rules can reach the same collective coarsening regime. For the MF-AVP, the recovery of Allen--Cahn scaling provides a successful benchmark of the constrained microscopic forces, while the simultaneous resolution of interfacial electronic states demonstrates the additional information retained beyond a phenomenological order-parameter description.

\subsection{CDW Coarsening on the Triangular Lattice}

\label{sec:tV-tri-CDW}

\begin{figure*}[t]
\centering
\includegraphics[width=1.99\columnwidth]{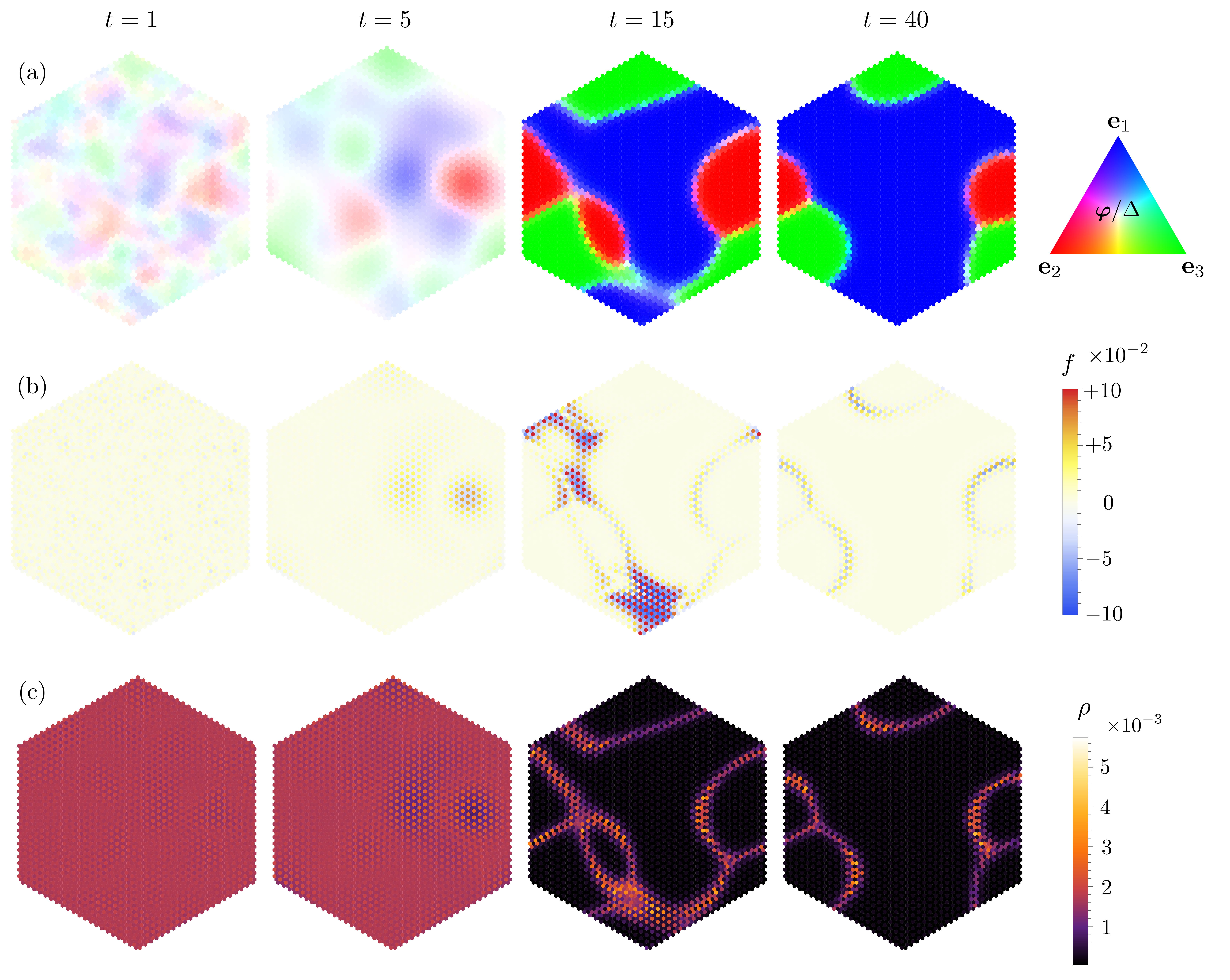}
\caption{
Phase-ordering dynamics of the $\sqrt{3}\times\sqrt{3}$ CDW in the triangular-lattice $t$--$V$ model at one-third filling, shown at $t=1$, $5$, $15$, and $40$. 
(a) Local two-component CDW order parameter $\boldsymbol{\phi}_i$, represented by the triangular color map whose vertices correspond to the three symmetry-related ordered states. The evolving configuration develops well-defined $Z_3$ domains separated by a network of domain walls and junctions.
(b) Conjugate thermodynamic force $f_i$ obtained from the constrained MF-AVP calculation, which becomes concentrated near the evolving domain boundaries and junctions.
(c) Local low-energy density of states $\rho_i$ of the constrained mean-field Hamiltonian, including the fields $f_i$, showing enhanced spectral weight associated with the domain-wall network.
}
\label{fig:tri_tV_coarsening}
\end{figure*}

We next consider the spinless $t$--$V$ model of Eq.~\eqref{eq:tV_model} on the triangular lattice at one-third filling, where three-sublattice charge ordering has been studied in this model and related interacting-fermion settings~\cite{Motrunich2004,Devakul2014}. Unlike the half-filled square lattice, the noninteracting Fermi surface is not perfectly nested, and charge order develops only beyond a finite interaction strength~\cite{Devakul2014}. The structure of the ordered phase is especially transparent in the strong-coupling limit. For $V/t_{\rm nn}\rightarrow\infty$, the nearest-neighbor repulsion is minimized by occupying one of three interpenetrating sublattices while leaving the other two empty, so that no occupied nearest-neighbor pairs occur. This $\sqrt{3}\times\sqrt{3}$ charge modulation triples the unit cell and has three symmetry-related realizations, corresponding to the choice of occupied sublattice. The ordered state therefore breaks a discrete $Z_3$ translation symmetry, with a symmetry-breaking structure analogous to that of the three-state Potts or clock model.

A convenient representation of this order is provided by a two-component field $\boldsymbol{\phi}=(\phi^1,\phi^2)$. Denoting the densities on the three sublattices by $n_A$, $n_B$, and $n_C$, we define
\begin{equation}
    \phi^1=\frac{n_A+n_B-2n_C}{\sqrt{6}}, \qquad \phi^2=\frac{n_A-n_B}{\sqrt{2}}.
    \label{eq:tri_phi}
\end{equation}
These combinations separate the charge modulation from the uniform density component. The three symmetry-related ordered states correspond to values $\boldsymbol{\Phi}_A$, $\boldsymbol{\Phi}_B$, and $\boldsymbol{\Phi}_C$ at the vertices of an equilateral triangle in the $(\phi^1,\phi^2)$ plane, with their distance from the origin set by the ordering amplitude. This representation connects the CDW to continuum descriptions of three-state ordering. For example, a phenomenological Ginzburg--Landau functional with the required threefold structure can be written as~\cite{Olejarz2013}
\begin{equation}
    F_{\rm GL}[\boldsymbol{\phi}]=\int d^2r\,\left[|\nabla\boldsymbol{\phi}|^2+\prod_{X=A,B,C}|\boldsymbol{\phi}-\boldsymbol{\Phi}_X|^2\right],
    \label{eq:tri_GL}
\end{equation}
with coefficients set to unity for illustration. Nonconserved relaxational dynamics generated by such a functional describes the growth of domains associated with the three degenerate minima~\cite{Lau1988,Olejarz2013}.

The defect structure is richer than in the two-state Ising case because three domain walls can meet at Y-shaped junctions. For equal, isotropic interfacial tensions, local mechanical balance gives $120^\circ$ angles between the walls~\cite{Olejarz2013}. Although the discrete manifold of the three ordered states does not support the usual $\pi_1$ classification of vortices~\cite{Mermin1979}, the junctions can be distinguished by the cyclic ordering of the surrounding domains: $A\to B\to C$ and $A\to C\to B$, traversed counterclockwise, define opposite chiralities. These junctions remain attached to domain walls and evolve as part of the wall network, introducing geometrical constraints and rearrangement processes absent in the Ising problem.

Studies of nonconserved three-state ordering provide a useful reference for the resulting coarsening dynamics~\cite{Safran1983,Sahni1983,Grest1988,Lau1988,Roland1990,Loureiro2010}. When domain walls and junctions remain mobile and their kinetic coefficients approach size-independent values, growth is expected to be governed primarily by curvature-driven domain-wall motion, which reduces the total interfacial energy and yields the conventional law $L(t)\sim t^{1/2}$. The approach to this regime can nevertheless be complicated by junction rearrangements, lattice pinning, and long-lived multidomain configurations; at zero temperature, trapping can even prevent continued coarsening~\cite{Safran1983,Sahni1983,Olejarz2013}. The three-state domain network can therefore exhibit substantial intermediate-time structure even when its eventual growth is consistent with Allen--Cahn scaling.

In the MF-AVP, the microscopic functional $F_{\rm AVP}[\bm\phi]$ is obtained from the constrained electronic problem, rather than prescribed through the empirical form in Eq.~\eqref{eq:tri_GL}. The threefold ordering landscape and the energetics of domain walls and junctions thus follow from the underlying lattice Hamiltonian within the mean-field approximation. We implement this construction using the same constrained-density dynamics as in Sec.~\ref{sec:tV-square-CDW}, taking the local densities $n_i$ as the slow variables. To characterize spatially inhomogeneous order, we construct a local two-component field on each upward-pointing elementary triangle by substituting its three vertex densities into Eq.~\eqref{eq:tri_phi}. The vertices are labeled according to a fixed global three-sublattice assignment, and the resulting vector is associated with the triangle center. This field resolves the local amplitude and orientation of the charge order, distinguishing the three symmetry-related domains and their intervening defects. Both components are combinations of commuting density observables, so their two-component representation does not introduce a reversible contribution. The constrained HF implementation is described in Appendix~\ref{app:tV_HF}.

\begin{figure}[t]
\centering
\includegraphics[width=0.98\columnwidth]{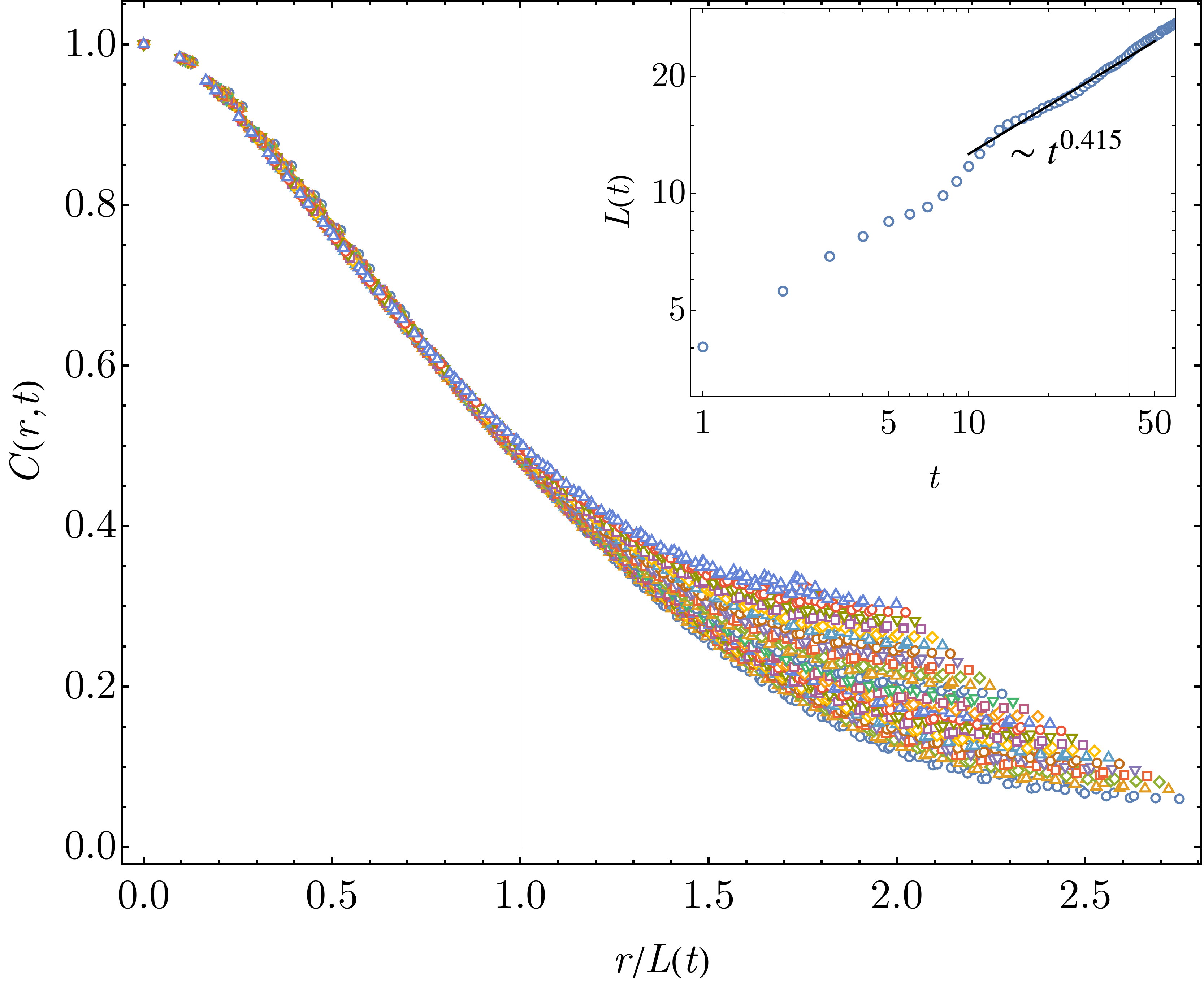}
\caption{
Dynamical scaling of the $\sqrt{3}\times\sqrt{3}$ CDW in the triangular-lattice $t$--$V$ model at one-third filling. Equal-time correlation functions $C(r,t)$ at different times are plotted as a function of the scaled distance $r/L(t)$, where the characteristic domain size is defined by $C(L(t),t)=1/2$. The approximate collapse of the correlation functions indicates dynamical scaling during the accessible coarsening regime. Inset: time dependence of the characteristic length $L(t)$. The solid line indicates an effective power-law fit $L(t)\sim t^{0.415}$ over the intermediate coarsening regime. }
\label{fig:tri_tV_scaling}
\end{figure}

We simulate the phase-ordering dynamics on a $63\times63$ triangular lattice at $V/t_{\rm nn}=2$ and electronic temperature $T/t_{\rm nn}=0.4$, with the nearest-neighbor hopping $t_{\rm nn}$ setting the energy unit. The chemical potential is chosen from the perfectly ordered ground-state CDW configuration and held fixed throughout the simulations. The initial density configurations are generated using random on-site potentials. Representative snapshots are shown in Fig.~\ref{fig:tri_tV_coarsening}. At early times, the CDW amplitude is weak and spatial correlations are short ranged. As the three-sublattice order develops, domains corresponding to the three symmetry-related $\sqrt{3}\times\sqrt{3}$ CDW states emerge and progressively coarsen. The triangular color representation of the local two-component field in Fig.~\ref{fig:tri_tV_coarsening}(a) distinguishes these states: order-parameter values within the domain interiors cluster near the three ordered-state values, while domain walls interpolate between them. Y-shaped junctions form where three different domains meet and persist through much of the simulated evolution.

Figure~\ref{fig:tri_tV_coarsening}(b) shows the conjugate forces $f_i$ acting on the local densities $n_i$. Once well-defined domains have formed, appreciable forces are concentrated primarily near domain boundaries and their junctions, identifying the interfacial network as the principal locus of continued relaxation. The forces are comparatively weak within mature domains, which have approached locally stationary CDW configurations. Complementary information is provided by the local low-energy density of states $\rho_i$ in Fig.~\ref{fig:tri_tV_coarsening}(c), evaluated with the constraint fields retained, as in the square-lattice calculation. Spectral weight is enhanced along the domain walls and near their junctions, while the ordered interiors become depleted. The microscopic calculation thus resolves both the forces driving the evolving domain network and the low-energy electronic states associated with its defects.

The domain growth is quantified by the correlation function $C(r,t)\propto\langle\bm\phi(\mathbf r_0,t)\cdot\bm\phi(\mathbf r_0+\mathbf r,t)\rangle$, where the average is over reference positions $\mathbf r_0$ and directions of $\mathbf r$. As in the square-lattice case, we define the characteristic length by $C(L(t),t)=1/2$, with $C(0,t)=1$. Figure~\ref{fig:tri_tV_scaling} shows an approximate collapse of the correlations plotted against $r/L(t)$, consistent with a single growing domain scale. The resulting $L(t)$, shown in the inset, follows the power law in Eq.~\eqref{eq:coarsening_powerlaw} with an effective exponent $\alpha\simeq0.415$ over the simulated interval, indicating slower growth than the Allen--Cahn behavior found on the square lattice.

This slower growth may reflect relaxation processes associated with the three-state domain network. Junction motion requires rearrangement of the adjoining walls, and slow junction kinetics or lattice pinning could impede interfacial relaxation. The persistence of junctions alone, however, does not establish such a mechanism, since junctions also occur in three-state systems exhibiting conventional curvature-driven growth. We therefore regard $\alpha\simeq0.415$ as an effective exponent over the accessible time and length scales, leaving open whether the dynamics eventually crosses over toward Allen--Cahn behavior or exhibits more persistent kinetic slowing. Resolving the long-time behavior requires larger systems and longer trajectories, which remain computationally demanding because each MF-AVP time step involves repeated solutions of the constrained electronic problem. This motivates the development of machine-learning surrogates for the microscopic thermodynamic forces, as discussed in the outlook.

\section{Adiabatic Spin-Density-Wave Dynamics}
\label{sec:Hubbard_SDW}

The CDW examples considered above involve commuting local density operators, for which the reversible tensor $\mathcal R$ vanishes. A qualitatively different situation arises for SDW order: the three components of the local magnetic moment inherit the noncommuting algebra of the electronic spin operators and generate reversible precession alongside dissipative relaxation. We demonstrate this aspect of the AVP using the repulsive Hubbard model on the two-dimensional square lattice. Both the orientations and magnitudes of the local moments are allowed to evolve, describing the development of magnetic correlations and the accompanying inhomogeneous spin textures.

The constrained electronic construction discussed in Sec.~\ref{sec:MF-AVP} has particularly close precedents in magnetism. Following the general formulation of constrained density-functional theory~\cite{Dederichs1984}, constraints on local magnetic moments have been used to characterize noncollinear configurations that are not stationary states of the unconstrained electronic problem~\cite{Stocks1998,Ujfalussy1999,Kurz2004,Singer2005}. These constraints can be enforced through Lagrange multipliers or through penalty functionals that penalize deviations from the prescribed configuration~\cite{MaDudarev2015,Cardias2021}. In both approaches, the constraining fields provide access to the electronic response associated with departures from magnetic equilibrium.

The use of electronic calculations to generate magnetization dynamics was developed in the pioneering {\em ab initio} spin-dynamics work of Antropov and collaborators~\cite{Antropov1995,Antropov1996}, together with related formulations of adiabatic local-moment dynamics~\cite{Halilov1998,Niu1999}. The electronic state adjusts to a slowly evolving magnetic configuration, and the resulting effective fields drive its motion. Subsequent implementations have employed constrained electronic states, using Lagrange-multiplier or penalty-based approaches, to follow magnetic relaxation and precession without relying on a fixed pairwise spin Hamiltonian~\cite{Ujfalussy2004,Cardias2021,Streib2022,Cardias2024}. The SDW application of the AVP shares this central construction, and the connection between constraining fields and effective magnetic fields has been examined explicitly within this literature~\cite{Streib2020,Brannvall2026}.

A separate issue concerns the dynamical variables retained in the magnetic description. Conventional fixed-length Landau--Lifshitz--Gilbert dynamics follows the orientations of local moments, while the formation and relaxation of itinerant SDW order can also involve substantial changes in their magnitudes. Established treatments of such longitudinal relaxation include Landau--Lifshitz--Bloch theory, which describes a thermally averaged magnetization of variable magnitude~\cite{Garanin1997,ChubykaloFesenko2006,Atxitia2017}, and generalized Langevin dynamics, which treats individual local-moment magnitudes as dynamical variables~\cite{MaDudarev2012}. Stochastic formulations incorporate the associated thermal fluctuations~\cite{Evans2012}, and longitudinal relaxation has also been included in semiclassical SDW dynamics~\cite{Chern2018}. These developments concern the structure of the magnetic equation of motion and are distinct from the electronic method used to determine its driving forces.

Here these two aspects meet in a microscopic description of magnetic ordering in itinerant electron systems. The AVP constraints identify the evolving local moments with the spin expectation values of the instantaneous electronic state, while the collective dynamics allows their orientations and amplitudes to evolve. Our focus is the formation and evolution of spatially extended magnetic textures, encompassing both the development of local moments and the growth of orientational correlations. Following magnetic ordering on these scales requires repeated constrained electronic calculations over the full lattice as the texture evolves. The MF-AVP simulations below address this computational challenge, extending the microscopic description developed for charge ordering to itinerant magnetism.

We consider the single-band Hubbard Hamiltonian
\begin{equation}
    \hat{\mathcal H}=-t_{\rm nn}\sum_{\langle ij\rangle,\alpha}\left(\hat c_{i,\alpha}^{\dagger}\hat c_{j,\alpha}^{\,}+{\rm H.c.}\right)+U\sum_i\hat n_{i,\uparrow}\hat n_{i,\downarrow},
    \label{eq:Hubbard_model}
\end{equation}
where $\alpha,\beta=\uparrow,\downarrow$ denote spin indices and $U>0$ is the on-site repulsion. At half filling on the square lattice, the ground state exhibits N\'eel antiferromagnetic order, connecting the weak-coupling itinerant regime continuously to the strong-coupling local-moment regime. At weak coupling, perfect nesting of the noninteracting Fermi surface by $\bm Q=(\pi,\pi)$ produces a divergent static spin susceptibility at zero temperature, making the electronic system unstable toward commensurate SDW order for arbitrarily weak repulsion. At strong coupling, electrons form local moments, and virtual hopping generates the antiferromagnetic exchange $J=4t_{\rm nn}^2/U$ to leading order. These limiting descriptions capture the same magnetic ordering tendency, making the Hubbard model a natural setting for studying its dynamics across itinerant and local-moment regimes.

The relevant collective operators are the local electronic spins,
\begin{equation}
    \hat s_i^a=\frac{1}{2}\sum_{\alpha\beta}\hat c_{i,\alpha}^{\dagger}\sigma^a_{\alpha\beta}\hat c_{i,\beta}^{\,},\qquad a=x,y,z,
    \label{eq:local_spin}
\end{equation}
where $\sigma^a$ are the Pauli matrices. We denote their prescribed expectation values by $\mathbf m_i=(m_i^x,m_i^y,m_i^z)$, using the conventional magnetic notation in place of the generic AVP variable $\boldsymbol{\phi}_i$. Similarly, we write the conjugate field as $\mathbf b_i$ rather than $\bm f_i$. The defining constraint $\langle\hat{\mathbf s}_i\rangle=\mathbf m_i$ identifies the evolving magnetic order parameter with the local spin expectation value of the constrained electronic state.

For a prescribed texture $\{\mathbf m_i\}$, the finite-temperature Hartree--Fock free-energy Lagrangian is
\begin{equation}
    \mathcal L_{\rm HF}[G,\{\mathbf b_i\};\{\mathbf m_i\}]=F_{\rm HF}[G]+\sum_i\mathbf b_i\cdot\left(\langle\hat{\mathbf s}_i\rangle_G-\mathbf m_i\right),
    \label{eq:Hubbard_constrained_F}
\end{equation}
where $G_{ij}^{\alpha\beta}=\langle\hat c_{j,\beta}^{\dagger}\hat c_{i,\alpha}^{\,}\rangle$ is the one-particle density matrix in real and spin space, and $\langle\cdots\rangle_G$ denotes an expectation value in the trial Gaussian state. The local charge density is $n_i=\sum_\alpha G_{ii}^{\alpha\alpha}$. Once the magnetic constraints are satisfied, the on-site density matrix takes the form $G_{ii}^{\alpha\beta}=(n_i/2)\delta_{\alpha\beta}+\mathbf m_i\cdot\boldsymbol{\sigma}_{\alpha\beta}$. During the constrained iteration, however, the spin expectation extracted from $G$ need not yet equal the prescribed $\mathbf m_i$.

When the prescribed spin constraints are satisfied, the constrained quadratic HF Hamiltonian takes the form
\begin{align}
    \hat{\mathcal H}_{\rm HF}={}&-t_{\rm nn}\sum_{\langle ij\rangle,\alpha}\left(\hat c_{i,\alpha}^{\dagger}\hat c_{j,\alpha}^{\,}+{\rm H.c.}\right)+\frac{U}{2}\sum_i n_i\hat n_i \nonumber\\
    &+\sum_i\left(\mathbf b_i-2U\mathbf m_i\right)\cdot\hat{\mathbf s}_i,
    \label{eq:Hubbard_HF}
\end{align}
where $\hat n_i=\sum_\alpha\hat n_{i,\alpha}$ and additive mean-field constants have been omitted from the quadratic Hamiltonian. During the constrained iteration, the exchange term instead involves the trial spin expectation $\mathbf s_i[G]=\langle\hat{\mathbf s}_i\rangle_G$, which approaches $\mathbf m_i$ as the spin constraints converge. The interaction generates a scalar Hartree potential and a local exchange field, while $\mathbf b_i$ enforces the prescribed spin density. The spin-dependent term resembles the exchange coupling in an $s$--$d$ or Kondo-lattice Hamiltonian; here, however, the magnetic configuration represents the spin density of the same electrons rather than an independent set of local moments.

The constrained HF problem is solved by adjusting $\mathbf b_i$ to enforce $\langle\hat{\mathbf s}_i\rangle_G=\mathbf m_i$, while determining the unconstrained charge densities and other mean fields self-consistently. Upon convergence, the constraining fields directly yield the thermodynamic forces through the AVP relation
\begin{equation}
    \mathbf b_i=-\frac{\partial F}{\partial\mathbf m_i},
    \label{eq:Hubbard_force}
\end{equation}
where $F[\{\mathbf m_i\}]$ is the constrained HF free energy at the chosen temperature and filling. These forces drive the subsequent evolution of the magnetic texture. Details of the self-consistent iteration are given in Appendix~\ref{app:SDW_MFAVP}.

The reversible dynamics follows from the local spin algebra, $[\hat s_i^a,\hat s_j^b]=i\delta_{ij}\epsilon_{abc}\hat s_i^c$. Substituting this commutator into Eq.~\eqref{eq:R_tensor} gives the reversible tensor
\begin{equation}
    \mathcal R_{ij}^{ab}=\delta_{ij}\epsilon_{abc}m_i^c,
    \label{eq:Hubbard_R_tensor}
\end{equation}
which reproduces the familiar Landau--Lifshitz precession, $\dot{\mathbf m}_i|_R=\mathbf m_i\times\mathbf b_i$. Thus, the precessional structure emerges directly from the spin commutation relations within the AVP formulation. Here we retain the convention $\hbar=1$ used in the general derivation; $\mathbf m_i$ is the electronic spin expectation value and $\mathbf b_i$ is its conjugate field in energy units.

To distinguish amplitude relaxation from orientational motion, we write
\begin{equation}
    \mathbf m_i=M_i\hat{\mathbf e}_i,\qquad M_i=|\mathbf m_i|,\qquad |\hat{\mathbf e}_i|=1,
    \label{eq:SDW_m_decomp}
\end{equation}
where $\hat{\mathbf e}_i$ is defined for $M_i>0$. The thermodynamic field $\mathbf b_i=-\partial F/\partial\mathbf m_i$ separates into longitudinal and transverse components,
\begin{equation}
    \mathbf b_{i,\parallel}=\hat{\mathbf e}_i(\hat{\mathbf e}_i\cdot\mathbf b_i),\qquad
    \mathbf b_{i,\perp}=\mathbf b_i-\mathbf b_{i,\parallel}.
    \label{eq:SDW_force_decomp}
\end{equation}
We adopt a local dissipative response with nonnegative kinetic coefficients $\Gamma_{\parallel}$ and $\Gamma_{\perp}$ for these two channels. Including thermal fluctuations, the generalized magnetic dynamics has the form
\begin{equation}
    \frac{d\mathbf m_i}{dt}=\mathbf m_i\times\mathbf b_i+\Gamma_{\parallel}\mathbf b_{i,\parallel}+\Gamma_{\perp}\mathbf b_{i,\perp}+\boldsymbol{\eta}_i(t).
    \label{eq:Hubbard_SDW_dynamics}
\end{equation}
The first term generates reversible precession, while the next two describe amplitude and orientational relaxation. The fluctuating term $\boldsymbol{\eta}_i$ represents the influence of the eliminated degrees of freedom. The kinetic coefficients are treated as phenomenological parameters in the present implementation.

Writing the local moment in terms of its amplitude and orientation gives a particularly transparent representation of the dynamics. The longitudinal equation is
\begin{equation}
    \frac{dM_i}{dt}=\Gamma_{\parallel}\hat{\mathbf e}_i\cdot\mathbf b_i+\xi_i(t),
    \label{eq:SDW_M_dynamics}
\end{equation}
where $\xi_i(t)$ is the amplitude noise. Since $\hat{\mathbf e}_i\cdot\mathbf b_i=-\partial F/\partial M_i$ at fixed orientation, the deterministic term has the Model-A form for a nonconserved amplitude~\cite{Hohenberg1977}.

The corresponding orientational equation takes the Landau--Lifshitz form,
\begin{equation}
    \frac{d\hat{\mathbf e}_i}{dt}=\hat{\mathbf e}_i\times\mathbf b_i-\frac{\Gamma_{\perp}}{M_i}\hat{\mathbf e}_i\times\left(\hat{\mathbf e}_i\times\mathbf b_i\right)+\boldsymbol{\zeta}_i(t).
    \label{eq:SDW_e_dynamics}
\end{equation}
Here the first term describes precession about the thermodynamic field, the double-cross-product term gives Landau--Lifshitz damping, and $\boldsymbol{\zeta}_i(t)$ represents orientational fluctuations. The factor $1/M_i$ converts transverse relaxation of the magnetic moment into angular motion. Both deterministic terms are perpendicular to $\hat{\mathbf e}_i$ and thus preserve its unit length, while damping drives relaxation toward lower free energy. The stochastic interpretation of these equations, the relation between the vector and amplitude--orientation noises, and the associated fluctuation--dissipation and numerical details are specified in Appendix~\ref{app:SDW_MFAVP}.

Although written separately, the amplitude and orientation equations remain coupled through the dependence of $\mathbf b_i$ on the full magnetic texture. Their distinct relaxation channels allow local-moment formation and the growth of orientational correlations to proceed on different time scales. As amplitude fluctuations become increasingly costly in the local-moment regime, $M_i$ becomes approximately fixed and the remaining dynamics approaches the familiar Landau--Lifshitz evolution of moment orientations. Retaining both channels allows the MF-AVP to follow itinerant SDW textures with evolving magnetic amplitudes, using the constrained electronic calculation to supply the driving fields throughout the trajectory.

\begin{figure*}[t]
\centering
\includegraphics[width=1.99\columnwidth]{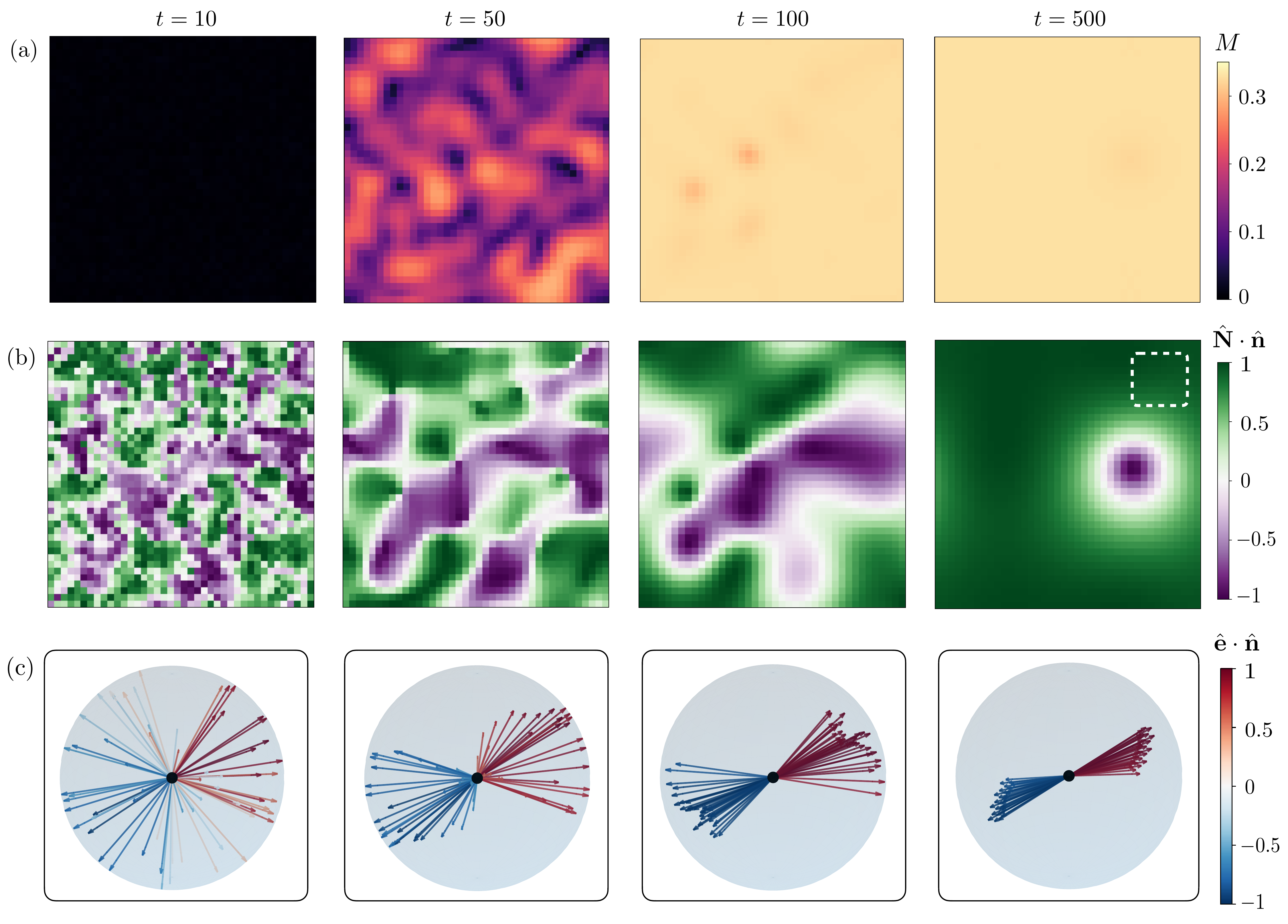}
\caption{Evolution of antiferromagnetic SDW correlations in the half-filled two-dimensional Hubbard model following a thermal quench, shown at $t=10$, $50$, $100$, and $500$ in units of $\hbar/U$. The local moment is $\mathbf m_i=M_i\hat{\mathbf e}_i$, and the staggered magnetization $\mathbf N_i=(-1)^{x_i+y_i}\mathbf m_i$ is defined in Eq.~\eqref{eq:local_Neel}. (a) Spatial distribution of $M_i$, showing the rapid development of local moments and their subsequent approach to a nearly uniform magnitude. (b) Projection $\hat{\mathbf N}_i\cdot\hat{\mathbf n}$, where $\hat{\mathbf N}_i=\mathbf N_i/|\mathbf N_i|$ and the fixed reference axis $\hat{\mathbf n}$ is chosen along the N\'eel direction selected at late times. Extended regions of correlated staggered orientations grow during the evolution. (c) Distribution of $\hat{\mathbf N}_i$ on the unit sphere for sites within the region indicated by the dashed square in the rightmost panel of (b), showing their increasing alignment as antiferromagnetic correlations develop.}
\label{fig:Hubbard_SDW}
\end{figure*}

We illustrate the MF-AVP dynamics following a thermal quench in the half-filled Hubbard model on a $40\times40$ square lattice, solving the constrained quadratic Hamiltonian in Eq.~\eqref{eq:Hubbard_HF} by exact diagonalization at each stage of the evolution. We use $t_{\rm nn}/U=0.25$ and a post-quench temperature $T/U=0.0005$, with dimensionless longitudinal and transverse dissipative coefficients $\Gamma_{\parallel}=0.1$ and $\Gamma_{\perp}=0.3$. Time is measured in units of $\hbar/U$. Unlike the discrete-symmetry CDW examples, the isotropic SDW has continuous spin-rotation symmetry, for which the Mermin--Wagner theorem precludes thermodynamic long-range order at finite temperature in two dimensions. We therefore examine the development of antiferromagnetic correlations on the finite lattice, focusing on the coupled evolution of magnetic amplitudes and orientations.

Figure~\ref{fig:Hubbard_SDW}(a) shows the local moment amplitude $M_i$ defined in Eq.~\eqref{eq:SDW_m_decomp}. Starting from small values, the moments grow rapidly, at first with substantial spatial variation, and then approach an approximately uniform magnitude. This development of local moments does not yet establish a coherent N\'eel pattern: their orientations remain spatially disordered and evolve more slowly. The subsequent growth of antiferromagnetic correlations is therefore governed primarily by orientational ordering.

To characterize the orientational ordering, we define the local staggered magnetization
\begin{equation}
    \mathbf N_i=(-1)^{x_i+y_i}\mathbf m_i,
    \label{eq:local_Neel}
\end{equation}
where $(x_i,y_i)$ are the integer lattice coordinates of site $i$. The sublattice factor removes the sign alternation of a N\'eel state, so neighboring moments in an ordered region give aligned $\mathbf N_i$. Figure~\ref{fig:Hubbard_SDW}(b) shows $\hat{\mathbf N}_i\cdot\hat{\mathbf n}$, where $\hat{\mathbf N}_i=\mathbf N_i/|\mathbf N_i|$ and the fixed reference axis $\hat{\mathbf n}$ is chosen along the N\'eel direction selected at late times. The initially disordered orientations develop extended correlated regions that grow and merge as the texture relaxes. By late times, these correlations span much of the simulated lattice, giving an apparently ordered N\'eel configuration on the finite spatial scale studied here.

Figure~\ref{fig:Hubbard_SDW}(c) provides a complementary view of orientational ordering: the staggered unit vectors $\hat{\mathbf N}_i$ within the dashed-square region of panel (b) are plotted on the unit sphere. Their initially broad distribution narrows around a common direction as the evolution proceeds. Panels (b) and (c) thus show the growth of correlated regions in real space and the alignment of moments within one such region, respectively. These changes arise from the combined precessional and dissipative dynamics of Eq.~\eqref{eq:Hubbard_SDW_dynamics}, driven by fields obtained from the instantaneous constrained electronic state.

A more systematic study of SDW ordering would track correlations of the staggered magnetization to extract a growing domain scale, alongside the staggered structure factor. These measures could distinguish continued domain growth from saturation caused by thermal fluctuations or the finite lattice size. In the isotropic two-dimensional system studied here, the equilibrium correlation length remains finite at nonzero temperature. Extending the calculations to three dimensions would allow the same framework to examine the growth of true N\'eel long-range order and the approach to its ordering transition.

The stochastic AVP dynamics also offers a route to the equilibrium statistical physics of itinerant magnetism. When the noise and dissipative response are chosen to yield the appropriate stationary distribution, long Langevin trajectories can sample spatially varying SDW configurations according to their constrained free energy. Such sampling incorporates fluctuations of the order-parameter field absent from a spatially uniform mean-field treatment and may be especially useful at intermediate $U$, where both the formation and orientation of local moments matter. Reaching the system sizes and sampling times needed for this program would require many constrained electronic calculations. In the Outlook, we discuss how machine-learning surrogates for the free energy and its forces could reduce this cost while maintaining the thermodynamic consistency of the dynamics.

\section{Discussion and Outlook}
\label{sec:discussion}

In this work, we have developed the Adiabatic Variational Principle (AVP) as a framework for connecting microscopic many-body calculations to the nonequilibrium dynamics of emergent electronic order. Under a separation of time scales, each instantaneous order-parameter configuration defines a constrained quasi-stationary electronic state, whose Lagrange multipliers provide the thermodynamic forces driving the collective evolution. Nonlinear Mori--Zwanzig projection theory places this construction within a statistical-mechanical framework: the intrinsic organized motion determines the reversible sector, while the eliminated degrees of freedom generate memory, dissipation, and fluctuations. In the Markovian adiabatic limit, the resulting dynamics provides a microscopic route to phase-field-like simulations without prescribing an empirical free-energy functional.

The mean-field applications demonstrate this construction for charge- and spin-density-wave dynamics in interacting lattice models. The square-lattice $t$--$V$ model recovers Allen--Cahn coarsening, while the triangular lattice exhibits slower intermediate-time growth within a network of threefold domains and persistent junctions. Both calculations resolve the evolving textures together with their conjugate forces and defect-associated electronic states. The Hubbard-model application extends this description to magnetic ordering, with Landau--Lifshitz precession emerging from the spin commutation algebra. The simulations capture rapid local-moment formation followed by slower orientational ordering, through which antiferromagnetic correlations spread across the finite two-dimensional lattice.

These demonstrations use a mean-field electronic solver, but the AVP construction is independent of this choice. Extending the constrained calculation to more correlated many-body states is therefore a natural next step. The present simulations inherit the limitations of fermionic Gaussian states, particularly in their treatment of strong correlations and the electronic structure of defects.  Variational Monte Carlo, matrix-product and other tensor-network states, and neural-network quantum states offer routes to retaining additional correlations within the same AVP construction. Density-functional implementations provide a complementary route toward material-specific applications. Constrained DFT and adiabatic spin dynamics already realize closely related constructions for magnetic degrees of freedom~\cite{Dederichs1984,Antropov1996,MaDudarev2015}. The broader AVP framework motivates extending this strategy to other collective electronic variables, including charge and orbital order, and to their coupled evolution.

Alongside improvements in the constrained electronic state, a more complete microscopic description requires a corresponding treatment of the kinetics. Although the thermodynamic forces are obtained from the constrained electronic problem, the dissipative coefficients in the present simulations remain phenomenological. Evaluating the transport kernels of the projected dynamics would connect these coefficients and their associated fluctuations to microscopic electronic processes and system--bath coupling~\cite{Grabert1978,teVrugt2020}. Retaining their temporal structure would also allow finite-memory corrections beyond the Markovian approximation. When additional degrees of freedom evolve on comparable time scales, they must instead be incorporated explicitly into the retained slow sector~\cite{Hohenberg1977}. These extensions would clarify how microscopic relaxation controls collective kinetics and broaden the range of nonequilibrium regimes accessible to the framework.

Machine learning (ML) offers a practical route to extending AVP simulations to the length and time scales needed to resolve collective textures. Even at the mean-field level, each dynamical time step requires an iterative electronic calculation, making repeated evaluations of the thermodynamic forces a major computational cost. ML surrogates have enabled substantial advances in {\em ab initio} molecular dynamics~\cite{behler07,bartok10,zhang18,Batzner2022,Batatia2025} and multiscale lattice-Hamiltonian simulations~\cite{zhang21,zhang22b,Cheng2023,Fan24,Ghosh24,Fan26}, where electronic calculations typically supply energies or forces for independently specified atomic or classical variables within an adiabatic description. The AVP provides the link needed to extend this ML strategy to emergent electronic order by defining the constrained microscopic states and conjugate thermodynamic forces associated with evolving order-parameter configurations. These calculations supply training data for ML models of the constrained free energy or its forces. Once trained, the ML surrogates could replace repeated constrained electronic calculations during the dynamics, enabling simulations of larger defect networks and longer collective relaxation times.

A further opportunity is to explore how preparation protocols control the formation and evolution of electronic textures. Extending AVP simulations beyond sudden thermal quenches to controlled cooling and annealing would allow studies of cooling-rate-dependent defect production and subsequent relaxation in the context of Kibble--Zurek physics~\cite{Kibble1976,Zurek1985,delCampo2014,delCampo2024}. Such protocols must remain compatible with rapid microscopic relaxation, even as the collective variables fall out of equilibrium near a transition. This separation between microscopic adjustment and collective relaxation is the regime targeted by the AVP. By resolving both the resulting textures and their electronic structure, these simulations could connect defect densities and spatial correlations measured by real-space imaging, including magnetic imaging~\cite{Milde2013}, with local spectroscopic signatures of defect-associated electronic states~\cite{Yazdani2016,Liu2017}.

The same ability to follow collective textures together with their electronic structure motivates applications to more complex forms of emergent order. Multicomponent and intertwined phases, including superconductivity coupled to charge or spin order~\cite{Fradkin2015,Fernandes2019,Agterberg2020}, provide settings in which several collective fields compete, coexist, and evolve through coupled reversible and dissipative channels.  Their dynamics could reveal how microscopic electronic structure influences the formation and transformation of heterogeneous phases. Extensions beyond conventional symmetry breaking, including Mott transitions and quantum spin liquids, present further intriguing possibilities, although they require suitable collective variables and constrained state manifolds. These directions would broaden the scope of the AVP from ordering dynamics to the nonequilibrium evolution of correlated quantum phases more generally.

\begin{acknowledgments}
This work was supported by the US Department of Energy Basic Energy Sciences under Contract No. DE-SC0020330. The authors acknowledge Research Computing at The University of Virginia for providing computational resources and technical support that have contributed to the results reported within this publication. 
\end{acknowledgments}

\appendix

\section{Nonlinear Projection Formalism}
\label{app:nonlinear_projection}

This Appendix summarizes the nonlinear projection formalism underlying Sec.~\ref{sec:nonlinear-MZ} and clarifies the role of bath centering in the organized motion. We first specify the relevant ensemble, projection operator, and equations for the macroscopic means and their fluctuations, then apply the organized-drift construction to the system--bath setting of the AVP.

\subsection{Relevant Ensemble and Projected Dynamics}

We follow the nonlinear projection construction of Ref.~\cite{Grabert1978}, with notation adapted to the present application; see also Ref.~\cite{teVrugt2020}. Consider a set of time-independent Hermitian observables $\{\hat{\mathcal A}_\mu\}$ whose expectation values $a_\mu(t)$ specify the macroscopic state. Maximizing the entropy subject to these expectation values defines the generalized canonical ensemble~\cite{Grabert1978,teVrugt2020}
\begin{equation}
	\hat{\rho}_{\rm rel}[a(t)]=\frac{1}{Z(t)}\exp\!\left[-\sum_\mu\lambda_\mu(t)\hat{\mathcal A}_\mu\right].
	\label{eq:app_rho_rel}
\end{equation}
The conjugate fields are determined by $a_\mu(t)={\rm Tr}\{\hat{\rho}_{\rm rel}[a(t)]\hat{\mathcal A}_\mu\}$. This ensemble reproduces the retained expectation values but need not coincide with the actual microscopic state $\hat{\rho}(t)$. Its time dependence enters through the macroscopic trajectory $a(t)$.

For an arbitrary microscopic operator $X$, the corresponding projection is~\cite{Grabert1978}
\begin{equation}
\begin{aligned}
	\mathcal P(t)X&={\rm Tr}\{\hat{\rho}_{\rm rel}(t)X\}\hat I\\
	&\quad+\sum_\mu\left(\hat{\mathcal A}_\mu-a_\mu(t)\hat I\right){\rm Tr}\!\left\{\frac{\partial\hat{\rho}_{\rm rel}(t)}{\partial a_\mu(t)}X\right\}.
\end{aligned}
	\label{eq:app_projection}
\end{equation}
The complementary projector is $\mathcal Q(t)=1-\mathcal P(t)$. Although $\mathcal P(t)$ is linear in $X$, its coefficients depend on the evolving macroscopic state, allowing nonlinear collective dynamics. To verify its projection properties, we differentiate the normalization ${\rm Tr}\,\hat\rho_{\rm rel}=1$ and the constraints ${\rm Tr}(\hat\rho_{\rm rel}\hat{\mathcal A}_\nu)=a_\nu$ with respect to $a_\mu$. This gives ${\rm Tr}(\partial\hat\rho_{\rm rel}/\partial a_\mu)=0$ and ${\rm Tr}[(\partial\hat\rho_{\rm rel}/\partial a_\mu)\hat{\mathcal A}_\nu]=\delta_{\mu\nu}$, which imply $\mathcal P(t)\hat I=\hat I$ and $\mathcal P(t)\hat{\mathcal A}_\nu=\hat{\mathcal A}_\nu$. Thus, $\mathcal P(t)$ acts as the identity on the span of the retained observables and $\hat I$, which is also its range. Since this span is independent of time,
\begin{equation}
	\mathcal P(t)\mathcal P(t')=\mathcal P(t'),
	\label{eq:app_projection_identity}
\end{equation}
including the usual idempotency condition at equal times.

Let the closed microscopic dynamics be generated by the time-independent total Hamiltonian $\hat{\mathcal H}_{\rm tot}$, with $\mathcal L X=[\hat{\mathcal H}_{\rm tot},X]$ and $\hbar=1$. In the system--bath setting, $\hat{\mathcal H}_{\rm tot}$ includes the electronic Hamiltonian, the bath Hamiltonian, and their coupling, consistent with the notation in the main text. We write $\hat{\mathcal A}_\mu(t)=e^{i\mathcal Lt}\hat{\mathcal A}_\mu$ and use $\dot{\hat{\mathcal A}}_\mu=i\mathcal L\hat{\mathcal A}_\mu$ for the microscopic velocity at the reference time. The organized drift is
\begin{equation}
	v_\mu[a(t)]={\rm Tr}\{\hat{\rho}_{\rm rel}[a(t)]\,\dot{\hat{\mathcal A}}_\mu\}.
	\label{eq:app_organized_v}
\end{equation}
For the relevant entropy $S_{\rm rel}=-{\rm Tr}\{\hat{\rho}_{\rm rel}\ln\hat{\rho}_{\rm rel}\}$, the maximum-entropy construction gives $dS_{\rm rel}=\sum_\mu\lambda_\mu\,da_\mu$. Moreover,
\begin{equation}
	\sum_\mu\lambda_\mu v_\mu
	=-i\,{\rm Tr}\{\hat{\rho}_{\rm rel}[\hat{\mathcal H}_{\rm tot},\ln\hat{\rho}_{\rm rel}]\}=0.
	\label{eq:app_entropy_org}
\end{equation}
Thus, the organized motion preserves the relevant entropy and constitutes the reversible part of the projected dynamics~\cite{Grabert1978}.

The complementary evolution is described by
\begin{equation}
	\mathcal G(s,t)=\mathcal T_{-}\exp\!\left[i\int_s^t du\,\mathcal L\mathcal Q(u)\right],
	\label{eq:app_orthogonal_propagator}
\end{equation}
where $\mathcal T_{-}$ orders operators from left to right in increasing time. With this convention, $\partial_t\mathcal G(s,t)=\mathcal G(s,t)i\mathcal L\mathcal Q(t)$ and $\mathcal G(s,s)=1$. The fluctuating force associated with reference time $s$ is
\begin{equation}
	\mathcal F_\mu(t,s)=e^{i\mathcal Ls}\mathcal Q(s)\mathcal G(s,t)\dot{\hat{\mathcal A}}_\mu.
	\label{eq:app_fluctuating_force}
\end{equation}
Its complementary character is expressed by $\mathcal P(s)e^{-i\mathcal Ls}\mathcal F_\mu(t,s)=0$. Both $\mathcal G$ and $\mathcal F_\mu$ depend on the macroscopic trajectory through the time-dependent projector.

The memory contribution to the mean dynamics is
\begin{equation}
	\mathcal K_\mu[t,s;a]
	={\rm Tr}\{\hat{\rho}_{\rm rel}(s)i\mathcal L\mathcal Q(s)\mathcal G(s,t)\dot{\hat{\mathcal A}}_\mu\}.
	\label{eq:app_memory}
\end{equation}
It describes the systematic feedback of the complementary dynamics and is a functional of the mean trajectory between $s$ and $t$. For an initial preparation $\hat{\rho}(0)=\hat{\rho}_{\rm rel}[a(0)]$, the random force has zero initial-ensemble average, ${\rm Tr}\{\hat{\rho}(0)\mathcal F_\mu(t,0)\}=0$. The exact mean transport equation is consequently~\cite{Grabert1978}
\begin{equation}
	\frac{da_\mu(t)}{dt}=v_\mu[a(t)]+\int_0^t ds\,\mathcal K_\mu[t,s;a].
	\label{eq:app_transport}
\end{equation}
For a general initial state, an additional mean-force term ${\rm Tr}\{\hat{\rho}(0)\mathcal F_\mu(t,0)\}$ remains. The preparation assumption removes this term, not the fluctuations themselves.

To display the latter explicitly, define $\delta\hat{\mathcal A}_\mu(t)=\hat{\mathcal A}_\mu(t)-a_\mu(t)\hat I$. Under the same initial-state assumption, these fluctuations obey
\begin{equation}
\begin{aligned}
	\frac{d\delta\hat{\mathcal A}_\mu(t)}{dt}
	&=\sum_\nu\Omega_{\mu\nu}(t)\delta\hat{\mathcal A}_\nu(t)\\
	&\quad+\int_0^t ds\sum_\nu\Phi_{\mu\nu}(t,s)\delta\hat{\mathcal A}_\nu(s)+\mathcal F_\mu(t,0),
\end{aligned}
	\label{eq:app_fluctuation_dynamics}
\end{equation}
where $\Omega_{\mu\nu}(t)=\partial v_\mu[a(t)]/\partial a_\nu(t)$ describes organized motion about the mean trajectory. The fluctuation memory kernel is
\begin{equation}
\begin{aligned}
	\Phi_{\mu\nu}(t,s)
	&={\rm Tr}\!\left\{\frac{\partial\hat{\rho}_{\rm rel}(s)}{\partial a_\nu(s)}
	i\mathcal L\mathcal Q(s)\mathcal G(s,t)\dot{\hat{\mathcal A}}_\mu\right\}\\
	&\quad-\sum_\kappa\dot a_\kappa(s)\,
	{\rm Tr}\!\left\{\frac{\partial^2\hat{\rho}_{\rm rel}(s)}
	{\partial a_\nu(s)\partial a_\kappa(s)}
	\mathcal G(s,t)\dot{\hat{\mathcal A}}_\mu\right\}.
\end{aligned}
	\label{eq:app_fluctuation_kernel}
\end{equation}
These equations separate the generally nonlinear evolution of the means from the linear, nonstationary evolution of fluctuations about that trajectory. The coefficients in the fluctuation equation are themselves functionals of the mean path.

The mean memory contribution and the fluctuating forces arise from the same complementary dynamics. Their connection is expressed by a generalized fluctuation--dissipation relation involving generalized canonical correlations of the random forces~\cite{Grabert1978}. Away from equilibrium, these correlations and the transport kernels depend on the evolving relevant state; linearization about a stationary equilibrium state recovers the familiar Mori structure. When complementary correlations decay rapidly on the collective time scale, a Markovian approximation yields local transport and short-correlated fluctuations. The resulting mean and fluctuation equations provide the basis for the effective stochastic description used in the main text.

For the AVP, the relevant observables include the system energy, bath energy, and order parameters. The electronic reduced state of Eq.~(\ref{eq:app_rho_rel}) then has the constrained Gibbs form in Eq.~(\ref{eq:rho_rel}). The following subsection establishes how the organized drift simplifies for this choice and the centered coupling used in Sec.~\ref{sec:nonlinear-MZ}.

\subsection{Bath Centering and Organized Motion}
\label{app:bath_centering}

For the system--bath construction used in the main text, the relevant ensemble factorizes as $\hat{\rho}_{\rm rel}^{\rm tot}=\hat{\rho}_{\rm rel}\otimes\hat{\rho}_{{\rm B},{\rm rel}}$, where $\hat{\rho}_{\rm rel}$ denotes the electronic relevant state. Define the bath-averaged interaction by $\overline{\mathcal H}_{\rm SB}={\rm Tr}_{\rm B}\{\hat{\rho}_{{\rm B},{\rm rel}}\hat{\mathcal H}_{\rm SB}\}$, an operator on the electronic Hilbert space. For any electronic observable $\hat A$, Eq.~(\ref{eq:app_organized_v}) gives
\begin{equation}
\begin{aligned}
	v_A&=i\,{\rm Tr}_{\rm SB}\{\hat{\rho}_{\rm rel}^{\rm tot}[\hat{\mathcal H}_{\rm tot},\hat A]\}\\
	&=i\,{\rm Tr}_{\rm S}\{\hat{\rho}_{\rm rel}[\hat{\mathcal H}+\overline{\mathcal H}_{\rm SB},\hat A]\},
\end{aligned}
	\label{eq:app_bath_organized}
\end{equation}
since $[\hat{\mathcal H}_{\rm B},\hat A]=0$. The direct bath contribution to organized motion is therefore determined by the average interaction.

In the main text, we retain the prescribed lattice Hamiltonian $\hat{\mathcal H}$ and assume that its coupling to the reference thermal bath satisfies $\overline{\mathcal H}_{\rm SB}=0$. This is a standard assumption in open-system treatments~\cite{Breuer2002,Rivas2012,Roepke2019}, satisfied naturally by linear coupling to harmonic-bath displacements about equilibrium, as illustrated below. It excludes an additional static bath-average field, but does not set the coupling operator or its fluctuations to zero. Under this assumption, the organized drift becomes
\begin{equation}
	v_A=i\,{\rm Tr}_{\rm S}\{\hat{\rho}_{\rm rel}[\hat{\mathcal H},\hat A]\}.
\end{equation}
Taking $\hat A=\hat{\mathcal H}$ gives $v_E=0$, while taking $\hat A=\hat A_i^\alpha$ yields the intrinsic order-parameter drift derived in the main text.

A thermal free-boson bath illustrates why the zero-average condition need not require fine-tuning. For $\hat{\mathcal H}_{\rm B}=\sum_q\omega_q\hat b_q^\dagger\hat b_q$, the averages satisfy $\langle\hat b_q\rangle_{\rm B}=\langle\hat b_q^\dagger\rangle_{\rm B}=0$. A linear coupling $\hat{\mathcal H}_{\rm SB}=\sum_q\hat S_q\otimes(g_q\hat b_q+g_q^*\hat b_q^\dagger)$, with Hermitian electronic operators $\hat S_q$, is therefore already centered, independently of the coefficients $g_q$ and the choice of $\hat S_q$. The condition follows from the vanishing mean displacement of the thermal bath, rather than from a cancellation imposed on the electronic state. Rapid decay of bath correlations is a separate dynamical requirement.

For more general couplings, the bath average need not vanish. At a fixed reference bath state, a nonzero average can be absorbed into the system Hamiltonian by the standard centering procedure~\cite{Breuer2002,Rivas2012,Roepke2019},
\begin{equation}
	\hat{\mathcal H}'=\hat{\mathcal H}+\overline{\mathcal H}_{\rm SB},\qquad
	\hat{\mathcal H}'_{\rm SB}=\hat{\mathcal H}_{\rm SB}-\overline{\mathcal H}_{\rm SB}\otimes\hat I_{\rm B}.
\end{equation}
The total Hamiltonian is unchanged, while ${\rm Tr}_{\rm B}\{\hat{\rho}_{{\rm B},{\rm rel}}\hat{\mathcal H}'_{\rm SB}\}=0$. This rearrangement redefines the electronic Hamiltonian and hence the energy and constrained variational problem associated with it. For example, a coupling $\hat{\mathcal H}_{\rm SB}=\sum_q g_q\hat S_q\otimes\hat b_q^\dagger\hat b_q$, with real $g_q$, has average $\sum_q g_q n_q\hat S_q$, where $n_q=(e^{\beta_{\rm B}\omega_q}-1)^{-1}$. Centering replaces $\hat b_q^\dagger\hat b_q$ by $\hat b_q^\dagger\hat b_q-n_q$ and includes the average contribution in $\hat{\mathcal H}'$. Thus, zero bath average is an assumption about the coupling to the prescribed $\hat{\mathcal H}$ in the present application, not a requirement of the AVP itself.

Centering does not eliminate fluctuations: for example, $\langle(\hat b_q+\hat b_q^\dagger)^2\rangle_{\rm B}=2n_q+1$ remains nonzero. Nor does factorization of the relevant ensemble imply factorization of the actual evolving state. Bath correlations and system--bath correlations generated by the full dynamics continue to contribute through memory and fluctuating forces. Thus, $v_E=0$ removes only the organized energy drift, not energy exchange or relaxation.

\section{Wave-Function Interpretation of the Reversible Dynamics}
\label{app:reversible_wavefunction}

The reversible tensor derived from nonlinear projection theory in Sec.~\ref{sec:nonlinear-MZ} also follows directly from the constrained wave-function formulation of Sec.~\ref{sec:AVP}. This complementary derivation makes transparent how the microscopic commutation algebra converts the conjugate thermodynamic fields into reversible order-parameter motion.

Consider an instantaneous constrained state $|\Psi[\bm\phi]\rangle$. Under unitary evolution generated by the electronic Hamiltonian, $i\,d|\Psi\rangle/dt=\hat{\mathcal H}|\Psi\rangle$, the Hamiltonian contribution to the order-parameter velocity is
\begin{equation}
	\left(\frac{d\phi_i^\alpha}{dt}\right)_{\!R}
	= i\langle\Psi|[\hat{\mathcal H},\hat A_i^\alpha]|\Psi\rangle.
	\label{eq:app_wavefunction_reversible}
\end{equation}
This expression evaluates the instantaneous Hamiltonian contribution at the constrained state; it does not assume that unitary evolution preserves the constrained manifold. The full AVP dynamics also includes relaxation and fluctuations.

The variational stationarity condition provides a simple way to evaluate Eq.~(\ref{eq:app_wavefunction_reversible}). From Eqs.~(\ref{eq:eigen-eq1}) and (\ref{eq:H_eff1}), the constrained state is an eigenstate of the deformed Hamiltonian $\hat{\mathcal H}^{*}[\bm f]$, so that $\langle\Psi|[\hat{\mathcal H}^{*},\hat A_i^\alpha]|\Psi\rangle=0$. Substituting $\hat{\mathcal H}=\hat{\mathcal H}^{*}-\sum_{j,\beta}f_j^\beta\hat A_j^\beta$ therefore gives
\begin{equation*}
	\left(\frac{d\phi_i^\alpha}{dt}\right)_{\!R}
	= i\sum_{j,\beta}f_j^\beta\langle\Psi|[\hat A_i^\alpha,\hat A_j^\beta]|\Psi\rangle.
\end{equation*}
Once the microscopic state has been optimized at fixed $\bm\phi$, the instantaneous reactive motion of the order parameters is thus determined by the conjugate fields acting through their operator commutators.

Using the closed algebra in Eq.~(\ref{eq:commutator_A}) and the relation $f_i^\alpha=-\partial F/\partial\phi_i^\alpha$ established in Eq.~(\ref{eq:force1}), we obtain
\begin{equation*}
	\left(\frac{d\phi_i^\alpha}{dt}\right)_{\!R}
	= \sum_{\beta,\gamma}C_{\alpha\beta\gamma}\,\phi_i^\gamma\frac{\partial F}{\partial\phi_i^\beta}.
\end{equation*}
Comparison with Eq.~(\ref{eq:dphi_dt1}) reproduces the reversible tensor in Eq.~(\ref{eq:R_tensor}), showing that its Poisson-bracket-like structure follows directly from the microscopic commutation algebra.

The nondissipative character of this motion is explicit in the corresponding rate of change of the constrained functional:
\begin{eqnarray}
	\left(\frac{dF}{dt}\right)_{\!R}
	&=& \sum_{i,\alpha}\frac{\partial F}{\partial\phi_i^\alpha}\left(\frac{d\phi_i^\alpha}{dt}\right)_{\!R} \nonumber\\
	&=& \sum_{i,\alpha,\beta,\gamma}C_{\alpha\beta\gamma}\,\phi_i^\gamma\frac{\partial F}{\partial\phi_i^\alpha}\frac{\partial F}{\partial\phi_i^\beta}=0.
	\label{eq:app_reversible_F}
\end{eqnarray}
The final equality follows from the antisymmetry of $C_{\alpha\beta\gamma}$ under $\alpha\leftrightarrow\beta$, whereas the product of the two thermodynamic forces is symmetric. The reversible flow therefore follows surfaces of constant $F$.

For spin order, this result has an especially simple interpretation. The SU(2) commutation algebra converts the thermodynamic field conjugate to the local magnetization into transverse precessional motion, giving the Landau--Lifshitz structure encountered in Sec.~\ref{sec:Hubbard_SDW}. The projection-theory formulation extends this interpretation to mixed relevant states and incorporates the memory and fluctuating contributions generated by the complementary microscopic sector.

\section{Constrained Hartree--Fock Implementation for the $t$--$V$ Model}
\label{app:tV_HF}

\begin{figure*}[t]
\centering
\includegraphics[width=1.99\columnwidth]{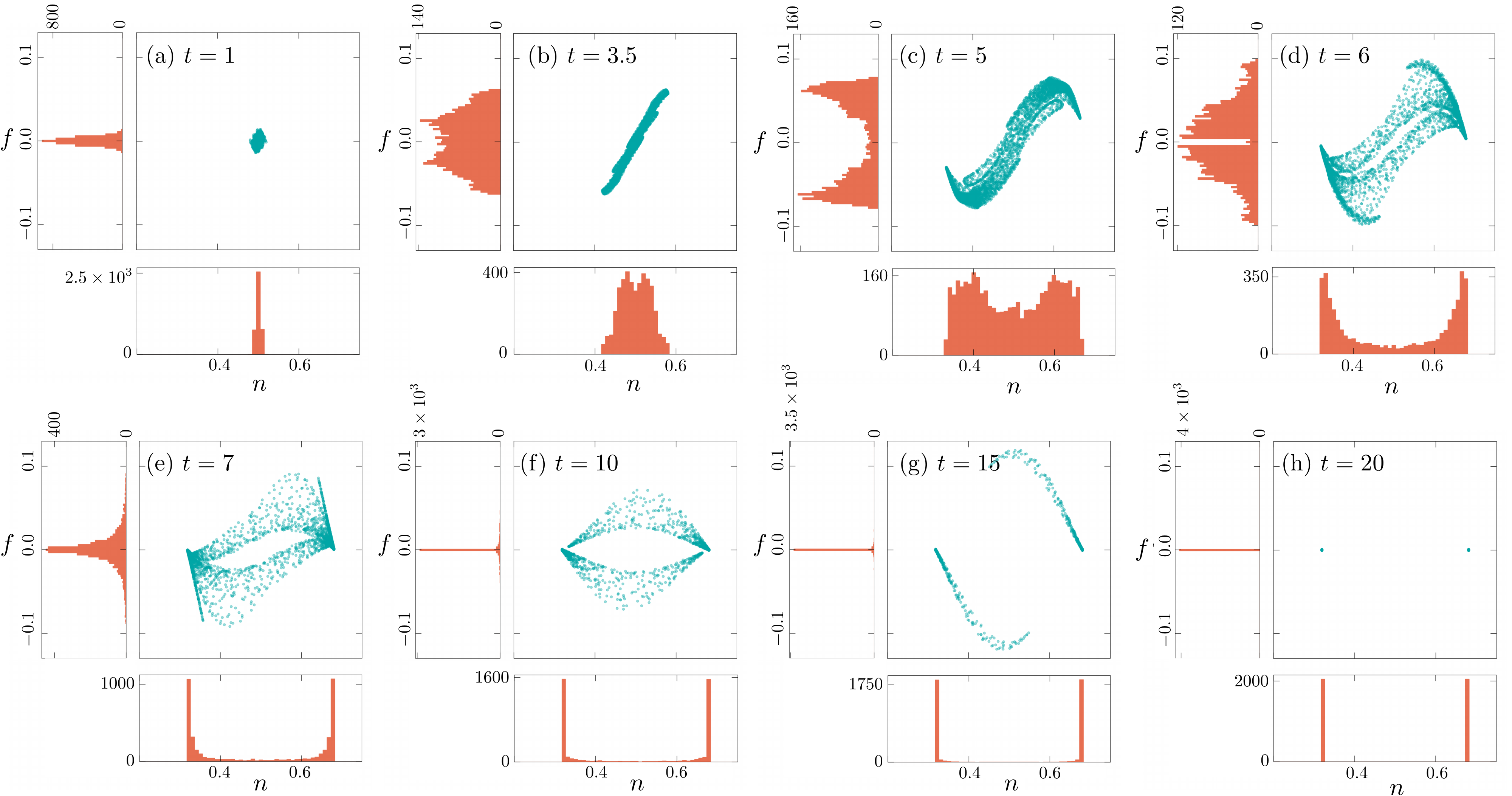}
\caption{Joint force--density distributions during square-lattice CDW ordering. Panels (a)--(h) show snapshots at the indicated times, measured in units of $(\Gamma t_{\rm nn})^{-1}$. Each point represents a lattice site with instantaneous density $n_i$ and conjugate force $f_i=-\partial F/\partial n_i$, where $F$ is the constrained grand potential. The histograms below and to the left of each scatter plot show the site counts in density and force bins, respectively. All distributions are obtained from the converged constrained HF states along the MF-AVP trajectory.}
\label{fig:f_vs_n}
\end{figure*}

Here we provide details of the constrained Hartree--Fock (HF) calculation used for the spinless $t$--$V$ model in Sec.~\ref{sec:CDW}. This calculation implements the inner loop of the MF-AVP workflow illustrated in Fig.~\ref{fig:MF_AVP_scheme}. At each step of the adiabatic dynamics, the instantaneous density variables $n_i$ are held fixed while the electronic density matrix $G$ and the conjugate fields $f_i$ are determined iteratively. Throughout this appendix, $G_{ii}$ denotes the on-site density of the trial electronic state; the equality $G_{ii}=n_i$ is enforced at convergence.

The constrained HF Hamiltonian in Eq.~\eqref{eq:tV_HF} contains both the Hartree contribution to the local potential and the Fock correction to the nearest-neighbor hopping. With $G_{ij}=\langle\hat c_j^\dagger\hat c_i^{\,}\rangle$, it is convenient to combine the Hartree and constraint contributions into a local effective potential,
\begin{equation}
	v_i=f_i+V\sum_{j\in\mathcal N(i)}G_{jj}.
	\label{eq:app_tV_vi}
\end{equation}
Writing the quadratic Hamiltonian as $\hat{\mathcal H}_{\rm HF}=\sum_{ij}H_{ij}\hat c_i^\dagger\hat c_j^{\,}$ gives the single-particle matrix
\begin{equation}
	H_{ij}=-\left(t_{\rm nn}+VG_{ij}\right)\delta_{\langle ij\rangle}+v_i\delta_{ij},
	\label{eq:app_tV_Hij}
\end{equation}
where $t_{\rm nn}$ is the nearest-neighbor hopping amplitude and $\delta_{\langle ij\rangle}=1$ when $i$ and $j$ are nearest neighbors and vanishes otherwise. Updating $v_i$ directly allows the density residual to control the on-site potential, while the Hartree contribution is subtracted after convergence to recover the constraint field $f_i$.

For a prescribed density profile $\bm n$, neither $v_i$ nor the unconstrained components of $G$ are known a priori. At iteration $k$, the current $G^{(k)}$ and $v_i^{(k)}$ define $H^{(k)}$, whose eigenstates satisfy
\begin{equation}
	\sum_j H_{ij}^{(k)}\psi_{m,j}^{(k)}=\epsilon_m^{(k)}\psi_{m,i}^{(k)}.
	\label{eq:app_tV_eigen}
\end{equation}
Here $m$ labels the single-particle eigenstates and $(k)$ labels the iteration. At finite temperature, the density matrix is updated using the Fermi--Dirac occupations,
\begin{equation}
	G_{ij}^{(k+1)}=\sum_m \psi_{m,i}^{(k)}n_{\rm F}\!\left(\epsilon_m^{(k)}-\mu\right) \psi_{m,j}^{(k)*},
	\label{eq:app_tV_Gupdate}
\end{equation}
where $n_{\rm F}(x)=(e^{x/T}+1)^{-1}$. For the square-lattice simulations discussed in the main text, $\mu$ is chosen beforehand at the center of the single-particle gap of the reference ordered CDW state and remains fixed throughout the constrained iteration and the dynamical evolution. The matrix $H$ excludes the chemical-potential term, which enters through the occupations in Eq.~\eqref{eq:app_tV_Gupdate}. At zero temperature, these occupations reduce to a projector onto states below $\mu$, apart from possible degeneracies at the chemical potential.

The updated diagonal elements $G_{ii}^{(k+1)}$ generally differ from the prescribed densities $n_i$. We feed this residual back into the local potentials through
\begin{equation}
	v_i^{(k+1)}=v_i^{(k)}+\alpha_v\left(G_{ii}^{(k+1)}-n_i\right),
	\label{eq:app_tV_vupdate}
\end{equation}
with $\alpha_v>0$. When the calculated density exceeds its prescribed value, increasing the local potential suppresses the occupation in the subsequent iteration, providing negative feedback. This update is the local realization of the constraint-update operation $\mathbb F$ in Fig.~\ref{fig:MF_AVP_scheme}. Its step size controls the stability and rate of convergence; more elaborate mixing, preconditioning, or response-based updates can be used within the same constrained formulation.

The density constraints must be satisfied together with HF self-consistency. In particular, the nearest-neighbor off-diagonal elements $G_{ij}$ remain variational and enter the Hamiltonian through the Fock term. The iteration is therefore continued until both the residuals $G_{ii}-n_i$ and the changes in the unconstrained mean fields are sufficiently small. Denoting the converged values by $\bar v_i$ and $\bar G_{ij}$, the conjugate fields are
\begin{equation}
	f_i=\bar v_i-V\sum_{j\in\mathcal N(i)}\bar G_{jj},
	\label{eq:app_tV_force}
\end{equation}
with $\bar G_{ii}=n_i$. Since the auxiliary Lagrangian includes $-\mu\,\mathrm{Tr}\,G$, these fields satisfy $f_i=-\partial F/\partial n_i$, where $F$ is the constrained grand potential. They enter the density dynamics in Eq.~\eqref{eq:tV_modelA} directly, without separate evaluation or numerical differentiation of $F$. Advancing $\bm n$ by one time step defines the next constrained electronic problem.

Figure~\ref{fig:f_vs_n} illustrates the forces obtained from this procedure through their joint distribution with the instantaneous densities. At $t=1$, the distribution is narrowly concentrated near half filling and zero force. By $t=3.5$, it develops a positive correlation between $f_i$ and $n_i-1/2$, indicating forces that amplify the emerging density modulation. As ordering proceeds through $t=5$--$7$, the scatter broadens and bends while the density histogram develops two peaks associated with the ordered sublattices. At $t=10$--$15$, these peaks are already sharp, although a smaller population of sites continues to experience appreciable forces. By $t=20$, the distribution is concentrated near the two ordered densities with nearly vanishing forces, indicating a nearly stationary configuration for this realization.

The intermediate distributions show that the instantaneous force is not determined by the on-site density alone: sites with comparable $n_i$ can experience different forces because their surrounding configurations differ. The persistence of nonzero forces after the two density populations have formed is consistent with the real-space maps in Fig.~\ref{fig:tV_coarsening}, where domain interiors relax while interfaces continue to evolve. These distributions thus complement the domain-growth analysis by resolving the changing relation between local density and thermodynamic driving force, with both quantities obtained from the constrained microscopic calculation.

\section{Constrained Hartree--Fock and Generalized Landau--Lifshitz Dynamics for the SDW}
\label{app:SDW_MFAVP}

This appendix provides details of the constrained electronic calculation and the stochastic magnetic dynamics used in Sec.~\ref{sec:Hubbard_SDW}. We first describe the spin-dependent Hartree--Fock iteration, distinguishing the prescribed magnetic texture from the moments calculated during the electronic minimization. We then establish the relation between the vector and amplitude--orientation descriptions and specify the associated noise correlations and numerical integration methods.

\subsection{Constrained Hartree--Fock calculation}

The electronic state is characterized by the single-particle density matrix $G_{ij}^{\alpha\beta}=\langle\hat c_{j,\beta}^{\dagger}\hat c_{i,\alpha}^{\,}\rangle$. To distinguish the prescribed magnetic moments $\mathbf m_i$ from those obtained during the constrained iteration, we denote the latter by $\mathbf s_i[G]=(1/2)\operatorname{tr}_{s}(\boldsymbol{\sigma}G_{ii})$, where $\operatorname{tr}_{s}$ is the trace over spin indices. The local density matrix therefore has the decomposition
\begin{equation}
    G_{ii}^{\alpha\beta}=\frac{n_i[G]}{2}\delta_{\alpha\beta}+\mathbf s_i[G]\cdot\boldsymbol{\sigma}_{\alpha\beta},
    \label{eq:app_SDW_localG}
\end{equation}
with $n_i[G]=\operatorname{tr}_{s}G_{ii}$. The constrained calculation requires $\mathbf s_i[G]=\mathbf m_i$ upon convergence, while the charge densities remain variational.

As in the CDW implementation of Appendix~\ref{app:tV_HF}, it is convenient to iterate the combined local field entering the single-particle Hamiltonian. For the Hubbard model, the exchange and constraint contributions appear through
\begin{equation}
    \mathbf h_i=\mathbf b_i-2U\mathbf s_i[G].
    \label{eq:app_SDW_h}
\end{equation}
The effective field $\mathbf h_i$ determines the spin-dependent electronic potential, whereas $\mathbf b_i$ is the thermodynamic field conjugate to the prescribed moment. Writing the quadratic Hamiltonian as $\hat{\mathcal H}_{\rm HF}=\sum_{ij,\alpha\beta}H_{i\alpha,j\beta}\hat c_{i,\alpha}^{\dagger}\hat c_{j,\beta}^{\,}$, its single-particle matrix becomes
\begin{equation}
    H_{i\alpha,j\beta}=-t_{\rm nn}\,\delta_{\langle ij\rangle}\delta_{\alpha\beta}+\delta_{ij}\left[\frac{U n_i}{2}\delta_{\alpha\beta}+\frac{1}{2}\mathbf h_i\cdot\boldsymbol{\sigma}_{\alpha\beta}\right],
    \label{eq:app_SDW_Hmatrix}
\end{equation}
where $\delta_{\langle ij\rangle}=1$ for nearest neighbors and vanishes otherwise. The factor of $1/2$ in the spin-dependent term follows from the definition of the electronic spin operator in Eq.~\eqref{eq:local_spin}.

For a prescribed magnetic texture $\{\mathbf m_i\}$, the effective fields $\mathbf h_i$ and unconstrained charge densities $n_i$ are determined iteratively. At iteration $k$, their current values define $H^{(k)}$, whose eigenstates satisfy
\begin{equation}
    \sum_{j,\beta}H_{i\alpha,j\beta}^{(k)} \psi_{\nu,j\beta}^{(k)}=\epsilon_\nu^{(k)} \psi_{\nu,i\alpha}^{(k)}.
    \label{eq:app_SDW_eigen}
\end{equation}
The updated density matrix is then
\begin{equation}
    G_{ij}^{(k+1),\alpha\beta}=\sum_\nu n_{\rm F}\!\left(\epsilon_\nu^{(k)}-\mu\right) \psi_{\nu,i\alpha}^{(k)} \psi_{\nu,j\beta}^{(k)*},
    \label{eq:app_SDW_G_eigen}
\end{equation}
where $\nu$ labels the single-particle eigenstates, $n_{\rm F}(x)=(e^{\beta x}+1)^{-1}$, and the chemical potential is adjusted to maintain half filling. The local blocks of $G^{(k+1)}$ determine the updated charge densities and calculated spin moments.

The discrepancy between the calculated and prescribed moments is fed back into the effective fields through
\begin{equation}
    h_i^{a,(k+1)}=h_i^{a,(k)}+\kappa_h\left(s_i^a[G^{(k+1)}]-m_i^a\right),\qquad a=x,y,z,
    \label{eq:app_SDW_b_update}
\end{equation}
with $\kappa_h>0$. Because the spin-dependent term is $+\mathbf h_i\cdot\hat{\mathbf s}_i$, increasing $h_i^a$ tends to suppress the corresponding spin expectation value, giving the required negative feedback. The three field components are coupled through the electronic response and must be converged together. During the same iteration, the charge densities entering the Hartree potential are updated from $n_i[G]$. Mixing successive charge densities or effective fields can be used to improve convergence.

The iteration continues until the charge densities are self-consistent and the prescribed magnetic moments are reproduced. Denoting the converged quantities by $\bar{\mathbf h}_i$ and $\bar G$, the thermodynamic fields returned by the constrained calculation are
\begin{equation}
    \mathbf b_i=\bar{\mathbf h}_i+2U\mathbf s_i[\bar G]=\bar{\mathbf h}_i+2U\mathbf m_i.
    \label{eq:app_SDW_force}
\end{equation}
The second equality follows from the converged spin constraints. Thus, the effective fields used to solve the electronic problem are converted into the conjugate fields $\mathbf b_i=-\partial F/\partial\mathbf m_i$, which drive both the precessional and dissipative magnetic dynamics.

\subsection{Relation between vector and amplitude--orientation dynamics}

The decomposition $\mathbf m_i=M_i\hat{\mathbf e}_i$ introduced in Eq.~\eqref{eq:SDW_m_decomp} separates changes in moment magnitude from rotations of its direction. For $M_i>0$, a variation satisfies $\delta\mathbf m_i=\hat{\mathbf e}_i\,\delta M_i+M_i\,\delta\hat{\mathbf e}_i$, with $\hat{\mathbf e}_i\cdot\delta\hat{\mathbf e}_i=0$. Consequently, the force conjugate to the amplitude is
\begin{equation}
    \frac{\partial F}{\partial M_i}=\hat{\mathbf e}_i\cdot\frac{\partial F}{\partial\mathbf m_i}=-\hat{\mathbf e}_i\cdot\mathbf b_i,
    \label{eq:app_dFdM}
\end{equation}
where the derivative is taken at fixed orientations and other amplitudes. The corresponding tangential gradient on the unit sphere is
\begin{equation}
    \nabla_{S^2,i}F=M_i\left(\frac{\partial F}{\partial\mathbf m_i}\right)_{\perp}=-M_i\mathbf b_{i,\perp}.
    \label{eq:app_dFde}
\end{equation}
Both driving forces therefore follow from the same conjugate field $\mathbf b_i$ supplied by the constrained electronic calculation.

The deterministic amplitude and orientation equations follow by projecting the vector dynamics along and perpendicular to the instantaneous moment:
\begin{equation}
    \frac{dM_i}{dt}=\hat{\mathbf e}_i\cdot\frac{d\mathbf m_i}{dt},\qquad
    \frac{d\hat{\mathbf e}_i}{dt}=\frac{1}{M_i}P_{i,\perp}\frac{d\mathbf m_i}{dt},
    \label{eq:app_dm_decomp}
\end{equation}
where $P_{i,\parallel}^{ab}=e_i^a e_i^b$ and $P_{i,\perp}^{ab}=\delta_{ab}-e_i^a e_i^b$. The longitudinal projection eliminates both precession and transverse damping, yielding the Model-A relaxation term $\Gamma_\parallel\hat{\mathbf e}_i\cdot\mathbf b_i=-\Gamma_\parallel\partial F/\partial M_i$. The transverse projection gives precession together with $\Gamma_\perp\mathbf b_{i,\perp}/M_i$. Using $\mathbf b_{i,\perp}=-\hat{\mathbf e}_i\times(\hat{\mathbf e}_i\times\mathbf b_i)$ recovers the Landau--Lifshitz damping term in Eq.~\eqref{eq:SDW_e_dynamics}. The factor $1/M_i$ converts a transverse change of the magnetic moment into a change of its orientation.

With the convention $\mathbf b_i=-\partial F/\partial\mathbf m_i$ and $\hbar=1$, the reversible orientational term is $+\hat{\mathbf e}_i\times\mathbf b_i$, consistently with the microscopic spin commutation algebra. Precession and transverse damping preserve the moment magnitude, while longitudinal relaxation changes it. Although these contributions separate under projection, their evolution remains coupled through the dependence of $\mathbf b_i$ on the full magnetic texture.

\subsection{Thermal fluctuations and numerical integration}

We implement thermal fluctuations directly in the amplitude and orientation equations, using independent longitudinal and transverse noise sources. The amplitude obeys
\begin{equation}
    \frac{dM_i}{dt}=\Gamma_\parallel\hat{\mathbf e}_i\cdot\mathbf b_i+\xi_i(t),
    \label{eq:app_M_dynamics}
\end{equation}
where $\xi_i(t)$ is a zero-mean Gaussian white noise with
\begin{equation}
    \left\langle\xi_i(t)\xi_j(t')\right\rangle=2T\Gamma_\parallel\delta_{ij}\delta(t-t').
    \label{eq:app_M_noise}
\end{equation}
Thus, the constrained electronic field determines the longitudinal relaxation, while the noise strength is set by the temperature and the same kinetic coefficient.

The orientational dynamics is
\begin{equation}
    \frac{d\hat{\mathbf e}_i}{dt}=\hat{\mathbf e}_i\times\mathbf b_i-\frac{\Gamma_\perp}{M_i}\hat{\mathbf e}_i\times\left(\hat{\mathbf e}_i\times\mathbf b_i\right)+\boldsymbol{\zeta}_i(t).
    \label{eq:app_e_dynamics}
\end{equation}
The noise is tangent to the unit sphere and is interpreted in the Stratonovich sense, as appropriate for the stochastic Landau--Lifshitz formulation~\cite{GarciaPalacios1998,Mentink2010}. An explicit representation is $\boldsymbol{\zeta}_i\,dt=\sqrt{2T\Gamma_\perp}\,P_{i,\perp}\circ d\mathbf W_i/M_i$, where $\mathbf W_i$ are independent three-component Wiener processes and $\circ$ denotes a Stratonovich differential. At a given instantaneous configuration, the corresponding covariance is
\begin{equation}
    \left\langle\zeta_i^a(t)\zeta_j^b(t')\right\rangle_{\mathbf m}
    =\frac{2T\Gamma_\perp}{M_i^2}P_{i,\perp}^{ab}\delta_{ij}\delta(t-t').
    \label{eq:app_e_noise}
\end{equation}
The subscript $\mathbf m$ indicates that the configuration-dependent noise amplitude is evaluated at the instantaneous magnetic configuration. The tangential projection and Stratonovich interpretation preserve $|\hat{\mathbf e}_i|=1$, and the transverse driving Wiener processes are independent of those generating $\xi_i$.

At the level of the local noise covariance, the longitudinal and transverse contributions combine as $\boldsymbol{\eta}_i=\hat{\mathbf e}_i\xi_i+M_i\boldsymbol{\zeta}_i$. Their covariance therefore has the fluctuation--dissipation form
\begin{equation}
    \left\langle\eta_i^a(t)\eta_j^b(t')\right\rangle_{\mathbf m}
    =2T\left(\Gamma_\parallel P_{i,\parallel}^{ab}+\Gamma_\perp P_{i,\perp}^{ab}\right)\delta_{ij}\delta(t-t').
    \label{eq:SDW_FDR}
\end{equation}
The factors of $M_i$ in the orientational equation and its noise covariance ensure that $\Gamma_\perp$ retains its meaning as the transverse mobility of the magnetic moment. The stochastic evolution used here is specified by the amplitude--orientation equations above.

The amplitude equation is integrated using a Heun predictor--corrector scheme, while the orientational equation is advanced using the semi-implicit B (SIB) method of Mentink et al.~\cite{Mentink2010}. SIB preserves the length of each orientation vector by construction. The same noise realization is used in the predictor and corrector stages of a given stochastic step. Both methods have second-order accuracy in the deterministic limit; this should be distinguished from their convergence order for stochastic trajectories or averaged observables~\cite{Mentink2010}.

In the simulations presented here, the moment amplitudes remain sharply distributed around their equilibrium magnitude once local moments have developed. We use a small longitudinal relaxation coefficient, with the corresponding noise strength fixed by Eq.~\eqref{eq:app_M_noise}. The physical upper bound $M_i\leq1/2$ is enforced numerically by setting $M_i=1/2$ whenever a proposed amplitude update exceeds this value. This clipping serves as a safeguard against excursions outside the allowed range; the trajectories studied here remain below the upper boundary.

\section{Classical Lattice-Gas Benchmarks for CDW Coarsening}
\label{app:classical_CDW}

\begin{figure*}
\centering
\includegraphics[width=1.99\columnwidth]{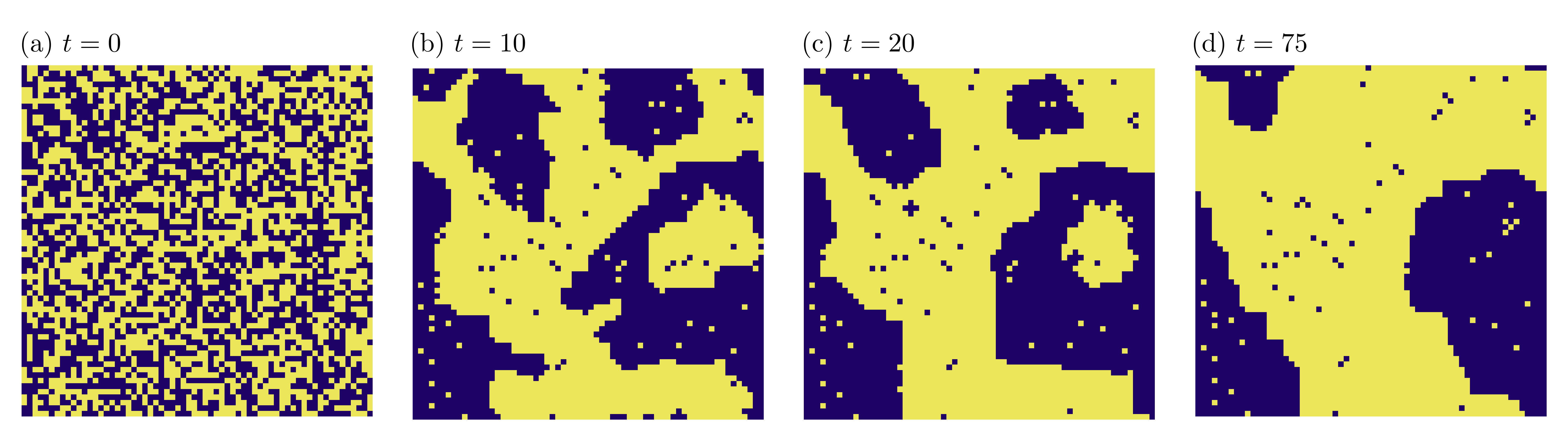}
\caption{Coarsening dynamics of the classical square-lattice model at $t=0$, $10$, $20$, and $75$. Shown is the local checkerboard order parameter $\phi_i$, defined from the local occupations in Eq.~\eqref{eq:local_phi}, including the sublattice factor $(-1)^{x_i+y_i}$. The snapshots illustrate the progressive growth of domains with the two symmetry-related checkerboard orders.}
    \label{fig:V-sq-coarsening}
\end{figure*}

\begin{figure}
\centering
\includegraphics[width=0.98\columnwidth]{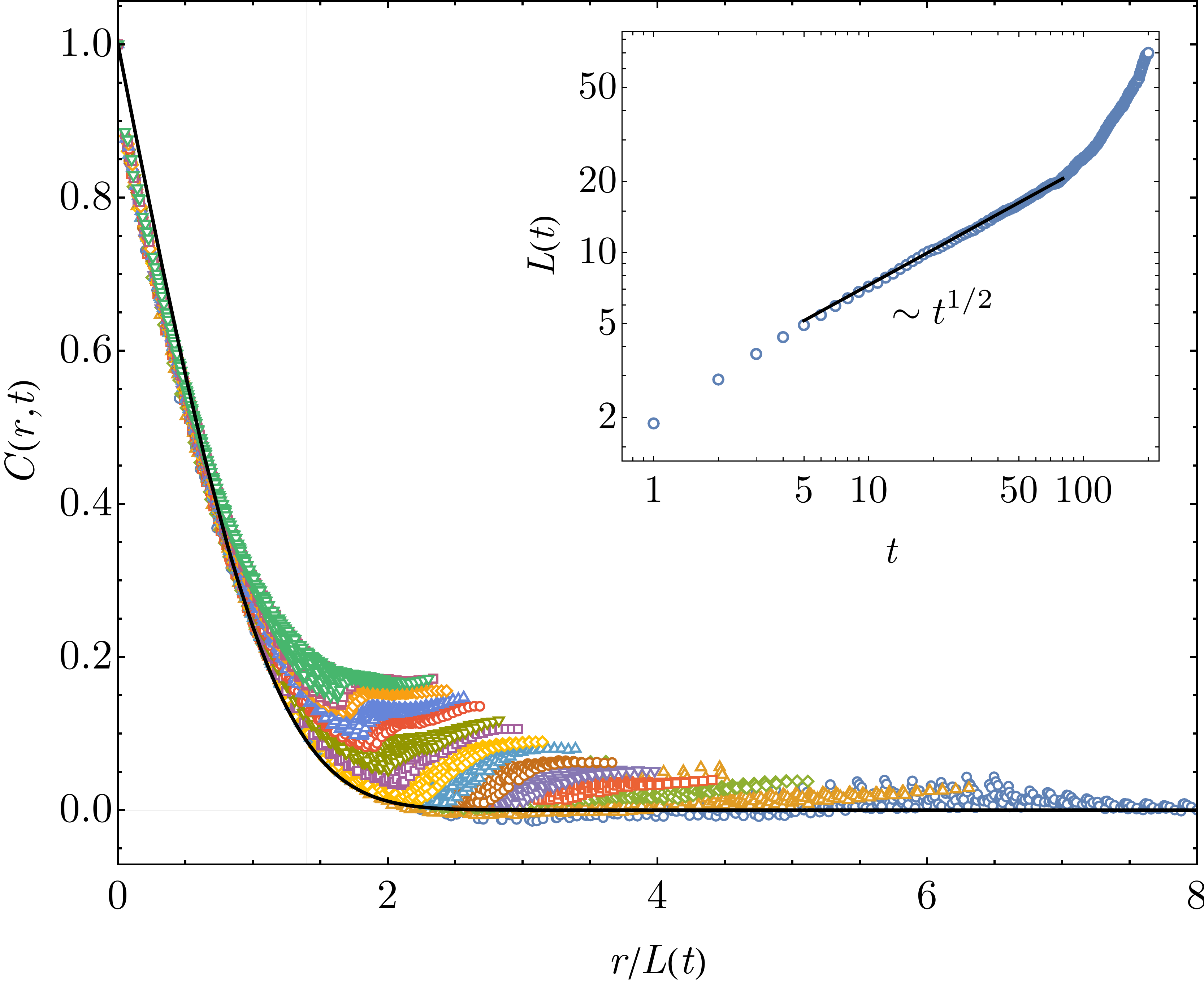}
\caption{Dynamical scaling of the CDW coarsening in the classical square-lattice model. The equal-time order-parameter correlation function $C(r,t)$ at different times is plotted against the rescaled distance $r/L(t)$, showing a good data collapse at short distances. The inset shows the characteristic correlation length $L(t)$, which follows the Allen--Cahn growth law $L(t)\sim t^{1/2}$ over the coarsening regime.}
    \label{fig:V-sq-scaling}
\end{figure}

\begin{figure*}[t]
\centering
\includegraphics[width=1.99\columnwidth]{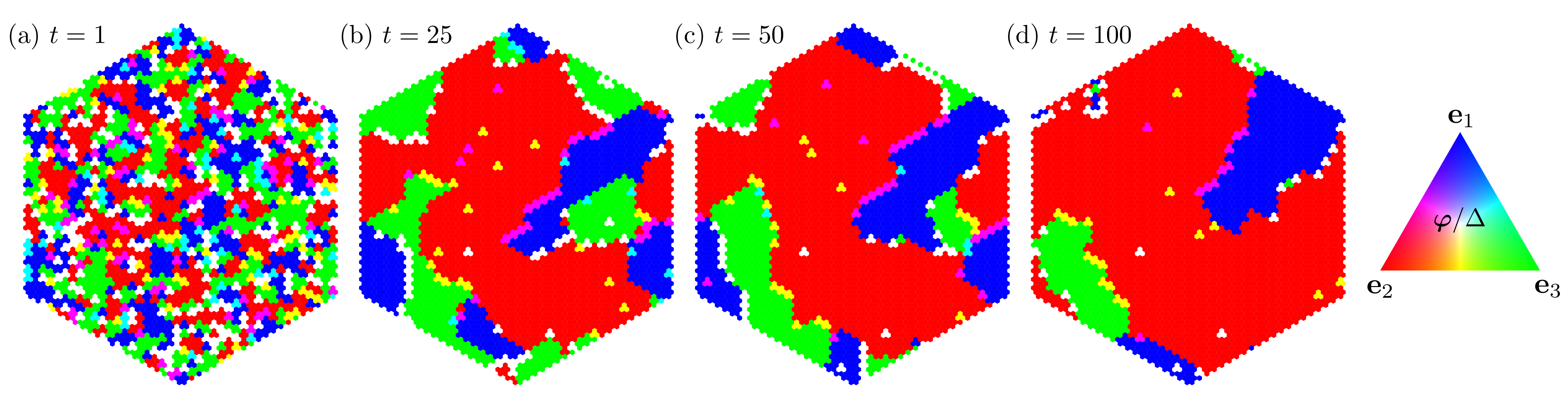}
\caption{Coarsening of the classical lattice-gas model on a $63\times63$ triangular lattice at one-third filling and $T/V=0.1$. Snapshots of the local two-component CDW order parameter are shown at (a) $t=1$, (b) $t=25$, (c) $t=50$, and (d) $t=100$, with time measured in sweeps over all electrons. The field is constructed on elementary triangles as described in Sec.~\ref{sec:tV-tri-CDW}. The triangular color map distinguishes the three symmetry-related $\sqrt{3}\times\sqrt{3}$ states and their local interpolations. Domains develop from the disordered initial configuration and coarsen through the evolution of a network of domain walls and Y-shaped junctions.}
\label{fig:V-tri-coarsening}
\end{figure*}

The nearest-neighbor repulsion responsible for CDW ordering in Sec.~\ref{sec:CDW} also defines a useful classical reference problem. Suppressing the electronic hopping leaves the lattice-gas Hamiltonian
\begin{equation}
    \mathcal H_{\rm cl}=V\sum_{\langle ij\rangle}n_i n_j,\qquad n_i=0,1.
    \label{eq:classical_lattice_gas}
\end{equation}
On the square lattice at half filling and the triangular lattice at one-third filling, this interaction favors the same checkerboard and three-sublattice ordering patterns discussed in Secs.~\ref{sec:tV-square-CDW} and \ref{sec:tV-tri-CDW}, respectively. The classical simulations use particle-conserving local hopping updates, whereas the MF-AVP calculations employ density dynamics at fixed chemical potential. The comparison therefore examines ordering morphology and growth laws under different microscopic kinetics.

The connection to classical spin models follows by introducing Ising variables $\sigma_i=2n_i-1=\pm1$. At fixed particle number, Eq.~\eqref{eq:classical_lattice_gas} is equivalent, up to a configuration-independent constant, to the antiferromagnetic Ising Hamiltonian
\begin{equation}
    \mathcal H_{\rm Ising}=\frac{V}{4}\sum_{\langle ij\rangle}\sigma_i\sigma_j.
    \label{eq:classical_Ising}
\end{equation}
The particle-number constraint becomes a fixed-magnetization condition: $\sum_i\sigma_i=0$ for the half-filled square lattice and $N^{-1}\sum_i\sigma_i=-1/3$ for the one-third-filled triangular lattice, where $N$ is the number of sites. The latter constraint is essential to the comparison. At this filling, the classical ground states are the three $\sqrt{3}\times\sqrt{3}$ configurations, rather than the extensively degenerate ground-state manifold of the unrestricted triangular-lattice antiferromagnetic Ising model.

We simulate the classical model using nearest-neighbor Kawasaki exchange dynamics with heat-bath acceptance. A proposed particle hop exchanges the occupations of a neighboring particle--hole pair, $10\leftrightarrow01$, and is accepted with probability
\begin{equation}
    P_{\rm acc}(\Delta E)=\frac{1}{1+\exp(\beta\Delta E)},
    \label{eq:classical_Glauber}
\end{equation}
where $\Delta E$ is the change in the classical interaction energy and $\beta=1/T$, with $k_{\rm B}=1$. The square and triangular lattices contain $64\times64$ and $63\times63$ sites, respectively, and both simulations use $V=1$ and $T=0.1$. One unit of simulation time corresponds to one sweep, during which every electron is considered once for a hopping update. These updates conserve total particle number but allow occupation to be transferred between the ordering sublattices, so neither the checkerboard nor the three-sublattice CDW order parameter is conserved.

For both lattices, we characterize domain growth using the corresponding order-parameter correlations defined in the main text. The equal-time correlation function is normalized such that $C(0,t)=1$, and the characteristic domain size is extracted from $C(L(t),t)=1/2$. Plotting the correlations against $r/L(t)$ then tests whether the evolving texture is described approximately by a single growing length scale.

\subsection{Square lattice at half filling}
\label{app:classical_square}

At half filling on the square lattice, the interaction energy is minimized by occupying one sublattice and leaving the other empty. The two resulting checkerboard states correspond to the symmetry-related N\'eel states of the Ising representation. Although the exchange dynamics conserves the total density, each particle hop changes the occupation of both sublattices and can therefore relax the checkerboard order. This permits curvature-driven coarsening of the nonconserved $Z_2$ order parameter despite conservation of the underlying particle number.

Figure~\ref{fig:V-sq-coarsening} shows representative configurations of the local checkerboard order parameter $\phi_i$, obtained from the occupation variables using Eq.~\eqref{eq:local_phi}. The sublattice factor $(-1)^{x_i+y_i}$ included in this definition makes the two ordered states distinguishable by the sign of $\phi_i$. The initially disordered state develops checkerboard domains that subsequently grow and merge. Their overall morphology resembles the MF-AVP configurations in Fig.~\ref{fig:tV_coarsening}, although the classical interfaces are microscopically rougher and isolated local defects persist within otherwise ordered regions over the displayed time interval. These defects are local occupation excitations whose removal requires particle rearrangements through the allowed nearest-neighbor exchanges. The differences in interface structure and defect relaxation reflect the distinct microscopic descriptions and kinetic rules of the two calculations.

The normalized correlation functions of $\phi_i$ in Fig.~\ref{fig:V-sq-scaling} exhibit an approximate collapse when plotted against $r/L(t)$. The characteristic length shown in the inset grows as $L(t)\sim t^{1/2}$ over the observed coarsening regime, consistent with Allen--Cahn behavior. Thus, the classical exchange dynamics and the microscopic MF-AVP evolution recover the same growth law despite their different update rules and interface structures. This comparison supports the interpretation of the square-lattice MF-AVP results as conventional coarsening of the checkerboard order generated by the repulsive interaction.

\begin{figure}[t]
\centering
\includegraphics[width=0.98\columnwidth]{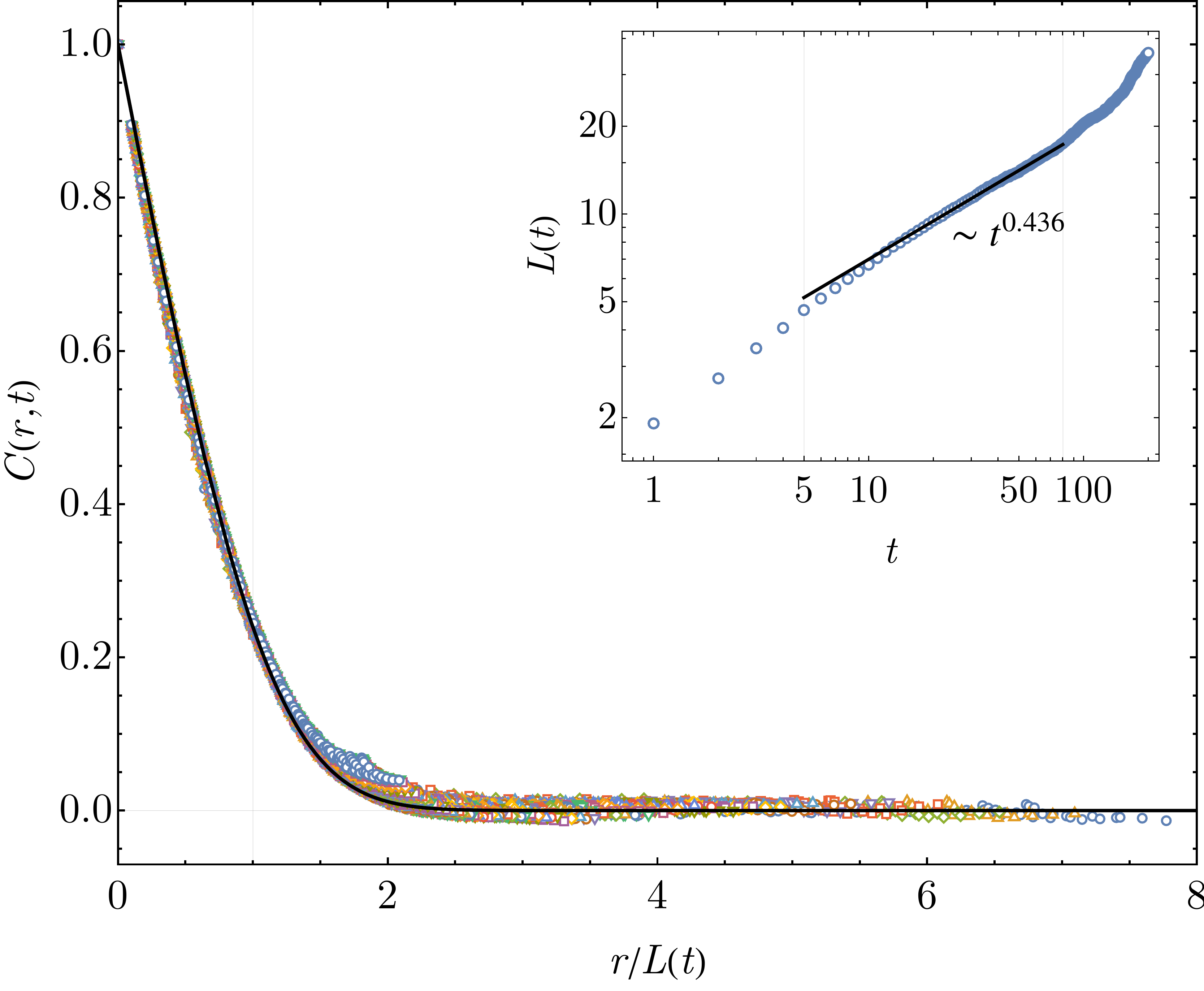}
\caption{Dynamical scaling of the classical triangular-lattice gas at one-third filling. The normalized equal-time correlation functions, $C(0,t)=1$, are plotted against $r/L(t)$, with $L(t)$ defined by $C(L(t),t)=1/2$. Their approximate collapse indicates scaling over the observed coarsening regime. Inset: characteristic domain size versus time, measured in sweeps over all electrons. The solid line is a power-law fit $L(t)\sim t^{0.436}$ over the intermediate-time regime. The system size and temperature are the same as in Fig.~\ref{fig:V-tri-coarsening}.}
\label{fig:V-tri-scaling}
\end{figure}

\subsection{Triangular lattice at one-third filling}
\label{app:classical_triangular}

At one-third filling on the triangular lattice, occupying any one of the three ordering sublattices eliminates all occupied nearest-neighbor pairs. The three possible choices give the symmetry-related $\sqrt{3}\times\sqrt{3}$ ground states and the associated $Z_3$ order. As on the square lattice, particle-conserving hopping updates can change the sublattice populations and therefore relax the CDW order. The additional ordered state allows three distinct domains to meet, producing the junction-rich textures characteristic of this problem.

Representative configurations are shown in Fig.~\ref{fig:V-tri-coarsening}. Local charge order develops from the initially disordered state, followed by the growth of domains separated by an extended network of walls and Y-shaped junctions. The local two-component field is reconstructed from the occupation variables using the same triangle-based definition as in Sec.~\ref{sec:tV-tri-CDW}. The resulting domain network closely resembles that of the electronic MF-AVP calculation, while the classical interfaces remain rougher and residual occupation defects persist within some ordered regions. These snapshots show that the repulsive lattice-gas interaction alone produces the principal geometrical structures of the three-state CDW texture.

Figure~\ref{fig:V-tri-scaling} quantifies this evolution through the normalized correlation functions. Their approximate collapse against $r/L(t)$ indicates that a single characteristic length describes the domain structure over the accessible coarsening regime. The inset gives an effective growth law $L(t)\sim t^{0.436}$, slower than the square-lattice result. This behavior parallels the reduced effective exponent $\alpha\simeq0.415$ found in the electronic triangular-lattice calculation, showing that slower intermediate-time growth can also occur in the absence of electronic hopping.

The similar trends suggest that the three-state domain geometry and the relaxation of its junction network are relevant to both calculations, although they do not establish a common microscopic slowing mechanism. Neither fitted exponent should be identified with a distinct asymptotic growth law on the basis of the present simulations. Junction rearrangements and long-lived configurations can contribute to an extended crossover regime, while finite system size limits the late-time analysis. The classical benchmark therefore provides a reference for interpreting the electronic results while leaving open whether their slower growth eventually crosses over to conventional Allen--Cahn behavior.

\bibliography{ref.bib}

\end{document}